\documentclass[a4paper,11pt]{article}
\usepackage{tablefootnote}
\usepackage[english]{babel}
\usepackage[utf8x]{inputenc}
\usepackage[T1]{fontenc}
\usepackage[flushmargin]{footmisc}
\usepackage{setspace}
\usepackage[comma]{natbib}
\usepackage{array}
\usepackage{float}
\usepackage{amsmath}
\usepackage{amsfonts}
\usepackage{amssymb}
\usepackage{ae}
\usepackage{adjustbox}
\usepackage{caption}
\usepackage{pgfplotstable}
\usepackage{booktabs}
\pgfplotsset{compat=1.10}

\usepackage{filecontents}

\usepackage{array}
\newcolumntype{H}{>{\setbox0=\hbox\bgroup}c<{\egroup}@{}}

\usepackage[a4paper,left=2.5cm,right=2.5cm,top=3cm,bottom=3cm]{geometry}
\usepackage{graphicx}
\usepackage[colorinlistoftodos]{todonotes}
\usepackage[colorlinks=true, allcolors=blue]{hyperref}
\begin{document} \doublespacing \pagestyle{plain}

\def\ci{\perp\!\!\!\perp}
\begin{center}

{\LARGE Double machine learning for sample selection models}

{\large \vspace{0.5cm}}

{\large Michela Bia*, Martin Huber**, and Luk\'{a}\v{s} Laff\'{e}rs+ }\medskip

{\small {*Luxembourg Institute of Socio-Economic Research and University of Luxembourg\\ **University of Fribourg, Dept.\ of Economics and\\Center for Econometrics
		and Business Analytics, St.\ Petersburg State University \\ +Matej Bel University, Dept. of Mathematics} \bigskip }
\end{center}

\smallskip

\noindent \textbf{Abstract:} {\footnotesize This paper considers the evaluation of discretely distributed treatments when outcomes are only observed for a subpopulation due to sample selection or outcome attrition. For identification, we combine a selection-on-observables assumption for treatment assignment with either selection-on-observables or instrumental variable assumptions concerning the outcome attrition/sample selection process. We also consider dynamic confounding, meaning that covariates that jointly affect sample selection and the outcome may (at least partly) be influenced by the treatment. To control in a data-driven way for a potentially high dimensional set of pre- and/or post-treatment covariates, we adapt the double machine learning framework for treatment evaluation to sample selection problems. We make use of (a) Neyman-orthogonal, doubly robust, and efficient score functions, which imply the robustness of treatment effect estimation to moderate regularization biases in the machine learning-based estimation of the outcome, treatment, or sample selection models and (b) sample splitting (or cross-fitting) to prevent overfitting bias. We demonstrate that the proposed estimators are asymptotically normal and root-n consistent under specific regularity conditions concerning the machine learners and investigate their finite sample properties in a simulation study. We also apply our proposed methodology to the Job Corps data for evaluating the effect of training on hourly wages which are only observed conditional on employment. The estimator is available in the \textit{causalweight} package for the statistical software \textsf{R}. 
}

{\small \smallskip }



{\small \smallskip {\tiny We have benefited from comments by Alyssa Carlson, David Kaplan, Peter Mueser, and seminar participants at the University of Missouri.
Addresses for correspondence: Michela Bia, Luxembourg Institute of Socio-Economic Research, 11 Porte des Sciences, Maison des Sciences Humaines, 4366 Esch-sur-Alzette/Belval, Luxembourg, michela.bia@liser.lu, michela.bia@ext.uni.lu; Martin Huber, University of Fribourg, Bd.\ de P\'{e}rolles 90, 1700 Fribourg, Switzerland, martin.huber@unifr.ch; Luk\'{a}\v{s} Laff\'{e}rs, Matej Bel University, Tajovskeho 40, 97411 Bansk\'{a} Bystrica, Slovakia, lukas.laffers@gmail.com.
Laff\'{e}rs acknowledges support provided by the Slovak Research and Development Agency under contract no. APVV-17-0329 and VEGA-1/0692/20.
}\thispagestyle{empty}\pagebreak  }

{\small \renewcommand{\thefootnote}{\arabic{footnote}} %
\setcounter{footnote}{0}  \pagebreak \setcounter{footnote}{0} \pagebreak %
\setcounter{page}{1} }

\section{Introduction}\label{intro}

In many studies aiming at evaluating the causal effect of a treatment or policy intervention, the empirical analysis is complicated by non-random outcome attrition or sample selection. Examples include the estimation of the returns to education when wages are only observed for the selective subpopulation of working individuals or the effect of educational interventions like vouchers for private schools on college admissions tests when students non-randomly abstain from the test. Furthermore, in observational studies, treatment assignment is typically not random, implying that the researcher faces a double selection problem, namely selection into the treatment and observability of the outcome. A large literature addresses treatment selection by assuming a selection-on-observables assumption, implying that treatment is as good as randomly assigned conditional on observed pre-treatment covariates, see for instance the reviews by \cite{Im04} and \cite{ImWo08}. Furthermore, a growing number of studies addresses the question of how to control for the crucial confounders in a potentially high-dimensional vector of covariates in a data-driven way based on machine learning algorithms, see for instance the double machine learning framework of \cite{Chetal2018}.

In this paper, we adapt the double machine learning framework to the evaluation of binary or multiply discrete treatments in the presence of sample selection or outcome attrition. In terms of identifying assumptions, we combine a selection-on-observables assumption for the treatment assignment with either selection-on-observables or instrumental variable assumptions concerning the outcome attrition/sample selection process. Such assumptions have previously been considered in \cite{Hu11} and \cite{Hu11b} for the estimation of the average treatment effect (ATE) based on inverse probability weighting, however, for pre-selected (or fixed) covariates. As methodological advancement, we derive doubly robust and efficient score functions for evaluating treatment effects under double selection and demonstrate that they satisfy so-called \cite{Neyman1959} orthogonality. The latter property permits controlling for covariates in a data-driven way by machine learning-based estimation of the treatment, outcome, and attrition models under specific conditions. Therefore, the subset of important confounders need not be known a priori (but must be contained in the  total set of covariates), which is particularly useful in high dimensional data with a vast number of covariates that could potentially serve as control variables. We also consider dynamic confounding based on a sequential selection-on-observables assumption that is closely related to assumptions found in the dynamic treatment effect literature as e.g.\ in  \cite{Ro86}, \cite{Robins1998}, and \cite{Lech09}. This assumption permits that covariates that jointly affect sample selection and the outcome may themselves be a function of the treatment, a scenario widely neglected in sample selection models despite its likely relevance in empirical applications. In particular when there is a substantial time lag between treatment assignment and the sample selection process, exploiting post-treatment covariates to tackle selection-outcome confounding seems more convincing than solely relying on pre-treatment covariates (as in conventional selection-on-observables assumptions) for addressing both treatment endogeneity and sample selection.

Following \cite{Chetal2018}, we show that treatment effect estimation based on our score functions (that are tailored to the various identifying assumptions) is root-$n$ consistent and asymptotically normal under particular regularity conditions, in particular the $n^{-1/4}$-convergence of the machine learners. A further condition in the double machine learning framework is the prevention of overfitting bias due to correlations between the various estimation steps. This is obtained by estimating the treatment, outcome, and selection models on the one hand and the treatment effect on the other hand in different parts of the data. As in \cite{Chetal2018}, we subsequently swap the roles of the data parts and average over treatment effects in order to prevent asymptotic efficiency losses, a procedure known as cross-fitting. We also provide a simulation study suggesting that our estimators perform decently in terms of the root mean squared error and coverage (by confidence intervals) in the simulation designs with several thousand observations considered. Finally, we present an empirical illustration considering the female sample of a study on Job Corps, a large training program for disadvantaged youth in the U.S. We apply our DML estimators to assess the effects of academic and vocational training on hourly wage, which is only observed conditional on employment, one and four years after program assignment and find some statistical evidence for positive longer-run impacts.

Our paper is related to a range of studies tackling sample selection and selective outcome attrition. One strand of the literature models the attrition process based on a selection-on-observables assumption also known as missing at random (MAR) condition. The latter imposes conditional independence of sample selection and the outcome given observed information like the covariates and the treatment. Examples include \cite{Ru76b}, \cite{LittleRubin87}, \cite{Carroll+95}, \cite{Shah+97}, \cite{FitzgeraldGottschalkMoffitt98a}, \cite{Abowd+01}, \cite{Wooldridge02}, and \cite{Wo07}. \cite{Robins+94}, \cite{RobinsRotnitzkyZhao95}, and \cite{BaRo05} discuss doubly robust estimators of the outcome that are consistent under MAR when either the conditional outcome or the attrition model are correctly specified. This approach satisfies Neyman orthogonality as required for double machine learning.\footnote{Relatedly, \cite{BarnwellChaudhuri2020} consider several outcome periods under a monotonic MAR assumption (i.e.\ outcome attrition being an absorbing state weakly increasing over time) and also discuss the evaluation of randomly assigned treatments in this context based on the efficient influence function. In contrast, our framework considers a single outcome period and permits selection into treatment to be related to observed confounders.} However, their framework does not consider double selection into treatment and the observability of the outcome at the same time as we do in this paper.

\cite{Negi2020} suggests an alternative estimator under double selection that falls into the weighted M-estimation framework described in \cite{slonwooldridge2018} and also satisfies doubly robustness, i.e.\ remains consistent under parametric misspecification of either the conditional outcome model  or the  treatment and selection models. This approach based on reweighting outcome models is nevertheless different to ours making use of efficient influence functions and to the best of our knowledge, \cite{Neyman1959} orthogonality (as required for double machine learning) has not been shown for weighted M-estimation (while we prove this property for our proposed estimators). A further difference is that \cite{Negi2020} focuses on treatment evaluation when controlling for pre-treatment covariates to tackle double selection, while we in addition consider identification based on both pre- and post-treatment covariates (dynamic confounding) or an instrument for sample selection.

In contrast to MAR-based identification, so-called sample selection or nonignorable nonresponse models allow for unobserved confounders of the attrition process and the outcome. Unless strong functional form assumptions as in \cite{Heck76}, \cite{Heckman79}, \cite{Hausman79}, and \cite{Little1995} are imposed, identification requires an instrumental variable (IV) for sample selection.  We refer to \cite{Daneve03}, \cite{Ne07}, \cite{Hu11}, and \cite{Hu11b} for nonparametric estimation approaches in this context. To the best of our knowledge, this study is the first one to propose a doubly robust treatment effect estimator under nonignorable outcome attrition and to consider machine learning techniques to control for (possibly high-dimensional) covariates in this context. Our estimators are available in the \textit{causalweight} package for \textsf{R} by \citet{BodoryHuber2018}.


This paper proceeds as follows. Using the potential outcome framework, Section \ref{marident} discusses the identification of the average treatment effect when outcomes are assumed to be missing at random (i.e.\ selection is on observables, as for the treatment) conditional on pre-treatment covariates. Section \ref{ivident} considers identification when outcome attrition is related to unobservables, known as nonignorable nonresponse, and an instrument is available for tackling this issue. Section \ref{dynsel} demonstrates identification under a sequential selection-on-observables which allows for dynamic confounding, meaning that outcomes are assumed to be missing at random conditional on pre- and post-treatment covariates. Section \ref{section:CrossFitting} proposes an estimator based on double machine learning and shows root-n consistency and asymptotic normality under specific regularity conditions.  Section \ref{Sim} provides a simulation study. Section \ref{Application} presents an empirical application to data from the US Job Corps Study.  Section \ref{conclusion} concludes.

\section{Identification under missingness at random}\label{marident}

Our target parameter is the average treatment effect (ATE) of a binary or multiply discretely distributed treatment variable $D$ on an outcome variable $Y$. To define the effect of interest, we use the potential outcome framework, see \cite{Rubin74}. Let  $Y(d)$ denote the potential outcome under hypothetical treatment assignment $d$ $\in$ $\{0,1,...,Q\}$, with $0$ indicating non-treatment and $1,...,Q$ the different treatment choices (where $Q$ denotes the number of non-zero treatments). The ATE when comparing two distinct treatment $d\neq d'$ then corresponds to $\Delta=E[Y(d)-Y(d')]$. Furthermore, let $Y$ denote the outcome realized under the treatment (f)actually assigned to a subject, i.e.\ $Y=Y(D)$. Therefore, $Y$ corresponds to the potential outcome under the treatment received, while the potential outcome under any counterfactual treatment assignment remains unknown.  A further complication in our evaluation framework is that $Y$ is assumed to be only observed for a subpopulation, i.e. conditional on $S=1$, where $S$ is a binary variable indicating whether $Y$ is observed/selected, or not.

Empirical examples with partially observed outcomes include wage regressions, with $S$ being an employment indicator, see for instance  \cite{Gr74}, or the evaluation of the effects of policy interventions in education on test scores, with $S$ being participation in the test, see \cite{AnBeKr04}. Throughout our discussion, $S$ is permitted to be a function of $D$ and $X$, i.e.\ $S=S(D,X)$. However, $S$ must neither be affected by nor affect $Y$.\footnote{See for instance \cite{Imai09} for alternative assumptions,  which imply that selection is associated with the outcome but is independent of the treatment conditional on the outcome and other observable variables.} Therefore, selection per se does not causally influence the outcome. The following nonparametric outcome and selection models satisfy this framework:
\begin{eqnarray}\label{npmodel}
Y=\phi(D,X,U),\quad S=\psi(D,X,V),
\end{eqnarray}
where $U,V$ are unobserved characteristics and $\phi, \psi$ are general functions.\footnote{Note that $Y(d)=\phi(d,X,U)$, which means that fixing the treatment yields the potential outcome.} Throughout the paper we assume that the stable unit treatment value assumption (SUTVA, \cite{Rubin80}) holds such that $\Pr(D = d \implies Y = Y(d))=1$  This rules out interaction or general equilibrium effects and implies that the treatment is uniquely defined.

We subsequently formalize the assumptions that permit identifying the average treatment effect when both selection into the treatment and outcome attrition is related to observed characteristics.
\vspace{5pt}\newline
\textbf{Assumption 1 (conditional independence of the treatment):}\newline
$Y(d)  \bot D | X=x$ for all $d \in \{0,1,...,Q\}$ and $x$ in the support of $X$.\vspace{5pt}\newline
By Assumption 1, there are no unobservables jointly affecting the treatment and the  outcome conditional on covariates $X$. For model \eqref{npmodel}, this implies that $U$ is not associated with unobserved terms affecting $D$ given $X$. In observational studies, the plausibility of this assumption crucially hinges on the richness of the data, while in  experiments, it is satisfied if the treatment is randomized within strata defined by $X$ or randomized independently of $X$.
\vspace{5pt}\newline
\textbf{Assumption 2 (conditional independence of selection):}\newline
$ Y \bot  S | D=d, X=x  $ for all $d \in \{0,1,...,Q\}$ and $x$ in the support of $X$. \vspace{5pt}\newline
By Assumption 2, there are no unobservables jointly affecting selection and the outcome conditional on $D,X$, such that outcomes are missing at random (MAR) in the denomination of \cite{Ru76b}. Put differently, selection is assumed to be selective w.r.t.\ observed characteristics only. For model \eqref{npmodel}, this implies that $U$ and $V$ are conditionally independent given $D,X$.
\vspace{5pt}\newline
\textbf{Assumption 3 (common support):}\newline
(a) $\Pr(D=d| X=x)>0$ and (b) $\Pr(S=1| D=d, X=x)>0$ for all $d \in \{0,1,...,Q\}$ and $x$ in the support of $X$.\vspace{5pt}\newline
Assumption 3(a) is a common support restriction requiring that the conditional probability to receive a specific treatment given $X$, henceforth referred to as treatment propensity score, is larger than zero in either treatment state. Assumption 3(b) requires that for any combination of $D,X$, the conditional probability to be observed, henceforth referred to as selection propensity score, is larger than zero. Otherwise, the outcome is not observed for some specific combinations of these variables implying yet another common support issue.

\begin{figure}[!htp]
	\centering \caption{\label{figuremar}  Causal paths under the missing at random assumption}
\begin{center}
\centering 	\includegraphics[width=1.2\textwidth]{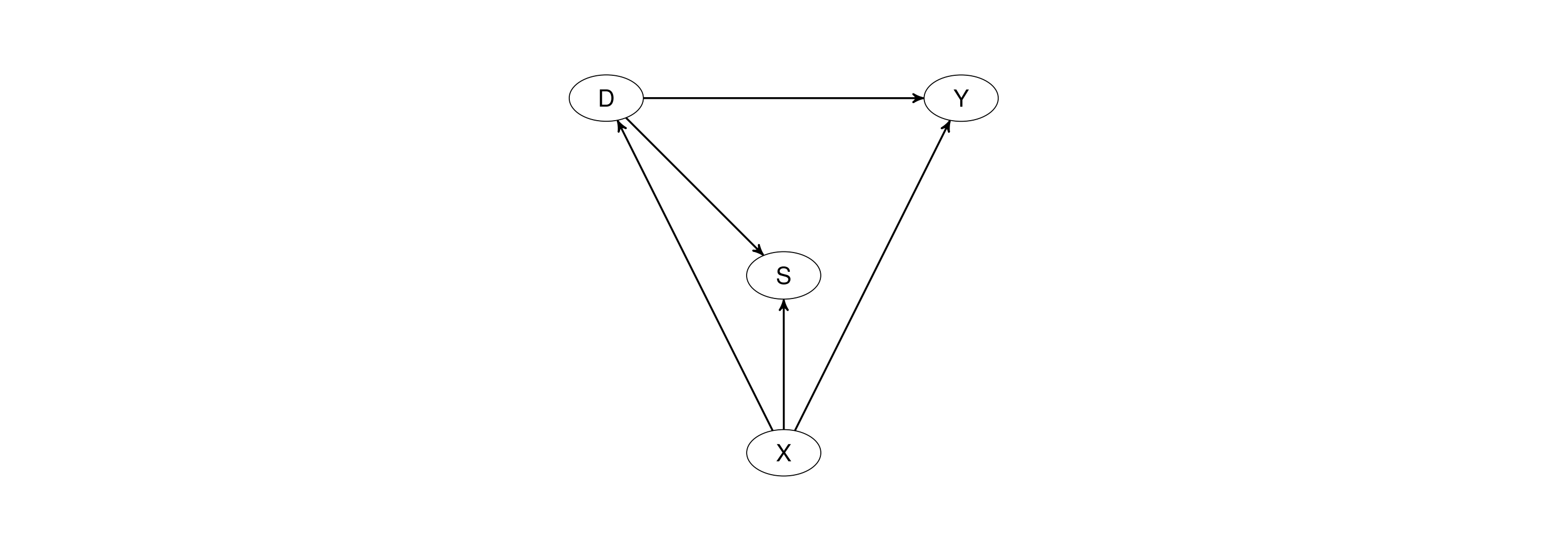}
\end{center}
\end{figure}

Figure \ref{figuremar} provides a graphical illustration of our identification setup using a directed acyclic graph, with arrows representing causal effects. Each of $D$, $S$, and $Y$  might be causally affected by distinct and statistically independent sets of unobservables not displayed in Figure \ref{figuremar}, but none of these unobservables may jointly affect $D$ and $Y$ given $X$ or $S$ and $Y$ given $D$ and $X$.\vspace{5pt}\newline

Our identifying assumptions imply that
\begin{eqnarray}\label{condindep}
E[Y(d)| X]=E[Y| D=d, X]=E[Y| D=d,S=1, X],
\end{eqnarray}
where the first equality follows from Assumption 1 and the second equality from Assumption 2. Therefore, the mean potential outcome identified by
\begin{eqnarray}\label{identreg}
E[Y(d)]=E[E[Y| D=d,S=1, X]],
\end{eqnarray}
or, using the fact that $E[Y| D=d,S=1, X]=E[I\{D=d\} \cdot S \cdot  Y | X]/\Pr(D=d, S=1|X)$, by
\begin{eqnarray}\label{identipw}
E[Y(d)]=E\Bigg[\frac{E[Y\cdot I\{D=d\} \cdot S | X]}{\Pr(D=d, S=1|X)}\Bigg]=E\Bigg[\frac{I\{D=d\} \cdot  S \cdot Y}{ \Pr(D=d|X) \cdot \Pr(S=1| D=d, X)}\Bigg],
\end{eqnarray}
where the second equality follows from the law of iterated expectations. $I\{\cdot\}$ denotes the indicator function, which is equal to one if its argument is satisfied and zero otherwise. Division by $\Pr(D=d|X) \cdot \Pr(S=1| D=d, X)$ in \eqref{identipw} also demonstrates the importance of Assumption 3 for nonparametric identification. For the sake of brevity, we henceforth denote by $\mu(D,S,X)=E[Y| D ,S , X]$ the conditional mean outcome and by $p^d(X)=\Pr(D=d|X)$ and $\pi(D,X)=\Pr(S=1| D, X)$ the propensity scores. Expressions \eqref{identreg} and \eqref{identipw} suggest that the mean potential outcomes (and thus, the ATE) are identified, either based on conditional mean outcomes or inverse probability weighting using the treatment and selection propensity scores.

Following the literature on doubly robust methods, see e.g.\ \cite{RobinsMarkNewey1992}, \cite{Robins+94}, and \cite{RobinsRotnitzkyZhao95}, we combine both approaches to obtain the following identification result:
\begin{eqnarray}\label{score}
E[Y(d)]&=&E\Big[\psi_{d} \Big],\textrm{ where}\notag\\
\psi_{d}&=&  \frac{I\{D=d\} \cdot S \cdot [Y- \mu(d,1,X)] }{ p_d(X)\cdot  \pi (d,X)}  + \mu(d,1,X).
\end{eqnarray}
The result in \eqref{score} is based on the so-called efficient score function, which is formally derived in Appendix \ref{effinf} following the approach outlined in \cite{Levy2019}. By noting that
\begin{eqnarray}\label{identDR}
&&E\Bigg[\frac{I\{D=d\} \cdot S \cdot [Y- \mu(d,1,X)] }{ p_d(X)\cdot  \pi (d,X)}\Bigg]=E\Bigg[\frac{E[I\{D=d\} \cdot S \cdot [Y- \mu(d,1,X)]|X] }{ p_d(X)\cdot  \pi (d,X)}\Bigg]\notag\\
&=&E [ E[ Y- \mu(d,1,X)|D=d, S=1, X] ]=E [ E[Y|D=d, S=1, X]- \mu(d,1,X) ]\notag\\
&=&E [ \mu(d,1,X)-\mu(d,1,X) ]=0,
\end{eqnarray}
it is easy to see that \eqref{score} is equivalent to \eqref{identreg} and thus, \eqref{identipw}. In contrast to \eqref{identreg} and \eqref{identipw}, however, expression \eqref{score} is doubly robust in the sense that it identifies $E[Y(d)]$ if either the conditional mean outcome $\mu(d,1,X)$ or the propensity scores  $p_d(X)$ and $\pi (d,X)$ are correctly specified. Furthermore, it  satisfies the so-called \cite{Neyman1959} orthogonality, i.e.\ is first-order insensitive to perturbations in  $\mu(D,S,X)$, $p_d(X)$, and $\pi (D,X)$, see Appendix \ref{Neyman1}. This entails desirable robustness properties when using machine learning to estimate the outcome, treatment, and selection models  in a data-driven way.

\section{Identification under nonignorable nonresponse}\label{ivident}

When sample selection or outcome attrition is related to unobservables even conditional on observables, identification generally requires an instrument for $S$. We therefore replace Assumptions 2 and 3, but maintain Assumption 1 (i.e.\ selection into treatment is on observables). \vspace{5pt}\newline
\textbf{Assumption 4 (Instrument for selection):}\newline
(a) There exists an instrument $Z$ that may be a function of $D$, i.e. $Z=Z(D)$, is conditionally correlated with $S$, i.e.\ $E[Z\cdot S | D,X]\neq 0$, and satisfies (i) $Y(d,z)=Y(d)$ and (ii) $Y \bot Z | D=d, X=x$  for all $d \in \{0,1,...,Q\}$ and $x$ in the support of $X$,\\
(b) $S=I\{V \leq \chi(D,X,Z)\}$, where $\chi$ is a general function and $V$ is a scalar (index of) unobservable(s) with a strictly monotonic cumulative distribution function conditional on $X$,\\
(c) $V \bot (D,Z)|X$. 
\vspace{5pt}\newline
Assumption 4 no longer imposes the conditional independence of $Y$ and $S$ given $D,X$. As the unobservable $V$ in the selection equation is allowed to be associated with unobservables affecting the outcome, Assumptions 1 and 2 generally do not hold conditional on $S=1$ due to the endogeneity of the post-treatment variable $S$. In fact, $S=1$ implies that $\chi(D,X,Z)> V$ such that conditional on $X$, the distribution of $V$ generally differs across values of $D$. This entails a violation of the conditional independence of $D$ and $Y(d)$ given $S=1$ and $X$ if the potential outcome distributions differ across values of $V$. We therefore require an instrumental variable denoted by $Z$, which must not affect $Y$ or be associated with unobservables affecting $Y$ conditional on $D$ and $X$, as invoked in 4(a).\footnote{As an alternative set of IV restrictions in the context of selection, \cite{Haultfoeuille2010} permits the instrument to be associated with the outcome, but assumes conditional independence of the instrument and selection given the outcome.} We apply a control function approach based on this instrument,\footnote{Control function approaches have been applied in semi- and nonparametric sample selection models, e.g. \cite{Ahpo93}, \cite{Daneve03}, \cite{Ne07}, \cite{Hu11}, and \cite{Hu11b}, as well as in nonparametric instrumental variable models, see for example \cite{NeweyPowellVella99}, \cite{BlPo04}, and \cite{ImNe09}.} which requires further assumptions.

By the threshold crossing model postulated in 4(b), $\Pr(S = 1 | D, X , Z )= \Pr(V \leq \chi(D, X , Z ) ) = F_{V}(\chi(D, X , Z ))$, where $F_{V}(v)$ denotes the cumulative distribution function of $V$ evaluated at $v$. We will henceforth use the notation $\Pi=\pi(D,X,Z)=\Pr(S=1|D,X,Z)$ for the sake of brevity. Again by Assumption 4(b), the selection probability $\Pi$ increases strictly monotonically in $\chi$, such that there is a one-to-one correspondence between the distribution function $F_{V}$ and specific values $v$ given $X$. 
By Assumption 4(c), $V$ is independent of $(D,Z)$ given $X$, implying that the distribution function of $V$ given $X$ is (nonparametrically) identified.
By comparing individuals with the same $\Pi$, we control for $F_V$ and thus for the confounding associations of $V$ with  $D$ and $Y(d)$ that occur conditional on $S=1,X$. In other words, $\Pi$ serves as control function where the exogenous variation comes from $Z$. Controlling for the distribution of $V$ based on the instrument is thus a feasible alternative to the (infeasible) approach of directly controlling for levels of $V$.

\begin{figure}[!htp]
	\centering \caption{\label{figureiv}  Causal paths under nonignorable nonresponse}
\begin{center}
\centering 	\includegraphics[width=1.2\textwidth]{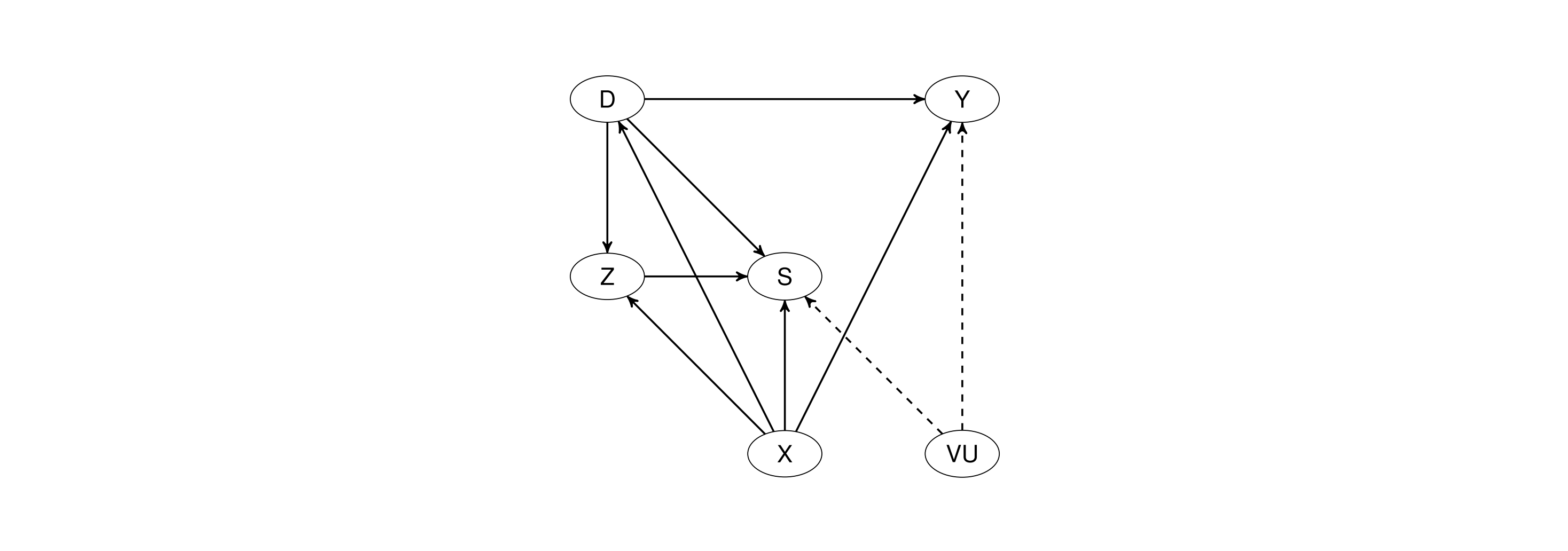}
\end{center}
\end{figure}

Figure \ref{figureiv} provides an acyclic graph of a causal model that can satisfy Assumptions 1 and 4. $U$ denotes unobservables affecting the outcome, which may be arbitrarily associated with $V$, the unobservable affecting selection. Note that the dashed lines indicate that $V,U$ are not observed in the data. Identification relies on instrument $Z$, which must not be associated with outcome $Y$ conditional on $D$ and $X$.\vspace{5pt}\newline

Furthermore, identification requires the following common support assumption, which is similar to Assumption 3(a), but in contrast to the latter also includes $\Pi$ as a conditioning \mbox{variable}.
\vspace{5pt}\newline
\textbf{Assumption 5 (common support):}\newline
$\Pr(D=d| X=x, \Pi=\pi )>0$ for all $d \in \{0,1,...,Q\}$ and $x,z$ in the support of $X,Z$.\vspace{5pt}\newline
This means that in fully nonparametric contexts, the instrument $Z$ must in general be continuous and strong enough to importantly shift the selection probability $\Pi$ conditional on $D,M,X$ in the selected population. Assumptions 1, 4, and 5 are sufficient for the identification of mean potential outcomes and the ATE in the selected population, denoted as $\Delta_{S=1}=E[Y(1)-Y(0)| S=1]$.

To see this, note that the identifying assumptions imply
\begin{eqnarray}\label{identivs1}
E[ Y(d)|S=1, X, F_V]=E[ Y(d)|S=1, X, \Pi]=E[ Y| D=d,S=1,X, \Pi]
\end{eqnarray}
The first equality follows from $\Pi=F_V$ under Assumption 4, the second from the fact that when controlling for $F_V$, conditioning on $S=1$ does not result in an association between $Y(d)$ and $D$ given $X$ such that $Y(d) \bot D |  X, \Pi, S=1$ holds by Assumptions 1 and 4. Therefore
\begin{eqnarray}\label{identreg2}
E[Y(d)|S=1]=E[E[ Y|D=d, S=1,X, \Pi]|S=1].
\end{eqnarray}
Denoting by $p_d(X, \Pi)=\Pr(D=d |X, \Pi )$  and $\mu (D,S,X,\Pi)=E[Y |D,S,X, \pi(D,X,Z)]$, an alternative expression for the mean potential outcome among the selected is obtained by
\begin{eqnarray}\label{score2}
E[Y(d)|S=1]&=&E\Big[\phi_{d,S=1}|S=1\Big],\textrm{ where}\notag\\
\phi_{d,S=1}&=& \frac{I\{D=d\} \cdot [Y- \mu(d,1,X,\Pi)] }{ p_d(X, \Pi)}  + \mu(d,1,X,\Pi),
\end{eqnarray}
where division by $p_d(X, \Pi)$ makes the reliance on Assumption 5 explicit. By applying the law of iterated expectations to replace $[Y- \mu(d,1,X,\Pi)]$ with $E[Y- \mu(d,1,X,\Pi)|D=d,S=1,X, \Pi]$ and noting that the latter expression is zero, one can see that \eqref{score2} is equivalent to \eqref{identreg2}. But in contrast to the latter, the identification result in \eqref{score2}  satisfies Neyman orthogonality and is based on the efficient influence function,\footnote{While the efficient influence function associated with \eqref{score2} is technically speaking doubly robust, i.e.\ consistent if either $\mu(d,1,X,\Pi)$ or $p_d(X, \Pi)$ is correctly specified, it is worth noting that this property can generally only hold if $\Pi$ is correctly specified because it enters both $\mu(d,1,X,\Pi)$ and $p_d(X, \Pi)$ as first step estimator. However, our approach does not rely on (global) doubly robustness but on Neyman orthogonality, which implies that DML is robust to local perturbations in $\Pi$ under specific regularity conditions.} see  Appendix \ref{effinf}.

The identification of the ATE in the total (rather than the selected) population is not feasible without further assumptions. The reason is that effects among selected observations cannot be extrapolated to the non-selected population if the effect of $D$ interacts with unobservables affecting the outcome, i.e.\ $U$ in \eqref{npmodel}, as the latter are in general distributed differently across $S=1,0$ even conditional on $(X, \Pi)$ or $(D, X, \Pi)$. To see this, note that conditional on $\Pi=\Pr(V \leq \chi(D, X , Z ) )$, the distribution of $V$ differs across the selected (satisfying  $V \leq \chi(D, X , Z )$) and the non-selected (satisfying  $V > \chi(D, X , Z )$),  such that the distribution of $U$ differs, too, if $V$ and $U$ are associated. This generally implies that $E[ Y(1)-Y(0)|S=1, X, \Pi]\neq E[ Y(1)-Y(0)|S=0, X, \Pi]$. While control function $\Pi$ ensures (together with $X$) that the treatment is unconfounded in the selected subpopulation, it does not permit extrapolating effects to the non-selected population with unobserved outcomes, see also \cite{HuberMelly} for further discussion.

Assumption 6 therefore imposes homogeneity in the average treatment effect across selected and non-selected populations conditional on $X,V$. A sufficient condition for effect homogeneity is the separability of observed and unobserved components in the outcome equation, i.e.\ $Y=\eta(D,X)+\nu(U)$, where $\eta,\nu$ are general functions. Furthermore, common support as postulated in Assumption 5 needs to be strengthened to hold in the entire population. In addition, the selection probability $\Pi$ must be larger than zero for any $d,x,z$ in their support. Otherwise, outcomes are not observed for some values of $D,X$. Assumption 7 formalizes this common support restriction.\vspace{5pt}\newline
\textbf{Assumption 6 (conditional effect homogeneity):}\newline
$E[Y(d)-Y(d')|S=1,X=x,V=v]=E[Y(d)-Y(d')|X=x,V=v]$ for all $d\neq d' \in \{0,1,...,Q\}$ and $x,v$ in the support of $X,V$.\vspace{5pt}\newline
\textbf{Assumption 7 (common support):}\newline
$\pi(d,x,z)>0$ for all $d \in \{0,1,...,Q\}$ and $x,z$ in the support of $X,Z$.\vspace{5pt}\newline

Under Assumptions 1,4,5,6, and 7, it follows that
\begin{eqnarray}\label{identiv}
\mu(d,1,X,\Pi)-\mu(d',1,X,\Pi)=E[ Y(d)-Y(d')|S=1, X, V]= E[ Y(d)-Y(d')|X, V],
\end{eqnarray}
where the first equality follows from Assumptions 1 and 4, see \eqref{identivs1}, and the second one from Assumption 6. Therefore, the ATE is identified by
\begin{eqnarray}\label{atefull}
\Delta=E[\mu(d,1,X,\Pi)-\mu(d',1,X,\Pi)].
\end{eqnarray}
An alternative expression for the ATE that is based on the efficient influence function  and respects Neyman orthogonality is given by
\begin{eqnarray}\label{score3}
\Delta&=&E\Big[\phi_{d}-\phi_{d'}\Big],\textrm{ where}\notag\\
\phi_{d}&=& \frac{I\{D=d\} \cdot S \cdot [Y- \mu(d,1,X,\Pi)] }{ p_d(X,\Pi)\cdot  \pi (d,X,Z)}  + \mu(d,1,X,\Pi),
\end{eqnarray}
where division by $p_d(X,\Pi)\cdot  \pi (d,X,Z)$ relies on the satisfaction of Assumptions 5 and 7.

\section{Identification under sequential conditional independence}\label{dynsel}

In many applications, it might appear unrealistic that one can control for all variables jointly affecting the sample selection indicator by conditioning only on baseline covariates measured prior to treatment assignment, in particular when no instrument is at hand. This is particularly the case when there is a substantial time lag between treatment assignment and sample selection/attrition, which raises concerns about dynamic confounding. The latter implies that some confounders influencing both the outcome and sample selection are themselves a function of the treatment. We subsequently reconsider the MAR framework, but not modify the identifying assumptions such that observed post-treatment confounders of $Y$ and $S$ are permitted. We will subsequently refer to observed post-treatment variables by $M$, in order to distinguish them from pre-treatment covariates $X$. Identification is based on a sequential conditional independence, which is based on maintaining Assumption 1 (conditional independence of $D$ given $X$), but replacing Assumption 2 by a modified conditional independence assumption for the selection indicator $S$ that allows for dynamic confounding due to $M=M(D)$, i.e.\ covariates possibly influenced by the treatment.\vspace{5pt}\newline
\textbf{Assumption 8 (conditional independence of selection):}\newline
$ Y \bot  S | D=d, X=x, M=m  $ for all $d \in \{0,1,...,Q\}$ and $x,m$ in the support of $X$ and $M$. \vspace{5pt}\newline
By Assumption 8, there are no unobservables jointly affecting selection and the outcome conditional on $D,X,M$, such that sample selection is selective w.r.t.\ observed characteristics only. When modifying the nonparametric outcome and selection models in  \eqref{npmodel} to $Y=\phi(D,X,M,U)$ and $S=\psi(D,X,M,V)$, Assumption 8 is satisfied if unobservables $U$ and $V$ are independent.
\vspace{5pt}\newline
\textbf{Assumption 9 (common support):}\newline
(a) $\Pr(D=d| X=x)>0$ and (b) $\Pr(S=1| D=d, X=x, M=m)>0$ for all $d \in \{0,1,...,Q\}$ and $x,m$ in the support of $X,M$.\vspace{5pt}\newline
Part (b) in Assumption 9 imposes a somewhat stronger common support restriction than part (b) in Assumption 3,  as it requires  the selection propensity score to be larger than zero for any combination of $D,X,M$ (rather $D,X$ only).

\begin{figure}[!htp]
	\centering \caption{\label{figuredyn}  Causal paths under sequential conditional independence}
\begin{center}
\includegraphics[width=1.2\textwidth]{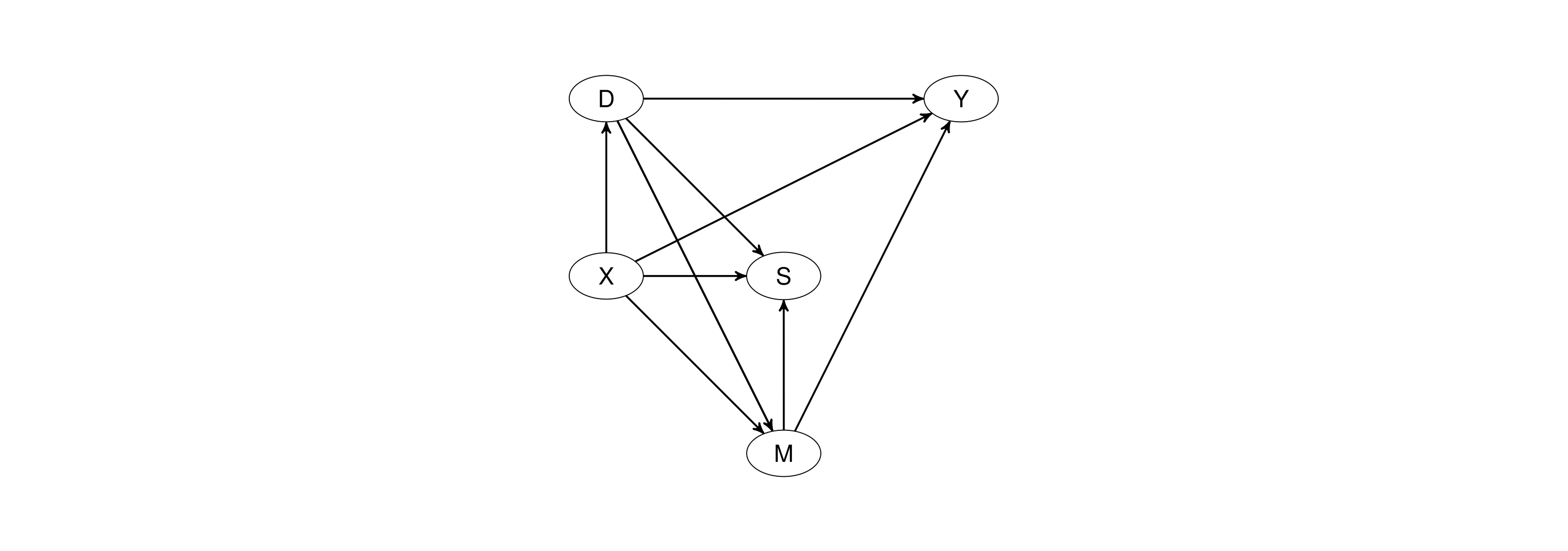}
\end{center}
\end{figure}

Figure \ref{figuredyn} provides an acyclic graph in which Assumptions 1 and 8 hold. Post-treatment covariates $M$ may be influenced by $D,X$ and might jointly affect $S$ and $Y$. Conditional on $D,X,M$, there are, however, no unobservables jointly influencing $S$ and $Y$.\vspace{5pt}\newline

Our identifying assumptions imply that
\begin{eqnarray}\label{condindep2}
E[Y(d)]&=&E[E[Y(d)|X]]=E[E[Y|D=d,X]]=E[E[E[Y| D=d, X,M]|D=d,X]]\notag\\
&=&E[E[E[Y| D=d, S=1, X,M]|D=d,X]].
\end{eqnarray}
where the first and third equalities follow from the law of iterated expectations, the second from Assumption 1, and the fourth from Assumption 8. Alternatively to this regression-based result using nested conditional mean outcomes, an IPW-based expression can be obtained, in which we use $\pi(D,X,M)=\Pr(S=1| D , X , M )$ as shortcut notation for the selection propensity score.
\begin{eqnarray}
&&E [ E[Y|D=d, S=1,X, M |D=d, X  ]] =E \Bigg[	E\Bigg[E\Bigg[\frac{  S  \cdot  Y}{ \pi(d,X,M)}\Bigg|D=d, X, M \Bigg] \Bigg|D=d, X\Bigg]\Bigg]\notag\\
&=&	E \Bigg[E\Bigg[\frac{  S  \cdot  Y }{\pi(d,X,M)}  \Bigg|D=d, X\Bigg]\Bigg]=E \Bigg[E\Bigg[\frac{I\{D=d\} \cdot S  \cdot  Y}{p_{d0}(X)\cdot \pi_0(d,X,M)}  \Bigg|X\Bigg]\Bigg]\notag\\
&=&E \Bigg[ \frac{I\{D=d\} \cdot S  \cdot  Y }{p_{d0}(X)\cdot \pi_0(d,X,M)}  \Bigg],
\end{eqnarray}
where the first and third equalities follow from basic probability theory and the second and last ones from the law of iterated expectations. Combining regression and IPW yields the following doubly robust identification result based on the efficient influence function, in which $\mu(d,1,X,M)=E[Y| D=d, S=1, X,M]$ and $\nu(d,1,X)=E[E[Y| D=d, S=1, X,M]|D=d,X]$ denote the conditional mean outcome and the nested conditional mean outcome, respectively:
\begin{eqnarray}\label{scoredyn}
E[Y(d)]&=&E\Big[\theta_{d} \Big],\textrm{ where}\notag\\
\theta_{d}&=&  \frac{I\{D=d\} \cdot S \cdot [Y- \mu(d,1,X,M)] }{ p_d(X)\cdot  \pi (d,X,M)}\notag\\
&+& \frac{I\{D=d\}\cdot  [\mu (d,1,X,M)-\nu(d,1,X)]}{p_d(X)} +\nu(d,1,X),
\end{eqnarray}
where division by $p_d(X)\cdot  \pi (d,X,M)$ relies on Assumption 9. The derivation of the efficient influence function is provided in Appendix \ref{effinf}.

\section{Estimation of the counterfactual with K-fold Cross-Fitting}\label{estimsec}
\label{section:CrossFitting}
We subsequently propose an estimation strategy for the counterfactual  $E[Y(d)]$ under MAR as discussed in Section \ref{marident} based on identification result \eqref{score} and show its root-n consistency under specific regularity conditions. 
Let to this end $\mathcal{W} = \{W_i|1\leq i \leq n\}$ with $W_i = (Y_{i}\cdot S_{i}, D_{i}, S_{i}, X_{i})$ for all $ i $ denote the set of observations in an i.i.d.\ sample of size $n$. $\eta$ denotes the plug-in (or nuisance) parameters, i.e.\ the conditional mean outcome, mediator density and treatment probability. Their respective estimates are referred to by $\hat{\eta} =  \{\hat\mu(D,1,X),\hat{p}_d(X), \hat{\pi} (D,X) \}$ and the true parameters by $\eta_0 =  \{\mu_0(D,1,X), p_{d0}(X), \pi_0(D,X) \}$. Finally, $\Psi_{d0}=E[Y(d)]$ denotes the true counterfactual.

We estimate $\Psi_{d0}$ by the following algorithm that combines the estimation of Neyman-orthogonal scores with sample splitting or cross-fitting and is root-$n$ consistent under conditions outlined further below. \vspace{8pt}\newline
\textbf{Algorithm 1: Estimation of $E[Y(d)]$ based on equation \eqref{score} }
\begin{enumerate}
	\item Split $\mathcal{W}$ in $ K $ subsamples. For each subsample $ k $, let $n_k$ denote its size, $\mathcal{W}_k$ the set of observations in the sample and $\mathcal{W}_k^{C}$ the complement set of all observations not in $k$.
	\item For each $k$, use  $\mathcal{W}_k^{C}$ to estimate the model parameters of the plug-ins $\mu(D,S=1,X)$, $p_d(X)$, $\pi (D,X)$ in order to predict these plug-ins in $\mathcal{W}_k$, where the predictions are denoted by $\hat\mu^k(D,1,X)$, $\hat{p}_d^k(X)$, and $\hat{\pi}^k(D,X)$.
	\item For each $ k $, obtain an estimate of the score function (see $\psi_d$ in \eqref{score}) for each observation $i$ in $\mathcal{W}_k$, denoted by $\hat \psi_{d,i} ^k$ :
	\begin{eqnarray}
	\hat \psi_{d,i}  ^k = \frac{I\{D_i=d\} \cdot S_i \cdot [Y_i- \hat\mu^k(d,1,X_i)] }{ \hat p^k_d(X_i)\cdot  \hat \pi^k (d,X_i)}  + \hat \mu^k(d,1,X_i).
	\end{eqnarray}
	\item Average the estimated scores $\hat \psi_{d,i}^k$ over all observations across all $ K $ subsamples to obtain an estimate of  $\Psi_{d0}=E[Y(d)]$ in the total sample, denoted by $\hat \Psi_d=1/n \sum_{k=1}^{K}  \sum_{i=1}^{n_k} \hat \psi_{d,i}^k$.\vspace{5pt}\newline
\end{enumerate}

In order to obtain root-n consistency for counterfactual estimation, we make the following assumption about the prediction qualities of machine learning for estimating the nuisance parameters. Following  \cite{Chetal2018}, we introduce some further notation: let $(\delta_n)_{n=1}^{\infty}$ and $(\Delta_n)_{n=1}^{\infty}$ denote sequences of positive constants with $\lim_{N\rightarrow \infty} \delta_n = 0 $ and $\lim_{N\rightarrow \infty} \Delta_n = 0.$ Furthermore, let $c, \epsilon, C$ and $q$ be positive constants such that $q>2,$ and let $K \geq 2$ be a fixed integer. Also, for any random vector $R = (R_1,...,R_l)$, let $\left\| R \right\|_{q} = \max_{1\leq j \leq l}\left\| R_l \right\|_{q},$ where
$\left \| R_l \right\|_{q}  =  \left( E\left[ \left| R_l \right|^q \right] \right)^{\frac{1}{q}}$.
In order to ease notation, we assume that $n/K$ is an integer. For the sake of brevity we omit the dependence of probability $\Pr_P,$ expectation $E_P(\cdot),$ and norm $\left\| \cdot  \right\|_{P,q}$ on the probability measure $P.$
\vspace{5pt}\newline

\textbf{Assumption 10 (regularity conditions and quality of plug-in parameter estimates):} \newline
For all probability laws $P \in \mathcal{P},$ where $\mathcal{P}$ is the set of all possible probability laws
 the following conditions hold for the random vector $(Y,D,S,X)$ for $d \in \{0,1,...,Q\}$:

\begin{enumerate}
\item[(a)] $  \left\| Y \right\|_{q} \leq C$,

$\left\|E[Y^2| D=d, S=1, X ] \right\|_{\infty} \leq C^2$,

\item[(b)]

$\Pr(\epsilon \leq p_{d0} (X) \leq 1-\epsilon) = 1,$

$\Pr(\epsilon \leq \pi_0 (d,X)) = 1,$

\item[(c)]
$\left\| Y-\mu_0(d,1,X)  \right\|_{2} = E_{ } \Big[\left(Y-\mu_0(d,1,X) \right)^2 \Big]^{\frac{1}{2}} \geq c$

\item[(d)] Given a random subset $I$ of $[n]$ of size $n_k=n/K,$ the nuisance parameter estimator $\hat \eta_0 = \hat \eta_0((W_i)_{i \in I^C})$ satisfies the following conditions. With $P$-probability no less than $1-\Delta_n:$
\begin{eqnarray}
\left\|  \hat \eta_0 - \eta_0 \right\|_{q} &\leq& C, \notag \\
\left\|  \hat \eta_0 - \eta_0 \right\|_{2} &\leq& \delta_n, \notag \\
\left\| \hat  p_{d0}(X)-1/2\right\|_{\infty}  &\leq& 1/2-\epsilon, \notag\\
 \left\| \hat  \pi_0(D,X)-1/2\right\|_{\infty} &\leq & 1/2-\epsilon, \notag \\
\left\|  \hat \mu_0(D,S,X)-\mu_0(D,S,X)\right\|_{2} \times \left\| \hat  p_{d0}(X)-p_0(X)\right\|_{2}  &\leq & \delta^{}_n n^{-1/2}, \notag \\
\left\|  \hat \mu_0(D,S,X)-\mu_0(D,S,X)\right\|_{2} \times \left\| \hat  \pi_0(D,X)-\pi_{0}(D,X)\right\|_{2} &\leq & \delta^{}_n n^{-1/2}.\notag
\end{eqnarray}
\end{enumerate}

The only non-primitive condition is the condition (d), which puts restrictions on the quality of the nuisance parameter estimators. Condition (a) states that the distribution of the outcome does not have unbounded moments. (b) refines the common support condition such that the treatment and selection propensity scores are bounded away from $0$and $1$ and $0$, respectively. (c) states that covariates $X$ do not perfectly predict the conditional mean outcome.

For demonstrating the root-n consistency of our estimator of the mean potential outcome, we show that it satisfies the requirements of the DML framework in \cite{Chetal2018}  by first verifying linearity and Neyman orthogonality of the score (see Appendix \ref{Neyman1}). As  $ \psi_{d}(W, \eta, \Psi_{d0}) $ is smooth in $ (\eta, \Psi_{d0}) $, it then suffices that the  plug-in estimators converge with rate $ n^{-1/4} $  for achieving $ n^{-1/2} $-convergence in the estimation of $ \hat{\psi}$, see Theorem 1. A rate of $ n^{-1/4} $ is achievable by many commonly used machine learners under specific conditions, such as lasso, random forests, boosting and neural nets, see for instance \cite{Bellonietal2014}, \cite{LuoSpindler2016}, \cite{WagerAthey2018}, and \cite{FarrellLiangMisra2018}.\vspace{5pt}\newline
\textbf{Theorem 1}\\
Under  Assumptions 1-3 and 10, it holds for estimating $\Psi_{d0}=E[Y(d)]$ based on Algorithm 1:  \\
$\sqrt{n} \Big(\hat \Psi_{d} -  \Psi_{d0} \Big) \rightarrow N(0,\sigma^2_{\psi_d})$,  where $\sigma^2_{\psi_{d}}= E[(\psi_d-\Psi_{d0})^2]$. \\	
The proof is provided in Appendix \ref{Neyman1}.

We subsequently discuss the estimation of $\Psi_{d0}$ based on \eqref{score3}. We note that in this case, one needs to estimate the nested nuisance parameters $\mu(d,1,X,\Pi)$ and  $p_d(X,\Pi)$, because they require the first-step estimation of $\Pi=\pi(D,X,Z)$. To avoid overfitting in the nested estimation procedure, the models for  $\Pi$ on the one hand and $\mu(d,1,X,\Pi), p_d(X,\Pi)$ on the other hand are estimated in different subsamples. The plug-in estimates are now denoted by $\hat{\eta} =  \{\hat\mu(D,1,X,\Pi),\hat{p}_d(X,\Pi), \hat{\pi} (D,X,Z) \}$ and the true plug-ins by $\eta_0 =  \{\mu_0(D,1,X,\Pi), p_{d0}(X,\Pi), \pi_0(D,X,Z) \}$.\vspace{5pt}\newline
\textbf{Algorithm 2: Estimation of $E[Y(d)]$ based on equation \eqref{score3}}
\begin{enumerate}
	\item Split $\mathcal{W}$ in $ K $ subsamples. For each subsample $ k $, let $n_k$ denote its size, $\mathcal{W}_k$ the set of observations in the sample and $\mathcal{W}_k^{C}$ the complement set of all observations not in $k$.
	\item Split  $\mathcal{W}_k^{C}$ into 2 nonoverlapping subsamples and estimate the model parameters of $\pi_0(D,X,Z)$ in one subsample and the model parameters of $\mu_0(D,1,X,\Pi)$ and $p_{d0}(X,\Pi)$   in the other subsample.  Predict the plug-in models in $\mathcal{W}_k$, where the predictions are denoted by $\hat \Pi^k$, $\hat{p}_d^k(X, \hat \Pi^k)$, and $\hat \mu(D,1,X,\hat \Pi)$.
	\item For each $ k $, obtain an estimate of the efficient score function (see $\phi_d$ in \eqref{score3}) for each observation $i$ in $\mathcal{W}_k$, denoted by $\hat \phi_{d,i} ^k$ :
	\begin{eqnarray}
	\hat \phi_{d,i}^{k} = \frac{I\{D_i=d\} \cdot S_i \cdot [Y_i- \hat \mu^k(d,1,X_i,\hat \Pi_i)] }{ \hat p_d(X_i,\hat \Pi_i)\cdot  \hat \pi (d,X_i,Z_i)}  + \hat\mu(d,1,X_i,\hat\Pi_i)
	\end{eqnarray}
	\item Average the estimated scores $\hat \phi_{d,i}^{k}$ over all observations across all $ K $ subsamples to obtain an estimate of  $\Psi_{d0}=E[Y(d)]$ in the total sample, denoted by $\hat \Phi_d=1/n \sum_{k=1}^{K}  \sum_{i=1}^{n_k} \hat \phi_{d,i}^{k}$.\vspace{5pt}\newline
\end{enumerate}

An estimator of $\Psi_{d0}^{S=1}=E[Y(d)|S=1]$ based on \eqref{score2} is obtained by two modifications in Algorithm 2. First, rather than relying on the total sample $n$, one merely uses the subsample with observed outcomes which of size $\sum_{i=1}^n S_i$ to split it into $K$ subsamples. Second, in step 3, $\hat \phi_{d,i}^{k}$ is to be replaced by
\begin{eqnarray}
	\hat \phi_{d,S=1,i}^{k} = \frac{I\{D_i=d\} \cdot [Y_i- \hat \mu^k(d,1,X_i,\hat \Pi_i)] }{ \hat p_d(X_i,\hat \Pi_i)}  + \hat\mu(d,1,X_i,\hat\Pi_i)
	\end{eqnarray}
to estimate $\Psi_{d0}^{S=1}$ in step 4 by $\hat \Phi_d^{S=1}=\frac{1}{\sum_{i=1}^n S_i} \sum_{k=1}^{K}  \sum_{i=1}^{n_k} \hat \phi_{d,S=1,i}^{k}$. As $\sum_{i=1}^n S_i$ is an asymptotically fixed proportion of $n$, also this approach can be shown to be root-$n$ consistent under particular regularity conditions outlined in Assumption 11, that are in analogy to those in Assumption 10, but now adapted our IV-dependent identifying assumptions. \vspace{5pt}\newline
\textbf{Assumption 11 (regularity conditions and quality of plug-in parameter estimates):} \vspace{5pt}\newline
For all probability laws $P \in \mathcal{P},$ where $\mathcal{P}$ is the set of all possible probability laws
 the following conditions hold for the random vector $(Y,D,S,X,Z)$ for $d \in \{0,1,...,Q\}$:

\begin{enumerate}
\item[(a)] $  \left\| Y \right\|_{q} \leq C$,

$\left\|E[Y^2| D=d, S=1, X, \Pi ] \right\|_{\infty} \leq C^2$,

\item[(b)]

$\Pr(\epsilon \leq p_{d0} (X,\Pi) \leq 1-\epsilon) = 1,$

$\Pr(\epsilon \leq \pi_0 (d,X,Z)) = 1,$

\item[(c)]
$\left\| Y-\mu_0(d,1,X,\Pi)  \right\|_{2} = E_{ } \Big[\left(Y-\mu_0(d,1,X,\Pi) \right)^2 \Big]^{\frac{1}{2}} \geq c$

\item[(d)] Given a random subset $I$ of $[n]$ of size $n_k=n/K,$ the nuisance parameter estimator $\hat \eta_0 = \hat \eta_0((W_i)_{i \in I^C})$ satisfies the following conditions. With $P$-probability no less than $1-\Delta_n:$
\begin{eqnarray}
\left\|  \hat \eta_0 - \eta_0 \right\|_{q} &\leq& C, \notag \\
\left\|  \hat \eta_0 - \eta_0 \right\|_{2} &\leq& \delta_n, \notag \\
\left\| \hat  p_{d0}(X,\hat \Pi)-1/2\right\|_{\infty}  &\leq& 1/2-\epsilon, \notag\\
 \left\| \hat  \pi_0(D,X,Z)-1/2\right\|_{\infty} &\leq & 1/2-\epsilon, \notag \\
\left\|  \hat \mu_0(D,S,X,\hat\Pi)-\mu_0(D,S,X,\Pi)\right\|_{2} \times \left\| \hat  p_{d0}(X,\hat\Pi)-p_0(X,\Pi)\right\|_{2}  &\leq & \delta^{}_n n^{-1/2}, \notag \\
\left\|  \hat \mu_0(D,S,X,\hat\Pi)-\mu_0(D,S,X,\Pi)\right\|_{2} \times \left\| \hat  \pi_0(D,X,Z)-\pi_{0}(D,X,Z)\right\|_{2} &\leq & \delta^{}_n n^{-1/2}.\notag
\end{eqnarray}
\end{enumerate}
Theorems 2 and 3 postulate the root-n consistency and asymptotic normality of the estimators of the mean potential outcomes in the selected and total populations, respectively. \vspace{5pt}\newline
\textbf{Theorem 2}\\
Under  Assumptions 1, 4, 6, 7, and 11, it holds for estimating $\Psi_{d0}=E[Y(d)]$ based on Algorithm 2:  \\
$\sqrt{n} \Big(\hat \Phi_{d} -  \Psi_{d0} \Big) \rightarrow N(0,\sigma^2_{\phi_d})$,  where $\sigma^2_{\phi_{d}}= E[(\phi_d-\Psi_{d0})^2]$. \\
\textbf{Theorem 3}\\ 
Under  Assumptions 1, 4, 5, and 11, it holds for estimating $\Psi_{d0}^{S=1}=E[Y(d)|S=1]$ based on Algorithm 2:  \\
$\sqrt{n} \Big(\hat \Phi_{d}^{S=1} -  \Psi_{d0}^{S=1} \Big) \rightarrow N(0,\sigma^2_{\phi_{d,S=1}})$,  where $\sigma^2_{\phi_{d,S=1}}= E[(\phi_{d,S=1}-\Psi_{d0}^{S=1})^2]$. \\	
The proofs are provided in Appendices \ref{Neyman2} and \ref{Neyman3}.

Next, we consider the estimation of $\Psi_{d0}$ based on \eqref{scoredyn}. Similarly to estimation based on \eqref{score3}, we are required to estimate a nested nuisance parameter, namely  $\nu(d,1,X)=E[\mu(d,1,X,M)|D=d,X]$. To avoid overfitting in the nested estimation procedure, the models for  $\mu(d,1,X,M)$  and $\nu(d,1,X)$ estimated in different subsamples.\vspace{5pt}\newline
\textbf{Algorithm 3: Estimation of $E[Y(d)]$ based on equation \eqref{scoredyn}}
\begin{enumerate}
	\item Split $\mathcal{W}$ in $ K $ subsamples. For each subsample $ k $, let $n_k$ denote its size, $\mathcal{W}_k$ the set of observations in the sample and $\mathcal{W}_k^{C}$ the complement set of all observations not in $k$.
	\item For each $k$, use  $\mathcal{W}_k^{C}$ to estimate the model parameters of $p_d(X)$ and $\pi (d,X,M)$. Split  $\mathcal{W}_k^{C}$ into 2 nonoverlapping subsamples and estimate the model parameters of the conditional mean $\mu(d,1,X,M)$ and the nested conditional mean $\nu(d,1,X)$   in the distinct subsamples.  Predict the models among $\mathcal{W}_k$, where the predictions are denoted by $\hat{p}^k_d(X)$, $\hat{\pi}^k(d,X,M)$, $\hat{\mu}^k(d,1,X,M)$, $\hat \nu^k(d,1,X)$.
	\item For each $ k $, obtain an estimate of the moment condition for each observation $i$ in $\mathcal{W}_k$, denoted by $\hat \theta^{k}_{d,i}$ :
	\begin{eqnarray}
	\hat \theta^{k}_{d,i}&=& \frac{I\{D_{i}=d\} \cdot S_{i} \cdot [Y_{i}-\hat\mu^k(d,1,X_i,M_i)]}{\hat p^k_{d}(X_{i})\cdot \hat \pi^k(d,X_i,M_i)} \notag\\
	& + &\frac{I\{D_{i}=d\}\cdot  [\hat \mu^k(d,1,X_i,M_i)-\hat \nu^k(d,1,X_i)]}{\hat p^k_{d}(X_{i})} +\hat \nu^k(d,1,X_i).\notag
	\end{eqnarray}
	\item Average the estimated scores $\hat \theta^{k}_{d,i}$ over all observations across all $ K $ subsamples to obtain an estimate of  $\Psi_{d0}=E[Y(d)]$ in the total sample, denoted by $\hat \Theta_{d}=1/n \sum_{k=1}^{K}  \sum_{i=1}^{n_k} \hat \theta^{k}_{d,i}$.
\end{enumerate}

To show root-$n$ consistency for this estimation approach, we  impose the following regularity conditions, where we again assume that $n/K$ is an integer and omit the dependence of probability $\Pr_P,$ expectation $E_P(\cdot),$ and norm $\left\| \cdot  \right\|_{P,q}$ on the probability measure $P$:
\vspace{5pt}\newline
\textbf{Assumption 12 (regularity conditions and quality of plug-in parameter estimates):}  \newline
For all probability laws $P \in \mathcal{P}$
the following conditions hold for the random vector $(Y,D,S,X,M)$ for all $d \in \{0,1,...,Q\}$:
\begin{enumerate}
	\item[(a)] $  \left\| Y \right\|_{q} \leq C,$
	
	$\left\|E[Y^2| D = d, S = 1, X, M ] \right\|_{\infty} \leq C^2$,
	
	\item[(b)]
	
	
	
	
	
	$\Pr(\epsilon \leq p_{d0} (X) \leq 1-\epsilon) = 1,$
	
	
	$\Pr(\epsilon \leq \pi_0(d,X,M) \leq 1-\epsilon) = 1,$
	
	
	\item[(c)]
	$\left\| Y -\mu_0(d,1,X,M)  \right\|_{2} = E \Big[\left(Y -\mu_0 (d,1,X,M)\right)^2 \Big]^{\frac{1}{2}} \geq c$
	
	\item[(d)] Given a random subset $I$ of $[n]$ of size $n_k=n/K,$ the nuisance parameter estimator $\hat \eta_0 = \hat \eta_0((W_i)_{i \in I^C})$ satisfies the following conditions. With $P$-probability no less than $1-\Delta_n:$
	\begin{eqnarray}
	\left\|  \hat \eta_0 - \eta_0 \right\|_{q} &\leq& C, \notag \\
	\left\|  \hat \eta_0 - \eta_0 \right\|_{2} &\leq& \delta_n, \notag \\
	\left\| \hat  p_{d0}(X)-1/2\right\|_{\infty}  &\leq& 1/2-\epsilon, \notag\\
	\left\| \hat  \pi_0 (D,X,M)-1/2\right\|_{\infty} &\leq & 1/2-\epsilon, \notag \\
	\left\|  \hat \mu_0(D,S,X,M)\right\|_{2} \times \left\| \hat  p_{d0}(X)-p_{d0}(X)\right\|_{2}  &\leq & \delta^{}_n n^{-1/2}, \notag \\
	\left\|  \hat \mu_0(D,S,X,M)-\mu_0(D,S,X,M)\right\|_{2} \times \left\| \hat  \pi_0(D,X,M)-\pi_0(D,X,M)\right\|_{2} &\leq & \delta^{}_n n^{-1/2},\notag \\
	\left\|  \hat \nu_0(D,S,X)-\nu_0(D,S,X)\right\|_{2} \times \left\|  \hat  p_{d0}(X)-p_{d0}(X) \right\|_{2} &\leq & \delta^{}_n n^{-1/2}.\notag
	\end{eqnarray}
\end{enumerate}
Under these regularity conditions and the sequential conditional independence assumption, estimation based on Algorithm 3 is asymptotically normal, as postulated in Theorem 4. \vspace{5pt}\newline
\textbf{Theorem 4}\\
Under  Assumptions 1, 8, 9, and 12, it holds for estimating $E[Y(d)]$ based on Algorithm 3:  \\
$\sqrt{n} \Big(\hat \Theta_{d} -  \Psi_{d0} \Big) \rightarrow N(0,\sigma^2_{\theta_{d}})$,  where $\sigma^2_{\theta_{d}}= E[(\theta_{d}-\Psi_{d0})^2]$. \\	
The proof of Theorem 4 is provided in Appendix \ref{Neyman4}.

\section{Simulation study}\label{Sim}

This section provides a simulation study to investigate the finite sample behavior of our estimation approaches either relying on a MAR assumption of an instrument for selection based on the following data generating process:
\begin{eqnarray*}
	Y &=& D + X'\beta   + U\text{ with }
	Y \text{ being}\text{ observed if }S=1, \\
	S &=& I\{  D + \gamma Z + X'\beta + V>0\},\quad
	D = I\{X'\beta + W >0\},\\
	X &\sim& N(0,\sigma^2_X),\quad Z \sim N(0,1), \quad (U,V) \sim N(0,\sigma^2_{U,V}), \quad W \sim N(0,1).
\end{eqnarray*}
Outcome $Y$ is a linear function of $D$ (whose treatment effect is one), covariates $X$ (for $\beta\neq 0$), and the unobservable $U$ and is only observed if the selection indicator $S$ is equal to one. Selection is a function of $D$, $X$, the unobservable $V$, and of instrument $Z$ if $\gamma\neq 0$. The treatment $D$ is a function of $X$ and the unobservable $W$. Both $Z$ and $W$ are random, standard normally distributed variables that are uncorrelated with $X$ or $(U,V)$. The correlation between the mean zero and normally distributed covariates in $X$ is determined by the covariance matrix $\sigma^2_X$. Similarly, $\sigma^2_{U,V}$ determines the correlation between the mean zero and normally distributed unobservables in the outcome and selection equation. In this setup, MAR is violated if the covariance between $U$ and $V$ is non-zero. We consider the perfromance of our estimators in $1000$ simulations with two sample sizes of $n=2000$ and $8000$.

In our simulations, we set the number of covariates $p$ to 100. $\sigma^2_X$ is defined based on setting the covariance of the $i$th and $j$th covariate in $X$  to $0.5^{|i-j|}$. $\beta$  gauges the impacts of the covariates on $Y$, $S$, and $D$, respectively, and thus, the magnitude of confounding. The $i$th element in the coefficient vector $\beta$ is set to $0.4/i^2$ for $i=1,...,p$, implying a squared decay of covariate importance in terms of confounding. In our first simulation design, we set $\gamma=0$ and $\sigma^2_{U,V}=\left(\begin{matrix} 1 & 0  \\ 0 & 1 \end{matrix}\right)$ such that MAR as discussed in Section \ref{marident} holds. We consider the performance of DML based on Theorem 1 (henceforth DML MAR), which does not make use of the instrument $Z$, as well as based on Theorem 2 (DML IV), which exploits the instrument despite the satisfaction of MAR.

The nuisance parameters, i.e.\ the linear and probit specifications of the outcome, selection, and treatment equations, are estimated by lasso regressions using the default options of the \textit{SuperLearner} package provided by \cite{vanderLaanetal2007} for the statistical software \textsf{R}. We use 3-fold cross-fitting for the estimation of the treatment effects. We drop observations whose products of estimated treatment and selection propensity scores are close to zero, namely smaller than a trimming threshold of $0.01$ (or 1\%). This avoids an explosion of the propensity score-based weights and, thus, of the variance when estimating the mean potential outcomes or ATE by the sample analogues of  \eqref{score} and \eqref{score3}, where the product of the propensity scores enters the respective denominators for reweighing the outcome. Our estimation procedure is available in the \textit{treatselDML} command of the \textit{causalweight} package for \textsf{R} by \citet{BodoryHuber2018}.

\begin{table}[htbp]
	\centering\normalsize
	\caption{Simulation results under MAR}
	\label{tab:mar}
	\begin{adjustbox}{max width=\textwidth}
		\begin{tabular}{c|cccccc}
			\hline\hline
			& true & bias & sd & RMSE & meanSE & coverage \\
			\hline
			$n$=2000	& \multicolumn{6}{c}{} \\
			DML MAR & 1.000 & 0.003 & 0.060 & 0.060 & 0.063 & 0.939 \\
			DML IV & 1.000 & 0.003 & 0.060 & 0.060 & 0.063 & 0.939 \\
			\hline
			$n$=8000	& \multicolumn{6}{c}{ } \\
			DML MAR & 1.000 & 0.012 & 0.031 & 0.033 & 0.034 & 0.934 \\
			DML IV & 1.000 & 0.012 & 0.031 & 0.033 & 0.034 & 0.939 \\
			\hline	
		\end{tabular}
	\end{adjustbox}
	\caption*{\scriptsize Notes: column `true' shows the true effect, `bias' the bias of the respective estimatior, `sd' the standard deviation, and `RMSE' the root mean squared error. Column `meanSE' displays the average standard error based on the asymptotic approximation across all simulations, `coverage' the covarage rate of the true effect based on 95\% confidence intervals.}
\end{table}

Table \ref{tab:mar} presents the simulation results. The biases (bias) of both DML MAR and DML IV are rather close to zero independent of the sample size. Furthermore, the estimators have virtually the same variance, despite the fact that DML IV unnecessarily relies on the control function approach and an irrelevant instrument. Both estimators appear to converge to the true effect with $\sqrt{n}$-rate, as the root mean squared error (RMSE) is roughly cut by half when quadrupling the sample size. The average standard error across simulations (meanSE) based on the asymptotic variance approximation comes close to the respective estimator's standard deviation (sd). Finally, the coverage rate (coverage), i.e.\ the share of simulations in which the 95\% confidence interval includes the true effect, is only slightly below the nominal level of 95\%.

\begin{table}[htbp]
	\centering\normalsize
	\caption{Simulation results under nonignorable selection}
	\label{tab:iv}
	\begin{adjustbox}{max width=\textwidth}
		\begin{tabular}{c|cccccc}
			\hline\hline
			& true & bias & sd & RMSE & meanSE & coverage \\
			\hline
			$n$=2000	& \multicolumn{6}{c}{ } \\
			DML MAR & 1.000 & -0.120 & 0.055 & 0.132 & 0.052 & 0.374 \\
			DML IV & 1.000 & -0.020 & 0.071 & 0.074 & 0.065 & 0.907 \\
			\hline
			$n$=8000	& \multicolumn{6}{c}{ } \\
			DML MAR & 1.000 & -0.116 & 0.028 & 0.119 & 0.027 & 0.009 \\
			DML IV & 1.000 & 0.006 & 0.040 & 0.040 & 0.036 & 0.915 \\
			\hline	
		\end{tabular}
	\end{adjustbox}
	\caption*{\scriptsize Notes: column `true' shows the true effect, `bias' the bias of the respective estimatior, `sd' the standard deviation, and `RMSE' the root mean squared error. Column `meanSE' displays the average standard error based on the asymptotic approximation across all simulations, `coverage' the covarage rate of the true effect based on 95\% confidence intervals.}
\end{table}

In a second simulation design, we set $\gamma=1$ and $\sigma^2_{U,V}=\left(\begin{matrix} 1 & 0.8  \\ 0.8 & 1 \end{matrix}\right)$, such that selection is nonignorable, i.e.\ related to unobservables  as discussed in Section \ref{ivident}, due to a strong correlation of $U$ and $V$. Table \ref{tab:iv} presents the results. DML MAR is no longer unbiased, while the bias of  DML IV appears to approach zero as the sample size increases, at the price of somewhat higher standard deviation than DML MAR. However, DML IV  dominates DML MAR under either sample size in terms of having a lower RMSE and has thus a more favorable bias-variance trade-off in the scenario considered. While coverage is quite satisfactory for DML IV, the 95\% confidence interval mostly fails to include the true effect in the case of DML MAR, in particular under the larger sample size.

\section{Application}\label{Application}

As an empirical illustration, we apply our method to the Job Corps (JC) training program. The data come from the National Job Corps Study (NJCS), a randomized social experiment conducted in the mid-to-late 1990s in the United States to evaluate the effectiveness of JC on different labor market outcomes. The JC is the largest and most comprehensive job training program for disadvantaged youth in the US, in which participants are exposed to different types of academic and vocational instruction. The data set contains very detailed pre-treatment information about program participants, such as: expectations, motivations for applying to JC, age, gender, number of children at the moment of treatment assignment, occupation, household income, hourly wage, educational level, marital status, whether the individual was previously attending a school, JC program, or some other academic or vocational training, health status, past employment, types of crimes committed, family support for attending the training, and information on the mother and the father (e.g. education and employment). Furthermore, a large range of variables for instance related to labor market status, employment, income, and education are reassessed in several follow-up interviews after JC assignment.  

\cite{SchBuGl01} and \cite{SchBuMc08} evaluate the impact of random program assignment on a wide range of labor market outcomes, showing positive effects on education, employment, and earnings in the longer run. Several studies focus on more specific program aspects of JC, e.g.\ on the effect of the length of exposure to training or of discrete sequences of training interventions on labor market outcomes like employment and earnings, as well as on JC's causal mechanisms, i.e.\ direct and indirect effects (operating via specific mediating variables) on labor market outcomes and health. \cite{FlGoNe12}, for instance, acknowledge the existence of several types of instruction, which, along with the self-paced nature of the program, creates selective heterogeneity in the number of weeks the participants are exposed to vocational or academic training. Considering continuously distributed treatment doses while controlling for baseline covariates, they find a positive effect of weeks in training on earnings, however, with decreasing marginal returns as a function of weeks already accomplished. \cite{FlFl09} and \cite{Hu14} investigate the causal mechanisms underlying JC when considering work experience or employment as mediators, respectively, and find positive direct effects on earnings and general health, respectively, when invoking a selection-on-observables assumption. \cite{FlFl10} avoid the latter assumption by suggesting a partial identification strategy based on which they estimate bounds the on causal mechanisms of JC when considering the achievement of a GED, high school degree, or vocational degree as mediators. Under their strongest set of bounding assumptions, the results suggest a positive effect on labor market outcomes even net of the indirect mechanism via obtaining a degree. \cite{FrHu17} base their analysis of causal mechanisms on an instrumental variable approach and find a positive indirect effect of JC training on earnings through an increase in the number of hours worked.

Estimation approaches specifically dealing with truncated outcomes such as wages, which are only observed and defined for a selective subpopulation like those in employment, have also been considered. For instance, \cite{FrMePaRu12} and \cite{ZhRuMe09} consider the principal stratification approach of \cite{FrangakisRubin02} to evaluate the effects of JC on employment and wages among specific groups, e.g.\ those finding employment irrespective of training participation, rather among the full population. \cite{FrMePaRu12} simultaneously address the several identification issues related to noncompliance of training participation with JC assignment as well as missingness in wage outcomes due to survey non-response  or non-employment based on a likelihood-based analysis using finite mixture models. They find positive effects on wages and evidence that the program should ideally have been designed differently for different subgroups of individuals depending on their personal characteristics.

\citet{Le05} considers a partial identification approach for bounding the effect of JC assignment on wage among those finding employment irrespective of the assignment. Rather than invoking MAR or IV assumptions for sample selection, this method merely relies on the monotonicity of employment in JC assignment (such that being randomized in never decreases the employment state), at the cost of giving up point identification. \citet{semenova2020} suggests a DML approach to tighten the bounds by controlling for covariates $X$ in a data-driven way. The results based on bounding generally point to a positive (intention-to-treat) effect of JC assignment on wages. Finally, \cite{BoHuLa2020} consider dynamic treatment evaluation based on DML, imposing a sequential conditional independence assumption to analyze discrete sequences of training in the first and second year after JC assignment. Controlling for both baseline characteristics and covariates measured after the first treatment period (i.e.\ one year after JC assignment) they find a positive effect of a sequence of vocational training on employment.

For our empirical analysis we similarly to \cite{FrHu17} consider female applicants to JC and aim at estimating the effects of academic or vocational training received in the first year of the program ($D$) on hourly wage ($Y$) in the short run measured in the last week of the first year or in the longer run, measured 4 years after random assignment to JC. Hourly wage is only observed conditional on employment ($S$) in the respective outcome period. Even though JC assignment is random, actual participation in training activities is likely selective and associated with individual factors, similarly to the selection into employment.  As for example discussed in \cite{LECHNER2013111} and \cite{BiFiOsPau14}, the previous labor market history and  socio-economic characteristics are likely important confounders when assessing the impact of training interventions,  which motivates our DML approach to account for a rich set of covariates in a data-driven way. To assess the short-run effects, we either assume MAR as discussed in Section \ref{marident} and use our DML approach based on Theorem 1 to control for our all in all $355$ baseline covariates ($X$), or we impose the IV assumptions of Section \ref{ivident} to estimate the ATE based on Theorem 3, considering the number of young children in the household at JC assignment as instrument ($Z$) for employment. Even though numerous studies in labor economics consider children as instrument for employment, the presence of small children might arguably be associated with personal characteristics also affecting the wage and thus violate IV validity - a concern we aim to mitigate by including a rich set of individual pre-treatment characteristics in $X$ that is likely associated with both fertility and wages.

For assessing the longer run effects, we invoke the sequential conditional independence assumption of Section \ref{dynsel} to apply DML based on Theorem 4. To this end, we additionally control for $619$ and $156$ post-treatment covariates ($M$) in the second and third year, respectively, after JC assignment, which include detailed information on the labor market participation after the first and prior to the second treatment. Appendix \ref{desstat} presents descriptive statistics for selected variables in $X$ and $M$.  We also refer to \cite{BoHuLa2020} for a more detailed description of the pre-and post-treatment covariates used in our application and note that all numeric variables have been standardized to have a mean equal to $0$ and standard deviation equal to $0.5$ to facilitate the machine learning-based estimation of the nuisance parameters. For estimation, we apply the \textit{treatselDML} and \textit{dyntreatDML} commands of the \textit{causalweight} package for \textsf{R}, using 3-fold cross-fitting and the random forest  (with default options of the \textit{SuperLearner} package) as machine learner. The latter is a nonparametric approach allowing for nonlinear associations between the outcome, treatment, and selection on the one hand and the covariates on the other hand.

\begin{table}[htbp]
	\centering\normalsize
	\caption{Treatment distribution}
	\label{tab:des}
	\begin{adjustbox}{max width=\textwidth}
		\begin{tabular}{cc}
			\hline\hline
			treatment    & observations \\
			\hline
			  randomized out of JC & 1698 \\
		 controls (no training) & 200  \\
			academic training & 830  \\
			 vocational training & 843  \\
			\hline	
		\end{tabular}
	\end{adjustbox}
\end{table}

Table \ref{tab:des} reports the total number of females randomized into JC for whom participation in either vocational ($843$) or academic training ($830$) in the first year after program assignment is registered in the data. $200$ females did not participate in any JC training activity in the first year and serve as the control group in our analysis. Furthermore, $1698$ were randomized out of JC. 

\begin{table}[htbp]
	\centering\normalsize
	\caption{ATE estimates}
	\label{tab:results1}
	\begin{adjustbox}{max width=\textwidth}
		\begin{tabular}{c|ccc}
			\hline\hline
			$D=1$ $\hspace{5mm}$ $D=0$ & ATE & standard error & p-value \\
			\hline
			\multicolumn{4}{c}{Theorem 1 (MAR) } \\
			\hline
			academic $\hspace{5mm}$ no training & -0.683   & 1.073   & 0.524  \\
			vocational $\hspace{5mm}$ no training & 0.611   &  0.629 &  0.331  \\
			\hline
			\multicolumn{4}{c}{Theorem 3 (IV) } \\
			\hline
			academic $\hspace{5mm}$ no training & -0.631 & 1.052 & 0.549  \\
			vocational $\hspace{5mm}$ no training & 0.586   & 0.645 & 0.364 \\
			\hline	
			\multicolumn{4}{c}{Theorem 4 (sequential) } \\
			\hline
			academic $\hspace{5mm}$ no training &  0.149  & 0.199 & 0.454  \\
			vocational $\hspace{5mm}$ no training &  0.567  & 0.208 & 0.007  \\
			\hline
		\end{tabular}
	\end{adjustbox}
\end{table}

Table \ref{tab:results1} reports the ATE estimates for academic and vocational training based on our various DML approaches. The upper panel provides the short-run effects on hourly wages in the last week of the first year when assuming MAR. The point estimate of academic training is negative ($-0.683$ US \$), while that of vocational training is positive ($0.611$ US \$), but neither effect is statistically significant at any conventional level. The findings are very similar when considering the IV-based estimates shown in the intermediate panel, with negative and positive effects for the academic and vocational training, respectively, which are again not statistically significant. The lower panel provides the longer-run effects on hourly wages 4 years after assignment based on the sequential conditional independence assumption. While now both ATE estimates are positive, only the effect of vocational training, which amounts to an hourly increase of $0.567$ \$, is highly statistically significant. Our findings therefore suggest that JC-based education may facilitate human capital accumulation in a way that increases hourly wages after several years, in particular through vocational training, while there is no clear-cut evidence for short term effects.

\section{Conclusion}\label{conclusion}

In this paper, we discussed the evaluation of average treatment effects in the presence of sample selection or outcome attrition based on double machine learning. In terms of identifying assumptions, we imposed a selection-on-observables assumption on treatment assignment, which was combined with either selection-on-observables or instrumental variable assumptions concerning the outcome attrition/sample selection process. We also considered a sequential selection-on-observables assumption allowing for dynamic confounding such that covariates jointly affecting the outcome and sample selection may be affected by the treatment, which avoids exclusively relying on pre-treatment covariates. We proposed doubly robust score functions and formally showed the satisfaction of Neyman orthogonality, implying that estimators based on these score functions are robust to moderate (local) regularization biases in the machine learning-based estimation of the outcome, treatment, or sample selection models. Furthermore, we demonstrated the root-n consistency and asymptotic normality of our double machine learning approach to average treatment effect estimation under specific regularity conditions. We also provided an empirical illustration to the US Job Corps data, in which we assessed the effects of training on hourly wage one and four years after program assignment and found some statistical evidence for positive longer-run impacts. Our estimation procedure is available in the \textit{causalweight} package for the statistical software \textsf{R}.
\pagebreak
\typeout{}
\bibliographystyle{econometrica}
\bibliography{research}

\bigskip

\renewcommand\appendix{\par
	\setcounter{section}{0}%
	\setcounter{subsection}{0}%
	\setcounter{table}{0}%
	\setcounter{figure}{0}%
	\renewcommand\thesection{\Alph{section}}%
	\renewcommand\thetable{\Alph{section}.\arabic{table}}}
\renewcommand\thefigure{\Alph{section}.\arabic{subsection}.\arabic{subsubsection}.\arabic{figure}}
\clearpage

\begin{appendix}
	
	\numberwithin{equation}{section}
	\noindent \textbf{\LARGE Appendices}

{\footnotesize

\section{Proofs of theorems}

For the proofs of Theorems 1, 2, and 3 it is sufficient to verify the conditions of Assumptions 3.1 and 3.2 of Theorems 3.1 and 3.2 as well as Corollary 3.2 in \cite{Chetal2018}. All bounds hold uniformly over $P \in \mathcal{P},$ where $\mathcal{P}$ is the set of all possible probability laws, and we omit $P$ for brevity.

\subsection{Proof of Theorem 1} \label{Neyman1}

Define the nuisance parameters to be the vector of functions $\eta=(p_d(X), \pi(D, X), \mu (D,S,X))$,
with $p_d(X)=\Pr(D=d |X )$, $\pi(D,X)=\Pr(S=1|D,X)$,  and $\mu (D,S,X)=E[Y |D,S,X]$. The Neyman-orthogonal score function for the counterfactual $\Psi_{d0}=E[Y(d)]$ is given by the following expression, with $W  = (Y \cdot S, D, S, X)$:
\begin{eqnarray}
\psi_{d}(W,  \eta,  \Psi_{d0}) &=& \frac{I\{D=d\} \cdot S \cdot [Y- \mu(d,1,X)] }{ p_d(X)\cdot  \pi (d,X)}  + \mu(d,1,X)  - \Psi_{d0}.
\end{eqnarray}

Let $\mathcal{T}_n$ be the set fo all $\eta=(p_d, \pi, \mu)$ consisting of $P$-square integrable functions $p_d$, $\pi$, $\mu$ such that
\begin{eqnarray}\label{Tn}
\left\|  \eta - \eta_0 \right\|_{q} &\leq& C,  \\
\left\|  \eta - \eta_0 \right\|_{2} &\leq& \delta_n, \notag \\
\left\|   p_d(X)-1/2\right\|_{\infty}  &\leq& 1/2-\epsilon, \notag\\
\left\|   \pi(D,X)-1/2)\right\|_{\infty} &\leq & 1/2-\epsilon, \notag \\
\left\|   \mu(D,S,X)-\mu_0(D,S,X)\right\|_{2} \times \left\|   p_{d}(X)-p_{d0}(X)\right\|_{2}  &\leq & \delta^{}_n n^{-1/2}, \notag \\
\left\|   \mu(D,S,X)-\mu_0(D,S,X)\right\|_{2} \times \left\|   \pi(D,X)-\pi_{0}(D,X)\right\|_{2} &\leq & \delta^{}_n n^{-1/2}.\notag
\end{eqnarray}

We furthermore replace the sequence $(\delta_n)_{n \geq 1}$ by $(\delta_n')_{n \geq 1},$ where $\delta_n' = C_{\epsilon} \max(\delta_n,n^{-1/2}),$ where $C_{\epsilon}$ is sufficiently large constant that only depends on $C$ and $\epsilon.$

\newpage
\textbf{Assumption 3.1:  Linear scores and Neyman orthogonality}
\vspace{5pt}\newline

\textbf{Assumption 3.1(a)}

\textbf{Moment Condition:} The moment condition $E\Big[\psi_{d}(W, \eta_0,  \Psi_{d0})\Big] = 0$ holds:
\begin{eqnarray}
E\Big[\psi_{d}(W,  \eta_0,  \Psi_{d0})\Big] &=& E\Bigg[ \overbrace{E\Bigg[\frac{I\{D=d\} \cdot S \cdot [Y- \mu_0(d,1,X)]}{ p_{d0}(X)\cdot  \pi_0(d,X)}\Bigg| X\Bigg]}^{=E[[Y- \mu_0(d,1,X)|D=d,S=1,X]=0}   + \mu_0(d,1,X)-\Psi_{d0}\Bigg] \notag\\
&=& E[ \mu_0(d,1,X)]- \Psi_{d0} =  0, \notag
\end{eqnarray}
where the first equality follows from the law of iterated expectations.

\textbf{Assumption 3.1(b)}
\textbf{Linearity:} The score $ \psi_{d}(W,  \eta_0,  \Psi_{d0}) $ is linear in $\Psi_{d0}$ :
$\psi_{d}(W, \eta_0, \Psi_{d0}) = \psi_{d}^a(W, \eta_0) \cdot\Psi^{d}_0 + \psi_{d}^b(W, \eta_0) $
with $\psi_{d}^a(W, \eta_0) = -1$ and
\begin{eqnarray}
\psi_{d}^b(W, \eta_0) &=&\frac{I\{D=d\} \cdot S \cdot [Y- \mu_0(d,1,X)] }{ p_{d0}(X)\cdot  \pi_0 (d,X)}  + \mu_0(d,1,X). \notag
\end{eqnarray}

\textbf{Assumption 3.1(c)}

\textbf{Continuity:}
The expression for the second Gateaux derivative of a map $\eta \mapsto E[\psi_{d}(W,  \eta,  \Psi_{d0})]$, given in (\ref{secondGatDer}), is continuous.

\textbf{Assumption 3.1(d)}

\textbf{Neyman Orthogonality}: For any $\eta \in \mathcal{T}_n,$ the Gateaux derivative in the direction $ \eta - \eta_0 =
(\pi(D ,X)-\pi_0 (D , X), p^{d }(X )-p_0^{d }(X ),  \mu(D ,S, X)-\mu_0 (D ,S, X )$ is given by:
\begin{align}
&\partial E \big[\psi_{d}(W, \eta, \Psi_{d})\big] \big[\eta - \eta_0 \big]  = \notag\\
& - E \Bigg[  \frac{I\{D=d\} \cdot S \cdot [\mu(d,1,X)-\mu_0(d,1,X)]}{p_{d0}(X)\cdot \pi_0(d,X)}  \Bigg]  \tag{$*$}\\
& + E [\mu(d,1,X)-\mu_0(d,1,X)]  \tag{$**$}\\
&- E \Bigg[ \frac{\overbrace{I\{D=d \} \cdot S \cdot  [Y -\mu_0 (d,1, X )]}^{E[\cdot|X]=E[ Y -\mu_0 (d,1, X )|D=d,S=1,X]=0}}{ p_{d0}(X)\cdot \pi_0(d,X) }\cdot\frac{p_{d}(X)-p_{d0}(X) }{p_{d0}(X)} \Bigg] \notag\\
&- E \Bigg[ \frac{\overbrace{I\{D=d \} \cdot S \cdot  [Y -\mu_0 (d,1, X )]}^{E[\cdot|X]=0}}{ p_{d0}(X)\cdot \pi_0(d,X) }\cdot\frac{\pi(d,X)-\pi_0(d,X) }{\pi_0(d,X)} \Bigg] =0.\notag
\end{align}
The Gateaux derivative is zero because expressions $(*)$ and $(**)$ cancel out. To see this, note that by the law of iterated expectations, $(*)$ corresponds to
\begin{eqnarray}
&&-E\Bigg[ E\Bigg[ \frac{I\{D=d\}}{p_{d0}(X)} \cdot E\Bigg[ \frac{ S \cdot [\mu(d,1,X)-\mu_0(d,1,X)] }{\pi_0(d,X)}\Bigg| D=d,  X\Bigg] \Bigg| X \Bigg]\Bigg]\notag\\
&=&-E\Bigg[ E\Bigg[ \frac{I\{D=d\}}{p_{d0}(X)} \cdot  \frac{ \overbrace{E[S| D=d,  X ]}^{=\pi_0(d,X)} \cdot [\mu(d,1,X)-\mu_0(d,1,X)] }{\pi_0(d,X)} \Bigg| X  \Bigg]\Bigg]\notag\\
&=& -E\Bigg[  \frac{\overbrace{E [I\{D=d\} | X  ]}^{=p_{d0}(X)}}{p^{d0}(X)} \cdot   [\mu(d,1,X)-\mu_0(d,1,X)] \Bigg]= -E [\mu(d,1,X)-\mu_0(d,1,X)], \notag
\end{eqnarray}
Therefore,
\begin{align}
&\partial E \big[\psi_{d}(W, \eta, \Psi_{d})\big] \big[\eta - \eta_0 \big] = 0 \notag
\end{align}
proving that the score function is orthogonal.

\bigskip

\textbf{Assumption 3.2:  Score regularity and quality of nuisance parameter estimators}
\vspace{5pt}\newline

\textbf{Assumption 3.2(a)}

This assumption directly follows from the construction of the set $\mathcal{T}_n$ and the regularity conditions (Assumption 10).

\textbf{Assumption 3.2(b)}

\textbf{Bound for $m_N$:}

Consider the following inequality
\begin{eqnarray}
\left\|  \mu_0(D,S,X)  \right\|_{q} &=& \left( E\left[ \left| \mu_0(D,S,X)  \right|^q \right] \right)^{\frac{1}{q}} \notag \\
&=&  \left( \sum_{d \in \{0,1,...,Q\}, s \in \{0,1\} } E\left[ \left| \mu_0(d,s,X) \right|^q \Pr(D=d,S=s|X)  \right]  \right)^{\frac{1}{q}} \notag \\
&\geq& \epsilon^{2/q} \left( \sum_{d \in \{0,1,...,Q\}, s \in \{0,1\} } E\left[ \left| \mu_0(d,s,X) \right|^q \right] \right)^{\frac{1}{q}} \notag \\
&\geq&  \epsilon^{2/q} \left( \max_{d \in \{0,1,...,Q\}, s \in \{0,1\} } E\left[ \left| \mu_0(d,s,X) \right|^q \right] \right)^{\frac{1}{q}} \notag  \\
&=& \epsilon^{2/q} \left( \max_{d \in \{0,1,...,Q\}, s \in \{0,1\} } \left\|  \mu_0(d,s,X) \right\|_{q}\right), \notag
\end{eqnarray}
where the first equality follows from definition, the second from the law of total probability, first inequality from the fact that $\Pr(D=d,S=1|X) = p_{d0}(X) \cdot \pi_0(d,X) \geq \epsilon^2$ and $\Pr(D=d,S=0|X) = p_{d0}(X) \cdot (1-\pi_0(d,X)) \geq \epsilon^2.$ Furthermore, by Jensen's inequality
$\left\|  \mu_0(D,S,X)  \right\|_{q} \leq \left\|  Y  \right\|_{q}$ and hence
$\left\|  \mu_0(d,1,X)  \right\|_{q} \leq C/\epsilon^{2/q}$ by conditions  (\ref{Tn}).
Using similar steps, for any $\eta \in \mathcal{T}_N:$
$\left\|  \mu(d,1,X) - \mu_0(d,1,X)  \right\|_{q} \leq C/\epsilon^{2/q}$ because
$\left\|  \mu(D,S,X) - \mu_0(D,S,X)  \right\|_{q} \leq C.$

Consider
\begin{eqnarray}
E\Big[ \psi_d(W, \eta, \Psi_{d0})\Big] &=& E\Bigg[ \underbrace{ \frac{ I\{D=d\} \cdot S }{p_d(X)\cdot \pi(d,X)} \cdot Y }_{=I_1}  +  \underbrace{  \bigg(1-   \frac{I\{D=d\} \cdot S }{p_d(X)\cdot \pi(d,X)}\bigg)  \mu(d,1,X)}_{=I_2} - \Psi_{d0}  \Bigg]  \notag
\end{eqnarray}
and thus
\begin{eqnarray}
\left\| \psi_d(W, \eta, \Psi_{d0}) \right\|_{q} &\leq&  \left\| I_1 \right\|_{q}  + \left\| I_2 \right\|_{q}  +  \left\| \Psi_{d0}  \right\|_{q} \notag \\
&\leq& \frac{1}{\epsilon^2}  \left\| Y \right\|_{q} \notag + \frac{1-\epsilon}{\epsilon} \left\|  \mu(d,1,X)  \right\|_{q}   +  | \Psi_{d0} | \notag \\
&\leq& C \left( \frac{1}{\epsilon^2} + \frac{2}{\epsilon^{2/q}} \cdot  \frac{1-\epsilon}{\epsilon}  + \frac{1}{\epsilon} \right),  \notag
\end{eqnarray}
because of triangular inequality and because the following set of inequalities hold:
\begin{eqnarray} \label{32b}
\left\|  \mu(d,1,X)  \right\|_{q} &\leq& \left\|  \mu(d,1,X) -  \mu_0(d,1,X)   \right\|_{q} + \left\|   \mu_0(d,1,X)   \right\|_{q} \leq 2C/\epsilon^{2/q},  \\
|\Psi^{\underline{d}_2}_{0} | &=& |E[  \mu_0(d,1,X) ] | \leq  E_{ } \Big[\left|  \mu_0(d,1,X)   \right|^1 \Big]^{\frac{1}{1}} =  \left\|  \mu_0(d,1,X)  \right\|_{P,1}  \notag \\
&\leq&  \left\|  \mu_0(d,1,X)  \right\|_{2} \leq  \left\| Y \right\|_{2}/\epsilon^{2/2} \overbrace{ \leq}^{q > 2}  \left\| Y \right\|_{q}/\epsilon \leq C /\epsilon.  \notag
\end{eqnarray}
which gives the upper bound on $m_n$ in Assumption 3.2(b) of \cite{Chetal2018}.

\textbf{Bound for $m'_n$:}

Notice that
$$\Big(E[ |\psi_d^{a}(W, \eta) |^q] \Big)^{1/q}=1$$
and this gives the upper bound on $m'_N$ in Assumption 3.2(b).

\textbf{Assumption 3.2(c)}

\textbf{Bound for $r_n$:}

For any $\eta = (p_d,\pi,\nu)$ we have
$$ \Big| E\Big( \psi^a_d(W, \eta) - \psi^a_{d0}(W, \eta_0) \Big) \Big| = |1-1| = 0 \leq \delta'_N,$$
and thus we have the bound on $r_n$ from Assumption 3.2(c).

In the following, we omit arguments for the sake of brevity and use $p_d= p_d(X), \pi = \pi(d,X),  \mu = \mu(d,1,X)$ and similarly for $p_{d0},\pi_0,\mu_0.$

\textbf{Bound for $r'_n$:}
\begin{eqnarray}\label{rprimeN}
&& \left\|   \psi_d(W, \eta, \Psi_{d0}) - \psi_d(W, \eta_0, \Psi_{d0})  \right\|_{2} \leq  \left\|    I\{D=d\} \cdot S \cdot Y \cdot \left( \frac{1}{p_d \pi } - \frac{1}{p_{d0} \pi_0} \right) \right\|_{2}   \\
&+&  \left\|     I\{D=d\} \cdot S \cdot \left( \frac{\mu}{p_d \pi} - \frac{\mu_0}{p_{d0} \pi_0} \right) \right\|_{2}  + \left\| \mu - \mu_0 \right\|_{2} \notag \\
&\leq&  \left\| Y \cdot \left( \frac{1}{p_d \pi } - \frac{1}{p_{d0} \pi_0} \right) \right\|_{2} +   \left\|  \frac{\mu}{p_d \pi} - \frac{\mu_0}{p_{d0} \pi_0}  \right\|_{2}  + \left\| \mu - \mu_0 \right\|_{2} \notag \\
&\leq&
\frac{C}{\epsilon^4} \delta_n \left(1 + \frac{1}{\epsilon} \right) + \delta_n \left(  \frac{1}{\epsilon^5} + C + \frac{C}{\epsilon} \right) +  \frac{\delta_n}{\epsilon}
\leq \delta_n' \notag
\end{eqnarray}
as long as $C_\epsilon$ in the definition of $\delta_n'$ is sufficiently large.  This gives the bound on $r'_n$ from Assumption 3.2(c).
Here we made use of the fact that $\left\| \mu - \mu_0 \right\|_{2} = \left\|  \mu(d,1,X) - \mu_0(d,1,X) \right\|_{2} \leq \delta_n/\epsilon,$ and $\left\| \pi - \pi_0 \right\|_{2} = \left\| \pi (d,X) - \pi _0(d,X) \right\|_{2} \leq \delta_n/\epsilon$ using similar steps as in Assumption 3.1(b).

The last inequality in (\ref{rprimeN}) holds because for the first term we have
\begin{eqnarray*}
	&&  \left\| Y \cdot \left( \frac{1}{p_d \pi } - \frac{1}{p_{d0} \pi_0} \right) \right\|_{2}  \leq   C  \left\| \frac{1}{p_d \pi } - \frac{1}{p_{d0} \pi_0}  \right\|_{2} \leq \frac{C}{\epsilon^4} \left\| p_{d0} \pi_0 - p_d \pi  \right\|_{2} \\
	&=& \frac{C}{\epsilon^4} \left\| p_{d0}\pi_0 - p_{d}\pi +  p_{d0} \pi  -  p_{d0} \pi  \right\|_{2} \leq \frac{C}{\epsilon^4}\left( \left\| p_{d0} (\pi_0 -\pi ) \right\|_{2} + \left\|\pi_0 (p_{d0} - p_{d})  \right\|_{2} \right) \\
	&\leq&  \frac{C}{\epsilon^4}\left( \left\| \pi_0 - \pi \right\|_{2} + \left\| p_{d0} - p_{d}  \right\|_{2} \right) \leq \frac{C}{\epsilon^4} \delta_n \left(1 + \frac{1}{\epsilon} \right),
\end{eqnarray*}
where the first inequality follows from the second inequality in Assumption 4(a). The second term in (\ref{rprimeN}) is bounded by
\begin{eqnarray*}
	&& \left\|  \frac{\mu}{p_{d}\pi} - \frac{\mu_0}{p_{d0}\pi_0} \right\|_{2} \leq \frac{1}{\epsilon^4} \left\| p_{d0}\pi_0 \mu -  p_{d}\pi \mu_0 \right\|_{2} =  \frac{1}{\epsilon^4}\left\|  p_{d0}\pi_0 \mu -  p_{d}\pi  \mu_0 + p_{d0}\pi_0 \mu_0 - p_{d0}\pi_0 \mu_0 \right\|_{2} \\
	&\leq&  \frac{1}{\epsilon^4} \left( \left\| p_{d0}\pi_0  (\mu - \mu_0) \right\|_{2} + \left\| \mu_0 ( p_{d0}\pi_0  - p_{d}\pi  ) \right\|_{2} \right) \leq   \frac{1}{\epsilon^4} \left( \left\| \mu - \mu_0 \right\|_{2} + C  \left\| p_{d0}\pi_0  - p_{d}\pi \right\|_{2} \right) \\
	&\leq& \frac{1}{\epsilon^4} \left( \frac{\delta_n}{\epsilon} + C \left\| p_{d0}\pi_0  - p_{d}\pi \right\|_{2} \right) \leq \delta_n \left(  \frac{1+C}{\epsilon^5} + \frac{C}{\epsilon^4}\right) ,
\end{eqnarray*}
where the third inequality follows from $E[Y^2| D=d, S=1, X] \geq (E[Y| D=d, S=1,X] )^2= \mu^2_0(d,1,X) $ by conditional Jensen's inequality and therefore $\left\| \mu_0 (d,1,X) \right\|_\infty \leq C^2.$

\textbf{Bound for $\lambda'_n$:}

Now consider


\begin{equation}
f(r) := E[\psi_d(W;\Psi_{d0},\eta + r(\eta-\eta_0)] \notag
\end{equation}
For any $r \in (0,1):$

\begin{eqnarray} \label{secondGatDer}
\frac{\partial^2 f(r)}{\partial r^2}&=& E\Bigg[ 2 \cdot I\{D=d\} \cdot S \cdot (Y - \mu_0 - r(\mu-\mu_0)) \frac{(p_d - p_{d0})^2 }{ \left(p_{d0} + r (p_d - p_{d0}) \right)^3 \left(\pi_{0} + r (\pi - \pi_{0}) \right)}   \Bigg]  \\ 
&+& E\Bigg[ 2 \cdot I\{D=d\} \cdot S \cdot (Y - \mu_0 - r(\mu-\mu_0)) \frac{(\pi - \pi_{0})^2 }{ \left(p_{d0} + r (p_d - p_{d0}) \right) \left(\pi_{0} + r (\pi - \pi_{0}) \right)^3}   \Bigg]  \notag \\  
&+& E\Bigg[ 2 \cdot I\{D=d\} \cdot S \cdot (Y - \mu_0 - r(\mu-\mu_0)) \frac{(p_d - p_{d0}) (\pi - \pi_{0}) }{ \left(p_{d0} + r (p_d - p_{d0}) \right)^2 \left(\pi_{0} + r (\pi - \pi_{0}) \right)^2}   \Bigg]  \notag \\ 
&+& E\Bigg[ 2 \cdot I\{D=d\} \cdot S \cdot  (\mu-\mu_0)  \frac{(p_d - p_{d0}) \left(\pi_{0} + r (\pi - \pi_{0}) \right) }{ \left(p_{d0} + r (p_d - p_{d0}) \right)^2 \left(\pi_{0} + r (\pi - \pi_{0}) \right)^2}   \Bigg]  \notag \\ 
&+& E\Bigg[ 2 \cdot I\{D=d\} \cdot S \cdot  (\mu-\mu_0)  \frac{\left(p_{d0} + r (p_d - p_{d0}) \right) (\pi - \pi_{0})}{ \left(p_{d0} + r (p_d - p_{d0}) \right)^2 \left(\pi_{0} + r (\pi - \pi_{0}) \right)^2}   \Bigg]  \notag  
\end{eqnarray}

Note that because
\begin{eqnarray}
E[Y-\mu_0(d,1,X)|D=d,S=1,X]  &= & 0, \notag \\
|p_d - p_{d0}|  \leq  2, \ \  \ \ \ \ |\pi - \pi_{0}|  &\leq & 2 \notag \\
\left\| \mu_0 \right\|_{q} \leq  \left\| Y \right\|_{q}/\epsilon^{1/q} &\leq & C/\epsilon^{2/q} \notag \\
\left\|  \mu-\mu_0\right\|_{2} \times \left\|  p_d-p_{d0}\right\|_{2}  &\leq & \delta^{}_n n^{-1/2}/\epsilon, \notag \\
\left\|  \mu-\mu_0\right\|_{2} \times \left\|  \pi- \pi_0\right\|_{2} &\leq & \delta^{}_n n^{-1/2}/\epsilon^2,\notag
\end{eqnarray}
we get that for some constant $C_{\epsilon}''$ that only depends on $C$ and $\epsilon$
\begin{equation}
\left|\frac{\partial^2 f(r)}{\partial r^2} \right| \leq C_{\epsilon}'' \delta_n n^{-1/2} \leq \delta_n' n^{-1/2} \notag
\end{equation}
and this gives the upper bound on $\lambda'_n$ in Assumption 3.2(c) of \cite{Chetal2018} as long as $C_{\epsilon} \geq C_{\epsilon}''$. We used the following inequalities
\begin{eqnarray}
\left\| \mu-\mu_0\right\|_{2} &=& \left\|\mu(d,1,X)-\mu_0(d,1,X)\right\|_{2} \leq  \left\|\mu(D,S,X)-\mu_0(D,S,X)\right\|_{2}/\epsilon \notag \\
\left\| \pi- \pi_0\right\|_{2} &=& \left\| \pi(d,X)-\pi_0(d,X)\right\|_{2} \leq  \left\| \pi(D,X)-\pi_0(D,X)\right\|_{2}/\epsilon, \notag
\end{eqnarray}
and these can be shown using similar steps as in Assumption 3.1(b).

To verify that $\left|\frac{\partial^2 f(r)}{\partial r^2} \right|  \leq C_{\epsilon}'' \delta_n n^{-1/2}$ holds, note that by the triangular inequality it is sufficient to bound the absolute value of each of the ten terms in (\ref{secondGatDer}) separately. We illustrate it for the first and last terms. For the first term:
\begin{eqnarray}
&& \left| E\Bigg[ 2 \cdot I\{D=d\} \cdot S \cdot (Y - \mu_0 - r(\mu-\mu_0)) \frac{(p_d - p_{d0})^2 }{ \left(p_{d0} + r (p_d - p_{d0}) \right)^3 \left(\pi_{0} + r (\pi - \pi_{0}) \right)}   \Bigg]  \right|  \notag \\
&\leq& \frac{2}{\epsilon^4} \left| E\Bigg[ I\{D=d\} \cdot S \cdot (Y - \mu_0 - r(\mu-\mu_0) )(p_d - p_{d0})^2 \Bigg] \right|  \notag \\
&\leq&  \frac{8}{\epsilon^4} \left| E\Bigg[ I\{D=d\} \cdot S \cdot (Y - \mu_0) \Bigg] \right|  +  \frac{2}{\epsilon^4}  \left| E\Bigg[r(\mu-\mu_0) (p_d - p_{d0})^2 \Bigg] \right| \notag \\
&\leq& \frac{2 \cdot 2}{\epsilon^4}  \left| E\Bigg[1\cdot(\mu-\mu_0) (p_d - p_{d0}) \Bigg] \right| \leq \frac{4}{\epsilon^4} \frac{\delta^{}_n}{\epsilon} n^{-1/2}. \notag
\end{eqnarray}
For the second inequality we used the fact that for  $1 \geq p_{d0} + r(p_d -p_{d0}) =  (1-r)p_{d0} + r p_{d} \geq (1-r)\epsilon + r \epsilon = \epsilon$ and similarly for $\pi$ and in the third Holder's inequality. Bounding of the second and third terms follows similarly.

For the fourth term, we get
\begin{eqnarray}
&& \left| E\Bigg[ 2 \cdot I\{D=d\} \cdot S \cdot  (\mu-\mu_0)  \frac{(p_d - p_{d0}) \left(\pi_{0} + r (\pi - \pi_{0}) \right) }{ \left(p_{d0} + r (p_d - p_{d0}) \right)^2 \left(\pi_{0} + r (\pi - \pi_{0}) \right)^2}   \Bigg] \right|  \notag \\
&\leq& \frac{2}{\epsilon^4} \left| E\Bigg[  I\{D=d\} \cdot S \cdot (\mu-\mu_0) (p_d - p_{d0})   \Bigg] \right|  \leq \frac{2}{\epsilon^4} \frac{\delta^{}_n}{\epsilon} n^{-1/2}  \notag
\end{eqnarray}
where in addition we made use of conditions (\ref{Tn}). The last term is bounded similarly.

\textbf{Assumption 3.2(d)} 
\begin{eqnarray}
E\Big[ (\psi^{d}(W, \eta_0, \Psi_{d0}) )^2\Big] &=& E\Bigg[ \Bigg( \underbrace{ \frac{ I\{D=d\} \cdot S \cdot [Y-\mu_0]}{p_{d0}\cdot \pi_{0}} }_{=I_1}
+  \underbrace{ \mu_0 - \Psi_{d0}}_{=I_2} \Bigg)^2 \Bigg]  \notag\\
& = & E[I_1^2 + I_2^2] \geq E[I^2_1]\notag\\
& = & E\Bigg[  I\{D=d\} \cdot S \cdot \Bigg( \frac{ [Y-\mu_0]}{p_{d0}\cdot \pi_{0} } \Bigg)^2 \Bigg] \notag\\
& \geq & \epsilon^2 E\Bigg[ \Bigg( \frac{ [Y-\mu_0]}{p_{d0}\cdot \pi_{0} } \Bigg)^2 \Bigg] \notag\\
&\geq& \frac{\epsilon^2 c^2}{(1-\epsilon)^4} > 0, \notag
\end{eqnarray}
because $\Pr(D = d, S=1  | X )= p_{d0}(X)\cdot \pi_{0}(d,X) \geq \epsilon^2, \  p_{d0}(X) \leq 1-\epsilon$ and $\pi_0(d,X) \leq 1-\epsilon$.



The second equality follows from
\begin{eqnarray}
E\Big[ I_1 \cdot I_2\Big] &=& E\Bigg[  \overbrace{\frac{ I\{D=d\} \cdot S \cdot [Y-\mu_0(d,1,X)]}{ p_{d0}(X)\cdot \pi_{0}(d,X)}  }^{E[\cdot| X]=0}   \cdot  [\mu_0(d,1,X) - \Psi_{d0}] \Bigg]. \notag
\end{eqnarray}

\subsection{Proof of Theorem 2} \label{Neyman2}

Define the nuisance parameters to be the vector of functions $\eta=(\pi(D, X, Z), p_d(X,\Pi), \mu (D,S,X,\Pi))$,
with  $\Pi=\pi(D,X,Z)=\Pr(S=1|D,X,Z)$, $p_d(X, \Pi)=\Pr(D=d |X, \pi(D,X,Z) )$, and $\mu (D,S,X,\Pi)=E[Y |D,S,X, \pi(D,X,Z)]$.

The shrinking neighbourhood $\mathcal{T}^*_n$ of nuisance parameter vector $\eta=(\pi, p_d, \mu)$ is defined analogously to  $\mathcal{T}_n$ from (\ref{Tn}) in the proof of theorem 1.

The score function for the counterfactual $\Psi_{d0}^{S=1}=E[Y(d)|S=1]$ is given by:
\begin{eqnarray}\label{phi}
\phi_{d,S=1}(W,  \eta,  \Psi_{d0}^{S=1}) &=& \frac{I\{D=d\} \cdot  [Y- \mu(d,1,X,\Pi)] }{ p_d(X)}  + \mu(d,1,X,\Pi)  - \Psi_{d0}^{S=1}.
\end{eqnarray}

\textbf{Assumption 3.1:  Linear scores and Neyman orthogonality}
\vspace{5pt}\newline

\textbf{Assumption 3.1(a)}

\textbf{Moment Condition:} The moment condition $E\Big[\phi_{d,S=1}(W,  \eta_0,  \Psi_{d0}^{S=1})|S=1\Big] = 0$ holds:
\begin{eqnarray}
E\Big[\phi_{d, S=1}(W,  \eta_0,  \Psi_{d0}^{S=1})\Big|S=1\Big] &=& E\Bigg[ \overbrace{E\Bigg[\frac{I\{D=d\} \cdot [Y-\mu_0(d,1,X,\Pi_0)] }{p_{d0}(X,\Pi_0)}\Bigg|S=1, X,\Pi_0\Bigg]}^{=E[Y-\mu_0(d,1,X,\Pi_0)|D=d,S=1,X,\Pi_0]=0} \notag\\
&+&\mu_0(d,1,X,\Pi_0)-\Psi_{d0}^{S=1}\Bigg|S=1 \Bigg] \notag\\
&=& E[ \mu_0(d,1,X,\Pi_0)|S=1]- \Psi_{d0}^{S=1} =  0, \notag
\end{eqnarray}
where the first equality follows from the law of iterated expectations.

\textbf{Assumption 3.1(b)}
\textbf{Linearity:} The score $ \phi_{d,S=1}(W,  \eta_0,  \Psi_{d0}^{S=1}) $ is linear in $\Psi_{d0}^{S=1}$ :
$\phi_{d, S=1}(W, \eta_0, \Psi_{d0}^{S=1}) = \phi_{d,S=1}^a(W, \eta_0) \cdot \Psi_{d0}^{S=1} + \phi_{d,S=1}^b(W, \eta_0) $
with $\phi_{d, S=1}^a(W, \eta_0) = -1$ and
\begin{eqnarray}
\phi_{d, S=1}^b(W, \eta_0) &=&\frac{I\{D=d\} \cdot [Y- \mu_0(d,1,X,\Pi_0)] }{ p_{d0}(X,\Pi_0)}  + \mu_0(d,1,X,\Pi_0). \notag
\end{eqnarray}

\textbf{Assumption 3.1(c)}

\textbf{Continuity:}
The expression for the second Gateaux derivative of a map $\eta \mapsto E[\phi_{d, S=1}(W,  \eta,  \Psi_{d0}^{S=1})]$, given in (\ref{phi}), is continuous.

\textbf{Assumption 3.1(d)}

\textbf{Neyman Orthogonality}: For any $\eta \in \mathcal{T}_n,$ the Gateaux derivative in the direction $ \eta - \eta_0 =
( \pi(D ,X, Z)-\pi_0 (D , X, Z), p_{d }(X, \Pi )-p_{d0}(X, \Pi ), \mu(D ,S, X, \Pi)-\mu_0 (D ,S, X, \Pi ))$ is given by:
\begin{align}
&\partial E \big[\phi_{d, S=1}(W, \eta, \Psi_{d}^{S=1}) |S=1 \big] \big[\eta - \eta_0 \big]  = \notag\\
& - E \Bigg[  \frac{I\{D=d\}  \cdot [\mu(d,1,X, \pi_0(d,X,Z))-\mu_0(d,1,X,\pi_0(d,X,Z))]}{p_{d0}(X, \pi_0(d,X,Z))} \Bigg|S=1 \Bigg]  \tag{$*$}\\
& + E [\mu(d,1,X,\pi_0(d,X,Z))-\mu_0(d,1,X,\pi_0(d,X,Z))|S=1]  \tag{$**$}\\
&- E \Bigg[ \frac{\overbrace{I\{D=d \} \cdot    [Y -\mu_0 (d,1, X,\pi_0(d,X,Z) )]}^{E[\cdot|S=1,X,\Pi_0]= E[Y-\mu_0(d,1,X,\Pi_0)|D=d,S=1,X,\Pi_0]=0}}{ p_{d0}(X, \pi_0(d,X,Z))  } \cdot\frac{p_{d}(X, \pi_0(d,X,Z))-p_{d0}(X, \pi_0(d,X,Z))}{ p_{d0}(X, \pi_0(d,X,Z))  } \Bigg|S=1 \Bigg] \notag\\
&- E \Bigg[ \frac{I\{D=d \} \cdot  \partial E[\mu_0 (d,1, X,\pi_0(d,X,Z) )]\cdot[\pi(d,X,Z)-\pi_0(d,X,Z)]}{ p_{d0}(X, \pi_0(d,X,Z)) } \Bigg|S=1 \Bigg]\tag{$***$}\\
&- E \Bigg[ \frac{\overbrace{I\{D=d \}  \cdot  [Y -\mu_0 (d,1, X,\pi_0(d,X,Z) )]}^{E[\cdot|S=1,X,\Pi_0]= E[Y-\mu_0(d,1,X,\Pi_0)|D=d,S=1,X,\Pi_0]=0}}{ p_{d0}(X, \pi_0(d,X,Z)) } \cdot   \frac{\partial E[p_{d0}^2(X, \pi_0(d,X,Z))]\cdot [\pi(d,X,Z)-\pi_0(d,X,Z)]}{p_{d0}(X, \pi_0(d,X,Z))} \Bigg|S=1 \Bigg]\notag\\
& + \partial E [\mu_0(d,1,X,\pi_0(d,X,Z))]\cdot [\pi(d,X,Z)-\pi_0(d,X,Z) | S=1 ] \tag{$****$}\\
&=0.\notag
\end{align}
The Gateaux derivative is zero because expressions $(*)$ and $(**)$ as well as $(***)$ and $(****)$, respectively, cancel out. To see this, note that by the law of iterated expectations and the fact that conditioning on $D,  X, \Pi$ is equivalent conditioning on $D,  X, \Pi$ (because $\Pi$ is deterministic in $Z$ conditional on $D,X$), $(*)$ corresponds to
\begin{eqnarray}
&& -E\Bigg[  \frac{\overbrace{E [I\{D=d\} | X, \Pi_0 ]}^{=p_{d0}(X, \pi_0(d,X,Z))}}{p_{d0}(X, \pi_0(d,X,Z))} \cdot   [\mu(d,1,X, \pi_0(d,X,Z))-\mu_0(d,1,X, \pi_0(d,X,Z))] \Bigg|S=1 \Bigg]\notag\\
&=& -E [\mu(d,1,X, \pi_0(d,X,Z))-\mu_0(d,1,X, \pi_0(d,X,Z))|S=1], \notag
\end{eqnarray}
which cancels out with $(**)$.
In an analogous way, it can be shown that $(***)$ corresponds to $$E[-\partial E [\mu_0(d,1,X,\pi_0(d,X,Z))]\cdot [\pi(d,X,Z)-\pi_0(d,X,Z)]|S=1],$$ which cancels out with $(****)$. Therefore,
\begin{align}
&\partial E \big[\phi_{d,S=1}(W, \eta, \Psi_{d}^{S=1})\big] \big[\eta - \eta_0 \big] = 0 \notag
\end{align}
proving that the score function is orthogonal.

\bigskip

\textbf{Assumption 3.2:  Score regularity and quality of nuisance parameter estimators}
\vspace{5pt}\newline
This proof follows in a similar way as the proof of Theorem 1 and is omitted for brevity.

\subsection{Proof of Theorem 3} \label{Neyman3}

The score function for the counterfactual $\Psi_{d0}=E[Y(d)]$ is given by:
\begin{eqnarray}\label{phi2}
\phi_{d}(W,  \eta,  \Psi_{d0}) &=& \frac{I\{D=d\} \cdot S \cdot [Y- \mu(d,1,X,\Pi)] }{ p_d(X)\cdot  \pi (d,X,Z)}  + \mu(d,1,X,\Pi)  - \Psi_{d0}.
\end{eqnarray}

\textbf{Assumption 3.1:  Linear scores and Neyman orthogonality}
\vspace{5pt}\newline

\textbf{Assumption 3.1(a)}

\textbf{Moment Condition:} The moment condition $E\Big[\psi_{d}(W, \eta_0,  \Psi_{d0})\Big] = 0$ holds:
\begin{eqnarray}
E\Big[\phi_{d}(W, \eta_0,  \Psi_{d0})\Big] &=& E\Bigg[ \overbrace{ E\Bigg[ \frac{I\{D=d\} \cdot S \cdot [Y-\mu_0(d,1,X, \Pi_0)]}{p_{d0}(X, \Pi_0)\cdot \pi_0(d,X,Z)} \Bigg| X, \Pi_0 \Bigg]}^{ =E[Y-\mu_0(d,1,X, \Pi_0)|D=d,S=1, X,\Pi_0]=0}  + \mu_0(d,1,X, \Pi_0)-\Psi_{d0}\Bigg] \notag\\
&=& E[ \mu_0(d,1,X, \Pi_0)]- \Psi_{d0} =  0, \notag
\end{eqnarray}
where the first equality follows from the law of iterated expectations.

\textbf{Assumption 3.1(b)}
\textbf{Linearity:} The score $ \phi_{d}(W,  \eta_0,  \Psi_{d0}) $ is linear in $\Psi_{d0}$ :
$\phi_{d}(W, \eta_0, \Psi_{d0}) = \phi_{d}^a(W, \eta_0) \cdot\Psi_{d0} + \phi_{d}^b(W, \eta_0) $
with $\phi_{d}^a(W, \eta_0) = -1$ and
\begin{eqnarray}
\phi_{d}^b(W, \eta_0) &=&\frac{I\{D=d\} \cdot S \cdot [Y- \mu_0(d,1,X,\Pi_0)] }{ p_{d0}(X,\Pi_0)\cdot  \pi_0 (d,X,Z)}  + \mu_0(d,1,X,\Pi_0). \notag
\end{eqnarray}

\textbf{Assumption 3.1(c)}

\textbf{Continuity:}
The expression for the second Gateaux derivative of a map $\eta \mapsto E[\phi_{d}(W,  \eta,  \Psi_{d0})]$, given in (\ref{phi2}), is continuous.
\newpage

\textbf{Assumption 3.1(d)}

\textbf{Neyman Orthogonality}: For any $\eta \in \mathcal{T}_n,$ the Gateaux derivative in the direction $ \eta - \eta_0 =
( \pi(D ,X, Z)-\pi_0 (D , X, Z), p_{d }(X, \Pi )-p_{d0 }(X, \Pi ), \mu(D ,S, X, \Pi)-\mu_0 (D ,S, X, \Pi ))$ is given by:
\begin{align}
&\partial E \big[\phi_{d}(W, \eta, \Psi_{d})\big] \big[\eta - \eta_0 \big]  = \notag\\
& - E \Bigg[  \frac{I\{D=d\} \cdot S \cdot [\mu(d,1,X, \pi_0(d,X,Z))-\mu_0(d,1,X,\pi_0(d,X,Z))]}{p_{d0}(X, \pi_0(d,X,Z))\cdot \pi_0(d,X,Z)}  \Bigg]  \tag{$*$}\\
& + E [\mu(d,1,X,\pi_0(d,X,Z))-\mu_0(d,1,X,\pi_0(d,X,Z))]  \tag{$**$}\\
&- E \Bigg[ \frac{\overbrace{I\{D=d \} \cdot S \cdot  [Y -\mu_0 (d,1, X,\pi_0(d,X,Z) )]}^{E[\cdot|X,\Pi_0]=E[Y -\mu_0 (d,1, X,\pi_0(d,X,Z) )|D=d,S=1,X,\Pi_0]=0}}{ p_{d0}(X, \pi_0(d,X,Z)) \cdot \pi_0(d,X,Z)  }  \cdot\frac{p_{d}(X, \pi_0(d,X,Z))-p_{d0}(X, \pi_0(d,X,Z))}{p_{d0}(X, \pi_0(d,X,Z))}\Bigg] \notag\\
&- E \Bigg[ \frac{I\{D=d \} \cdot S \cdot  \partial E[\mu_0 (d,1, X,\pi_0(d,X,Z) )]\cdot[\pi(d,X,Z)-\pi_0(d,X,Z)]}{ p_{d0}(X, \pi_0(d,X,Z))\cdot \pi_0(d,X, Z) } \Bigg]\tag{$***$}\\
&- E \Bigg[ \frac{\overbrace{I\{D=d \} \cdot S \cdot  [Y -\mu_0 (d,1, X,\pi_0(d,X,Z) )]}^{E[\cdot|X,\Pi_0]=E[Y -\mu_0 (d,1, X,\pi_0(d,X,Z) )|D=d,S=1,X,\Pi_0]=0}}{ p_{d0}(X, \pi_0(d,X,Z))\cdot \pi_0(d,X, Z) } \cdot\frac{\pi(d,X,Z)-\pi_0(d,X,Z)}{\pi_0(d,X, Z)} \Bigg]\notag\\
&- E \Bigg[ \frac{\overbrace{I\{D=d \} \cdot S \cdot  [Y -\mu_0 (d,1, X,\pi_0(d,X,Z) )]}^{E[\cdot|X,\Pi_0]=E[Y -\mu_0 (d,1, X,\pi_0(d,X,Z) )|D=d,S=1,X,\Pi_0]=0}}{ p_{d0}(X, \pi_0(d,X,Z))\cdot \pi_0(d,X, Z) } \cdot \frac{   \partial E[p_{d0}^2(X, \pi_0(d,X,Z))]\cdot [\pi(d,X,Z)-\pi_0(d,X,Z)]}{ p_{d0}(X, \pi_0(d,X,Z))} \Bigg]\notag\\
& + \partial E [\mu_0(d,1,X,\pi_0(d,X,Z))]\cdot [\pi(d,X,Z)-\pi_0(d,X,Z)] \tag{$****$}\\
&=0.\notag
\end{align}
The Gateaux derivative is zero because expressions $(*)$ and $(**)$ as well as $(***)$ and $(****)$, respectively, cancel out. To see this, note that by the law of iterated expectations and the fact that conditioning on $D,  X, Z$ is equivalent conditioning on $D,  X, \Pi$ (because $\Pi$ is deterministic in $Z$ conditional on $D,X$), $(*)$ corresponds to
\begin{eqnarray}
&&-E\Bigg[ E\Bigg[ \frac{I\{D=d\}}{p_{d0}(X, \pi_0(d,X,Z))} \cdot E\Bigg[ \frac{ S \cdot [\mu(d,1,X, \pi_0(d,X,Z))-\mu_0(d,1,X, \pi_0(d,X,Z))] }{\pi_0(d,X,Z)}\Bigg| D=d,  X , Z\Bigg] \Bigg| X , \Pi_0 \Bigg]\Bigg]\notag\\
&=&-E\Bigg[ E\Bigg[ \frac{I\{D=d\}}{p_{d0}(X, \pi_0(d,X,Z))} \cdot  \frac{ \overbrace{E[S| D=d,  X , Z]}^{=\pi_0(d,X,Z)} \cdot [\mu(d,1,X, \pi_0(d,X,Z))-\mu_0(d,1,X, \pi_0(d,X,Z))] }{\pi_0(d,X,Z)} \Bigg| X , \Pi_0 \Bigg]\Bigg]\notag\\
&=& -E\Bigg[  \frac{\overbrace{E [I\{D=d\} | X, \pi_0(d,X,Z) ]}^{=p_{d0}(X, \pi_0(d,X,Z))}}{p_{d0}(X, \pi_0(d,X,Z))} \cdot   [\mu(d,1,X, \pi_0(d,X,Z))-\mu_0(d,1,X, \pi_0(d,X,Z))] \Bigg]\notag\\
&=& -E [\mu(d,1,X, \pi_0(d,X,Z))-\mu_0(d,1,X, \pi_0(d,X,Z))], \notag
\end{eqnarray}
which cancels out with $(**)$.
In an analogous way, it can be shown that $(***)$ corresponds to $$E[-\partial E [\mu_0(d,1,X,\pi_0(d,X,Z))]\cdot [\pi(d,X,Z)-\pi_0(d,X,Z)] ],$$ which cancels out with $(****)$. Therefore,
\begin{align}
&\partial E \big[\phi_{d}(W, \eta, \Psi_{d})\big] \big[\eta - \eta_0 \big] = 0 \notag
\end{align}
proving that the score function is orthogonal.

\bigskip

\textbf{Assumption 3.2:  Score regularity and quality of nuisance parameter estimators}
\vspace{5pt}\newline
This proof follows in a similar manner to the proof of theorem 1 and is omitted for brevity.

\subsection{Proof of Theorem 4} \label{Neyman4}

Define the nuisance parameters to be the vector of functions $\eta=(p_d(X), \pi(D, X, M), \mu (D,S,X,M)$,
$ \nu(D,S,X,M)$, with $p_d(X)=\Pr(D=d |X )$, $\pi(D,X,M)=\Pr(S=1|D,X,M)$,   $\mu (D,S,X,M)=E[Y |D,S,X,M]$, and $\nu(D,S,X,M)=\int E[Y|D,S,X,M=m] dF_{M=m|D,X}$, where $F_{M=m|D,X}$ denotes the conditional distribution function of $M$ at value $m$. The score function for the counterfactual $\Psi_{d0}=E[Y(d)]$ is given by the following expression, with $W  = (Y \cdot S, D, S, X,M)$:
\begin{eqnarray}
\theta_d(W, \eta, \Psi_{d0}) &=& \frac{I\{D=d\} \cdot S \cdot [Y-\mu (d,1,X,M)]}{p_d(X)\cdot \pi(d, X, M)} \notag\\
& + & \frac{I\{D=d\}\cdot  [\mu (d,1,X,M)-\nu(d,1,X)]}{p_d(X)} \notag\\
& + &\nu(d,1,X) - \Psi_{d0}. \notag
\end{eqnarray}

Let $\mathcal{T}_n$ be the set fo all $\eta=(p_{d}, \pi,\mu, \nu)$ consisting of $P$-square integrable functions $p_d, \pi,\mu$, and $\nu$ such that
\begin{eqnarray} \label{Tn}
\left\|  \eta - \eta_0 \right\|_{q} &\leq& C,  \\
\left\|  \eta - \eta_0 \right\|_{2} &\leq& \delta_n, \notag \\
\left\|   p_d(X)-1/2\right\|_{\infty}  &\leq& 1/2-\epsilon, \notag\\
\left\|   \pi(D,X,M)-1/2)\right\|_{\infty} &\leq & 1/2-\epsilon, \notag \\
\left\|  \mu(D,1,X,M)-\mu_0(D,1,X,M)\right\|_{2} \times \left\|  p_d(X)-p_{d0}(X)\right\|_{2}  &\leq & \delta^{}_n n^{-1/2}, \notag \\
\left\|  \mu(D,1,X,M)-\mu_0(D,1,X,M)\right\|_{2} \times \left\|  \pi(D,X,M)-\pi_0(D,X,M)\right\|_{2} &\leq & \delta^{}_n n^{-1/2},\notag \\
\left\|  \nu(d,1,X)-\nu_0(d,1,X)\right\|_{2} \times \left\|  p_d(X)-p_{d0}(X)\right\|_{2} &\leq & \delta^{}_n n^{-1/2}.\notag
\end{eqnarray}
We furthermore replace the sequence $(\delta_n)_{n \geq 1}$ by $(\delta_n')_{n \geq 1},$ where $\delta_n' = C_{\epsilon} \max(\delta_n,n^{-1/2}),$ where $C_{\epsilon}$ is sufficiently large constant that only depends on $C$ and $\epsilon.$

\textbf{Assumption 3.1:  Linear scores and Neyman orthogonality}
\vspace{5pt}\newline

\textbf{Assumption 3.1(a)}

\textbf{Moment Condition:} The moment condition $E\Big[\theta_{d}(W, \eta_0, \Psi_{d0})\Big] = 0$ is satisfied:
\begin{eqnarray}
E\Big[\theta_{d}(W, \eta_0,  \Psi_{d0})\Big] &=& E\Bigg[ \overbrace{ E\Bigg[\frac{I\{D=d\} \cdot S \cdot  [Y-\mu_0(d,1,X,M)] }{ p_{d0}(X)\cdot \pi_0(d,X,M)} \Bigg|X,M\Bigg] }^{=E[E[Y-\mu_0(d,1,X,M)|D=d,S=1,X,M]|D=d,X]= 0} \Bigg] \notag\\
&+& \ E\Bigg[\overbrace{ E\Big[ \frac{I\{D=d\}\cdot [\mu_0(d,1,X,M)-\nu_0(d,1,X)]}{p_{d0}(X)} \Big|   X \Big]}^{ =\int E\big[  \mu_0(d,1,X,M)-\nu_0(d,1,X) \big| D=d,  X=x, M=m \big] dF_{M=m|D=d,X}=0} \Bigg] \notag\\
&+&  \  E\big[ \nu_0(d,1,X) \big] \ \ - \ \ \Psi_{d0} \ \ = \ \ \Psi_{d0}\ \ - \ \ \Psi_{d0} \ \ =  0, \notag
\end{eqnarray}
where the first equality follows from the law of iterated expectations. To better see this result, note that
\begin{eqnarray}
&&E\Bigg[\frac{I\{D=d\} \cdot S }{p_{d0}(X)\cdot \pi_0(d,X,M)} \cdot  [Y- \mu_0(d,1,X,M)]\Bigg|X\Bigg]\notag\\
&=&	E\Bigg[\frac{  S }{\pi_0(d,X,M)} \cdot  [Y- \mu_0(d,1,X,M)]\Bigg|D=d, X\Bigg]\notag\\
&=&	E\Bigg[E\Bigg[\frac{  S }{ \pi_0(d,X,M)} \cdot  [Y-\mu_0(d,1,X,M)]\Bigg|D=d, X, M \Bigg] \Bigg|D=d, X\Bigg]\notag\\
&=&E [ E[Y- \mu_0(d,1,X,M)|D=d, S=1,X, M |D=d, X  ]] \notag \\
&=&E [ \mu_0(d,1,X,M) -  \mu_0(d,1,X,M) |D=d, X   ]=0, \notag
\end{eqnarray}
where the first and third equalities follow from basic probability theory and the second from the law of iterated expectations. Furthermore,
\begin{eqnarray}
&&E\Bigg[ \frac{I\{D=d\} \cdot [ \mu_0(d,1,X,M)-\nu_0(d,1,X)]}{p_{d0}(X)} \Bigg|   X =  x \Bigg]\notag\\
&=&E\big[   \mu_0(d,1,X,M)- \nu_0(d,1,X) \big| D=d,  X = x \big]\notag\\
&=& \int E\big[   \mu_0(d,1,X,M)-\nu_0(d,1,X) \big| D=d,  X=x, M=m \big] dF_{M=m|D=d,X=x}\notag\\
&=& \int E\big[  \mu_0(d,1,X,M) \big| D=d,  X=x,M=m \big] dF_{M=m|D=d,X=x}-\nu_0(d,1,x)\notag\\
&=&\nu(d,1,x)-\nu(d,1,x)=0.\notag
\end{eqnarray}
where the first equality follows from basic probability theory, the second from conditioning on and integrating over $M$, and the third from the fact that $\nu_0(d,1,X)$ is not a function of $M$.

\textbf{Assumption 3.1(b)}

\textbf{Linearity:} The score $ \theta_{d}(W, \eta_0, \Psi_{d0}) $ is linear in $\Psi_{d0}$ :
$\theta_{d}(W, \eta_0, \Psi_{d0}) = \theta_{d}^a(W, \eta_0) \cdot\Psi_{d0} + \theta_{d}^b(W, \eta_0) $
with $\theta_{d}^a(W, \eta_0) = -1$ and
\begin{eqnarray}
\theta_{d}^b(W, \eta_0) &=&\frac{I\{D=d\} \cdot S \cdot [Y-\mu_0(d,1,X,M)]}{p_{d0}(X)\cdot \pi_0(d,X,M)}\notag\\
& + & \frac{I\{D=d\}\cdot  [\mu(d,1,X,M)-\nu(d,1,X)]}{p_d(X)} + \nu(d,1,X). \notag
\end{eqnarray}

\textbf{Assumption 3.1(c)}

\textbf{Continuity:}
The expression for the second Gateaux derivative of a map $\eta \mapsto E[\theta_{d}(W, \eta, \Psi_{d})]$ is continuous.
\newpage

\textbf{Assumption 3.1(d)}

\textbf{Neyman Orthogonality}: For any $\eta \in \mathcal{T}_N,$ the Gateaux derivative in the direction $ \eta - \eta_0 =
(p_d(X)-p_{d0}(X), \pi(d,X,M)-\pi_0(d,X,M),\mu(D,1,X,M)-\mu_0(D,1,X,M), \nu(d,1,X)-\nu_0(d,1,X)) $ is given by:
\begin{align}
&\partial E \big[\theta_{d}(W, \eta, \Psi_{d})\big] \big[\eta - \eta_0 \big]  = \notag\\
& - E \Bigg[  \frac{I\{D=d\} \cdot S \cdot [\mu(d,1,X,M)-\mu_0(d,1,X,M)]}{p_{d0}(X)\cdot \pi_0(d,X,M)}  \Bigg]  \tag{$*$}\\
& + E \Bigg[  \frac{I\{D=d\}  \cdot [\mu(d,1,X,M)-\mu_0(d,1,X,M)]}{p_{d0}(X)}  \Bigg]  \tag{$**$}\\
&- \  E \Bigg[ \frac{\overbrace{I\{D=d\} \cdot S \cdot  [Y-\mu_0(d,1,X,M)]}^{E[\cdot|X]=E[E[Y-\mu_0(d,1,X,M)|D=d,S=1,X,M]|D=d,X]= 0}}{ p_{d0}(X)\cdot \pi_0(d,X,M) }\cdot\frac{p_d(X)-p_{d0}(X) }{p_{d0}(X)} \Bigg] \notag\\
& - \  E \Bigg[  \overbrace{\frac{I\{D=d\}\cdot  [\mu_0(d,1,X,M)-\nu_0(d,1,X)]}{p_{d0}(X)}}^{E[\cdot| X]=\int E\big[  \mu_0(d,1,X,M)-\nu_0(d,1,X) \big| D=d,  X=x,M=m \big] dF_{M=m|D=d,X=x}=0} \cdot \frac{p_d(X)-p_{d0}(X) }{p_{d0}(X)} \Bigg]  \notag \\
&- \  E \Bigg[ \frac{\overbrace{I\{D=d\} \cdot S \cdot  [Y-\mu_0(d,1,X,M)]}^{E[\cdot|X]=E[E[Y-\mu_0(d,1,X,M)|D=d,S=1,X,M]|D=d,X]= 0}}{ p_{d0}(X)\cdot \pi_0(d,X,M) }\cdot\frac{\pi(d,X,M)-\pi_0(d,X,M) }{\pi_0(d,X,M)} \Bigg] \notag\\
&\underbrace{ - \  E \Bigg[  \underbrace{\frac{I\{D=d\}  \cdot [\nu(d,1,X)-\nu_0(d,1,X)]}{p_{d0}(X)}}_{ E[\cdot|X]=\frac{p_{d0}(X)}{p_{d0}(X)}\cdot[\nu(d,1,X)-\nu_0(d,1,X)]}   \Bigg]+  E[\nu(d,1,X)-\nu_0(d,1,X)]}_{=0} =0\notag
\end{align}
The Gateaux derivative is zero because expressions $(*)$ and $(**)$ cancel out. To see this, note that
\begin{eqnarray}
&&E\Bigg[ \frac{I\{D=d\} \cdot [\mu(d,1,X,M)-\mu_0(d,1,X,M)]}{p_d(X)} \Bigg|   X =  x \Bigg]\notag\\
&=&E\big[  \mu(d,1,X,M)-\mu_0(d,1,X,M) \big| D=d,  X = x \big]\notag\\
&=& \int E\big[  \mu(d,1,X,M)-\mu_0(d,1,X,M) \big| D=d,  X=x, M=m \big] dF_{M=m|D=d,X=x}, \notag 
\end{eqnarray}
where the first equality follows from basic probability theory and the second from conditioning on and integrating over $M$. Furthermore,
\begin{eqnarray}
&&E\Bigg[ \frac{I\{D=d\}  \cdot S  \cdot [\mu(d,1,X,M)-\mu_0(d,1,X,M)]}{p_d(X)  \cdot \pi_0(d,X,M) } \Bigg|   X =  x \Bigg]\notag\\
&=&E\Bigg[   \frac{S  \cdot[\mu(d,1,X,M)-\mu_0(d,1,X,M)]} {\pi_0(d,X,M) } \Bigg| D=d,  X = x \Bigg]\notag\\
&=& \int E\Bigg[  \frac{S  \cdot [\mu(d,1,X,M)-\mu_0(d,1,X,M)]}{\pi_0(d,X,M) }  \Bigg| D=d,  X=x, M=m \Bigg] dF_{M=m|D=d,X=x}\notag\\
&=& \int E\big[  \mu(d,1,X,M)-\mu_0(d,1,X,M) \big| D=d, S=1,  X=x, M=m \big] dF_{M=m|D=d,X=x}\notag\\
&=& \int E\big[  \mu(d,1,X,M)-\mu_0(d,1,X,M) \big| D=d,   X=x,M=m \big] dF_{M=m|D=d,X=x},\notag
\end{eqnarray}
where the first equality follows from basic probability theory, the second from conditioning on and integrating over $M$, the third from basic probability theory, and the fourth from simplification as $\mu(d,1,X,M)=E[Y|D=d,S=1,X=x,M=m]$ is already conditional on $S=1$.
\begin{align}
&\partial E \big[\theta_{d}(W, \eta, \Psi_{d})\big] \big[\eta - \eta_0 \big] = 0 \notag
\end{align}
proving that the score function is orthogonal.

\textbf{Assumption 3.1(e)}

\textbf{Singular values of $E[\theta_{d}^a(W;\eta_0)]$ are bounded:}
This holds trivially, because $\theta_{d}^a(W;\eta_0) = -1.$

\bigskip

\textbf{Assumption 3.2:  Score regularity and quality of nuisance parameter estimators}

\textbf{Assumption 3.2(a)}

This assumption directly follows from the construction of the set $\mathcal{T}_n$ and the regularity conditions (Assumption 12).

\textbf{Assumption 3.2(b)} 

\textbf{Bound for $m_n$:}
\begin{eqnarray}
\left\|  \mu_0(D,S,X,M)  \right\|_{q} &=& \left( E\left[ \left| \mu_0(D,S,X,M) \right|^q \right] \right)^{\frac{1}{q}} \notag \\
&=&  \left( \sum_{d \in \{0,1,...,Q\}, s \in \{0,1\}} E\left[ \left| \mu_0(d,s,X,M) \right|^q \Pr  _{P}(D=d, S=s|X,M)  \right]  \right)^{\frac{1}{q}} \notag \\
&\geq& \epsilon^{2/q} \left( \sum_{d \in \{0,1,...,Q\}, s \in \{0,1\}} E\left[ \left| \mu_0(d,s,X,M) \right|^q \right] \right)^{\frac{1}{q}} \notag \\
&\geq& \epsilon^{2/q} \left( \max_{d \in \{0,1,...,Q\}, s \in \{0,1\}} E\left[ \left| \mu_0(d,s,X,M) \right|^q \right] \right)^{\frac{1}{q}} \notag \\
&=& \epsilon^{2/q} \left( \max_{d \in \{0,1,...,Q\}, s \in \{0,1\}} \left\|  \mu_0(d,s,X,M)  \right\|_{q}\right), \notag
\end{eqnarray}
where the first equality follows from definition, the second from the law of total probability, and the third line from the fact that $\Pr(D = d, S=1|X,M) = p_{d0}(X) \cdot \pi_0 (d,X,M) \geq \epsilon^2$ and $\Pr(D = d, S=0|X,M) = p_{d0}(X) \cdot (1-\pi_0 (d,X,M)) \geq \epsilon^2$.  Similarly,
\begin{equation}
\left\|  \nu_0(D,S,X)  \right\|_{q} \geq \epsilon^{2/q} \left( \max_{d \in \{0,1,...,Q\}, s \in \{0,1\}} \left\|  \nu_0(d,s,X)  \right\|_{q}\right). \notag
\end{equation}

Notice that by Jensen's inequality $\left\|  \mu_0(D,S,X,M)  \right\|_{q} \leq \left\|  Y  \right\|_{q}$ and
$\left\|  \nu_0(D,S,X)  \right\|_{q} \leq \left\|  Y  \right\|_{q}$ and hence
$\left\|  \mu_0(d,1,X,M)  \right\|_{q} \leq C/\epsilon^{2/q}$ and
$\left\|  \nu_0(d,1,X)  \right\|_{q} \leq C/\epsilon^{2/q},$
by conditions (\ref{Tn}). Similarly, for any $\eta \in \mathcal{T}_N:$
$\left\|  \mu(d,1,X,M) - \mu_0(d,1,X,M)  \right\|_{q} \leq C/\epsilon^{2/q}$ and
$\left\|    \nu(d,1,X) - \nu_0(d,1,X)  \right\|_{q} \leq C/\epsilon^{2/q},$ because
$\left\|  \mu(D,S,X,M) - \mu_0(D,S,X,M)  \right\|_{q} \leq C$ and
$\left\|    \nu(D,S,X) - \nu_0(D,S,X)  \right\|_{q} \leq C.$

Consider
\begin{eqnarray}
E\Big[ \theta_{d}(W, \eta, \Psi_{d0})\Big] &=& E\Bigg[ \underbrace{ \frac{ I\{D=d\} \cdot S}{p_d(X)\cdot \pi(d,X,M)} \cdot Y }_{=I_1}  \notag\\
& + & \underbrace{ \frac{I\{D=d\}}{p_d(X)} \cdot \bigg(1- \frac{S}{\pi(d,X,M) } \bigg) \cdot \mu(d,1,X,M)  }_{=I_2}  \notag\\
& + & \underbrace{  \bigg(1-   \frac{I\{D=d\}}{p_d(X)}\bigg)  \nu(d,1,X)}_{=I_3} - \Psi_{d0}  \Bigg]  \notag
\end{eqnarray}
and thus
\begin{eqnarray}
\left\| \theta_{d}(W, \eta, \Psi_{d0}) \right\|_{q} &\leq&  \left\| I_1 \right\|_{q}  + \left\| I_2 \right\|_{q}  + \left\| I_3 \right\|_{q} +  \left\| \Psi_{d0}  \right\|_{q} \notag \\
&\leq& \frac{1}{\epsilon^2}  \left\| Y \right\|_{q} \notag + \frac{1- \epsilon}{\epsilon^2} \left\|  \mu(d,1,X,M)  \right\|_{q} + \\
&+& \frac{1-\epsilon}{\epsilon} \left\|  \nu(d,1,X)  \right\|_{q}   +  | \Psi_{d0} | \notag \\
&\leq& C \left( \frac{1}{\epsilon^2} + \frac{2(1-\epsilon)}{\epsilon^{2/q}} \left(\frac{1}{\epsilon^2} + \frac{1}{\epsilon} \right) + \frac{1}{\epsilon} \right),  \notag
\end{eqnarray}
because of triangular inequality and because the following set of inequalities hold:
\begin{eqnarray} \label{32b}
\left\|  \mu(d,1,X,M)  \right\|_{q} &\leq& \left\|  \mu(d,1,X,M) -  \mu_0(d,1,X,M)  \right\|_{q} + \left\|  \mu_0(d,1,X,M)  \right\|_{q} \leq 2C/\epsilon^{2/q},  \\
\left\|  \nu(d,1,X)  \right\|_{q} &\leq& \left\|  \nu(d,1,X) -  \nu_0(d,1,X)  \right\|_{q} + \left\|  \nu_0(d,1,X)  \right\|_{q} \leq 2C/\epsilon^{2/q}, \notag \\
|\Psi_{d0} | &=& |E[ \nu_0(d,1,X)] | \leq  E_{ } \Big[\left|  \nu_0(d,1,X)  \right|^1 \Big]^{\frac{1}{1}} =  \left\| \nu_0(d,1,X) \right\|_{1}  \notag \\
&\leq&  \left\| \nu_0(d,1,X) \right\|_{2} \leq  \left\| Y \right\|_{2}/\epsilon^{2/2} \overbrace{ \leq}^{q > 2}  \left\| Y \right\|_{q}/\epsilon \leq C /\epsilon.  \notag
\end{eqnarray}
which gives the upper bound on $m_n$ in Assumption 3.2(b) of \cite{Chetal2018}.

\textbf{Bound for $m'_n$:}

Notice that
$$\Big(E[ |\theta^a_{d}(W, \eta) |^q] \Big)^{1/q}=1$$
and this gives the upper bound on $m'_n$ in Assumption 3.2(b).

\textbf{Assumption 3.2(c)}

In the following, we omit arguments for the sake of brevity and use $p_d= p_d(X),\pi= \pi(d,X,M),\nu = \nu(d,1,X),  \mu = \mu(d,1,X,M)$ and similarly for $p_{d0},\pi_0,\nu_0,\mu_0.$

\textbf{Bound for $r_n$:}

For any $\eta = (p_d, \pi,\mu, \nu)$ we have
$$ \Big| E\Big( \theta^a_d(W, \eta) - \theta^a_d(W, \eta_0) \Big) \Big| = |1-1| = 0 \leq \delta'_N,$$
and thus we have the bound on $r_n$ from Assumption 3.2(c).

\textbf{Bound for $r'_n$:}
\begin{eqnarray}\label{rprimeN}
&& \left\|   \theta_{d}(W, \eta, \Psi_{d0}) - \theta_{d}(W, \eta_0, \Psi_{d0})  \right\|_{2} \leq  \left\|    I\{D=d\} \cdot S \cdot Y \cdot \left( \frac{1}{p_d \pi} - \frac{1}{p_{d0} \pi_{0}} \right) \right\|_{2}   \\
&+&  \left\|    I\{D=d\} \cdot S \cdot  \left( \frac{\mu}{p_d \pi} - \frac{\mu_0}{p_{d0} \pi_{0}} \right) \right\|_{2}  + \left\|    I\{D=d\}\cdot   \left( \frac{\mu}{p_d} - \frac{\mu_0}{p_{d0}} \right) \right\|_{2} \notag \\
&+& \left\|    I\{D=d\}  \left( \frac{\nu}{p_d} - \frac{\nu_0}{p_{d0}} \right) \right\|_{2} + \left\| \nu - \nu_0 \right\|_{2} \notag \\
&\leq&  \left\| Y \cdot \left( \frac{1}{p_d \pi} - \frac{1}{p_{d0} \pi_{0}} \right) \right\|_{2} +   \left\| \frac{\mu}{p_d \pi} - \frac{\mu_0}{p_{d0} \pi_{0}}  \right\|_{2} +  \left\|  \frac{\mu}{p_d} - \frac{\mu_0}{p_{d0}}  \right\|_{2} +  \left\|   \frac{\nu}{p_d} - \frac{\nu_0}{p_{d0}}  \right\|_{2} + \left\| \nu - \nu_0 \right\|_{2} \notag \\
&\leq&
\frac{C}{\epsilon^4} \delta_n \left(1 + \frac{1}{\epsilon} \right) + \delta_n \left(  \frac{1}{\epsilon^5} + C + \frac{C}{\epsilon} \right) +\delta_n \left( \frac{1}{\epsilon^3} + \frac{C}{\epsilon^2} \right) + \delta_n \left( \frac{1}{\epsilon^3} + \frac{C}{\epsilon^2} \right)+  \frac{\delta_n}{\epsilon}
\leq \delta_n' \notag
\end{eqnarray}
as long as $C_\epsilon$ in the definition of $\delta_n'$ is sufficiently large.  This gives the bound on $r'_n$ from Assumption 3.2(c).
Here we made use of the fact that $\left\| \mu - \mu_0 \right\|_{2} = \left\| \mu(d,1,X,M) - \mu_0(d,1,X,M) \right\|_{2} \leq \delta_n/\epsilon,$ $\left\| \nu - \nu_0 \right\|_{2} = \left\| \nu(d,1,X) - \nu_0(d,1,X) \right\|_{2} \leq \delta_n/\epsilon$ and $\left\| \pi - \pi_0 \right\|_{2} = \left\|\pi (d,X) - \pi _0(d,X) \right\|_{2} \leq \delta_n/\epsilon$ using similar steps as in Assumption 3.1(b).

The last inequality in (\ref{rprimeN}) is satisfied because we can bound the first term by
\begin{eqnarray*}
	&&  \left\| Y \cdot \left( \frac{1}{p_d \pi} - \frac{1}{p_{d0} \pi_{0}} \right) \right\|_{2}  \leq   C  \left\|  \frac{1}{p_d \pi} - \frac{1}{p_{d0} \pi_{0}}  \right\|_{2} \leq \frac{C}{\epsilon^4} \left\| p_{d0} \pi_{0} - p_d \pi  \right\|_{2} \\
	&=& \frac{C}{\epsilon^4} \left\| p_{d0} \pi_{0} - p_d \pi +  p_{d0}  \pi  -  p_{d0}  \pi  \right\|_{2} \leq \frac{C}{\epsilon^4}\left( \left\| p_{d0} (\pi_{0} - \pi ) \right\|_{2} + \left\| \pi_{0} (p_{d0} - p_d)  \right\|_{2} \right) \\
	&\leq&  \frac{C}{\epsilon^4}\left( \left\| \pi_{0} - \pi  \right\|_{2} + \left\| p_{d0} - p_d  \right\|_{2} \right) \leq \frac{C}{\epsilon^4} \delta_n \left(1 + \frac{1}{\epsilon} \right),
\end{eqnarray*}
where the first inequality follows from the second inequality in Assumption 4(a). The second term in (\ref{rprimeN}) is bounded by
\begin{eqnarray*}
	&& \left\|  \frac{\mu}{p_d \pi} - \frac{\mu_0}{p_{d0} \pi_{0}} \right\|_{2} \leq \frac{1}{\epsilon^4} \left\| p_{d0} \pi_{0} \mu -  p_d \pi \mu_0 \right\|_{2} =  \frac{1}{\epsilon^4}\left\|  p_{d0} \pi_{0} \mu -  p_d \pi  \mu_0 + p_{d0} \pi_{0} \mu_0 - p_{d0} \pi_{0} \mu_0 \right\|_{2} \\
	&\leq&  \frac{1}{\epsilon^4} \left( \left\| p_{d0} \pi_{0}  (\mu - \mu_0) \right\|_{2} + \left\| \mu_0 ( p_{d0} \pi_{0}  - p_d \pi  ) \right\|_{2} \right) \leq   \frac{1}{\epsilon^4} \left( \left\| \mu - \mu_0 \right\|_{2} + C  \left\| p_{d0} \pi_{0}  - p_d \pi \right\|_{2} \right) \\
	&\leq& \frac{1}{\epsilon^4} \left( \frac{\delta_n}{\epsilon} + C \left\| p_{d0} \pi_{0}  - p_d \pi \right\|_{2} \right) \leq \delta_n \left(  \frac{1}{\epsilon^5} + C + \frac{C}{\epsilon} \right) ,
\end{eqnarray*}
where the third inequality follows from $E[Y^2| D  =d , S=s, X,M] \geq (E[Y| D  =d , S=s, X,M])^2= \mu^2_0(d,s, X,M) $ by the conditional Jensen's inequality and therefore $\left\| \mu_0 (d,s,X,M) \right\|_\infty \leq C^2.$

For the third term we get
\begin{eqnarray*}
	&& \left\|  \frac{\mu}{p_d} - \frac{\mu_0}{p_{d0}}  \right\|_{2} = \frac{1}{\epsilon^2}  \left\| p_{d0} \mu-  p_d \mu_0  \right\|_{2} =  \frac{1}{\epsilon^2} \left\| p_{d0} \mu-  p_d \mu_0 + p_{d0} \mu_0 - p_{d0} \mu_0   \right\|_{2} \\
	&\leq&\frac{1}{\epsilon^2} \left( \left\|  p_{d0}(\mu - \mu_0) \right\|_{2} +  \left\| \mu_0 (p_{d0} - p_d) \right\|_{2} \right) \leq \frac{1}{\epsilon^2} \left( \left\| \mu - \mu_0 \right\|_{2} + C \left\| p_{d0} - p_d \right\|_{2} \right) \leq \delta_n \left( \frac{1}{\epsilon^3} + \frac{C}{\epsilon^2} \right),
\end{eqnarray*}
and similarly, for the fourth term we obtain
\begin{eqnarray*}
	&& \left\|  \frac{\nu}{p_d} - \frac{\nu_0}{p_{d0}}  \right\|_{2}  \leq \delta_n \left( \frac{1}{\epsilon^3} + \frac{C}{\epsilon^2} \right),
\end{eqnarray*}
where we used Jensen's inequality twice to get $\left\| \nu_0 (\underline{d}_2,X) \right\|_\infty \leq C^2$.

\textbf{Bound for $\lambda'_n$:}

Now consider


\begin{equation}
f(r) := E[\theta(W;\Psi_{d0},\eta + r(\eta-\eta_0)]. \notag
\end{equation}
For any $r \in (0,1):$
\begin{eqnarray} \label{secondGatDer}
\frac{\partial^2 f(r)}{\partial r^2}&=& E\Bigg[  I\{D=d\} \cdot S \cdot(-2)\cdot \frac{(\mu-\mu_0)(p_d - p_{d0})}{\left(p_{d0} + r(p_d -p_{d0})\right)^2\left(\pi_0 + r(\pi -\pi_0)\right)}  \Bigg]  \\
&+& E\Bigg[  I\{D=d\} \cdot S \cdot(-2)\cdot \frac{(\mu-\mu_0)(\pi - \pi_0)}{\left(p_{d0} + r(p_d -p_{d0})\right)\left(\pi_0 + r(\pi -\pi_0)\right)^2}  \Bigg]  \notag \\
&+& E\Bigg[  I\{D=d\} \cdot S \cdot2\cdot \frac{(Y - \mu_0 - r(\mu-\mu_0) )(p_d - p_{d0})^2}{\left(p_{d0} + r(p_d -p_{d0})\right)^3\left(\pi_0 + r(\pi -\pi_0)\right)}  \Bigg]  \notag \\
&+& E\Bigg[  I\{D=d\} \cdot S \cdot2\cdot \frac{(Y - \mu_0 - r(\mu-\mu_0) )(\pi - \pi_0)^2}{\left(p_{d0} + r(p_d -p_{d0})\right)\left(\pi_0 + r(\pi -\pi_0)\right)^3}  \Bigg]  \notag \\
&+& E\Bigg[  I\{D=d\} \cdot S \cdot2\cdot \frac{(Y - \mu_0 - r(\mu-\mu_0) )(p_d - p_{d0})(\pi - \pi_0)}{\left(p_{d0} + r(p_d -p_{d0})\right)^2\left(\pi_0 + r(\pi -\pi_0)\right)^2}  \Bigg]  \notag \\
&+& E\Bigg[  I\{D=d\} \cdot(-2)\cdot \frac{(\mu-\mu_0)(p_d - p_{d0})}{\left(p_{d0} + r(p_d -p_{d0})\right)^2}  \Bigg] + E\Bigg[  I\{D=d\} \cdot2\cdot \frac{(\nu-\nu_0)(p_d - p_{d0})}{\left(p_{d0} + r(p_d -p_{d0})\right)^2}  \Bigg]  \notag \\
&+& E\Bigg[  I\{D=d\} \cdot2\cdot \frac{r(\mu-\mu_0)(p_d - p_{d0})^2}{\left(p_{d0} + r(p_d -p_{d0})\right)^3}  \Bigg] + E\Bigg[  I\{D=d\}\cdot 2 \cdot \frac{(r(\nu-\nu_0)(p_d - p_{d0})^2}{\left(p_{d0} + r(p_d -p_{d0})\right)^3}  \Bigg] \notag \\
&+& E\Bigg[  I\{D=d\}\cdot 2 \cdot \frac{(\mu_0 -  \nu_0  )(p_d - p_{d0})^2}{\left(p_{d0} + r(p_d -p_{d0})\right)^3}  \Bigg]   \notag
\end{eqnarray}

Note that because
\begin{eqnarray}
E[Y-\mu_0(d,1,X,M)|D=d,S=1,X,M]  &= & 0, \notag \\
|p_d - p_{d0}|  \leq  2, \ \  \ \ \ \ |\pi - \pi_0|  &\leq & 2 \notag \\
\left\| \mu_0 \right\|_{q} \leq  \left\| Y \right\|_{q}/\epsilon^{1/q} &\leq & C/\epsilon^{2/q} \notag \\
\left\| \nu_0 \right\|_{q} \leq  \left\| Y \right\|_{q}/\epsilon^{1/q} &\leq & C/\epsilon^{2/q} \notag \\
\left\|  \mu-\mu_0\right\|_{2} \times \left\|  p_d-p_{d0}\right\|_{2}  &\leq & \delta^{}_n n^{-1/2}/\epsilon, \notag \\
\left\|  \mu-\mu_0\right\|_{2} \times \left\|  \pi-\pi_0\right\|_{2} &\leq & \delta^{}_n n^{-1/2}/\epsilon^2,\notag \\
\left\|  \nu-\nu_0\right\|_{2} \times \left\|  p_d-p_{d0}\right\|_{2} &\leq & \delta^{}_n n^{-1/2}/\epsilon.\notag
\end{eqnarray}
we get that for some constant $C_{\epsilon}''$ that only depends on $C$ and $\epsilon$
\begin{equation}
\left|\frac{\partial^2 f(r)}{\partial r^2} \right| \leq C_{\epsilon}'' \delta_n n^{-1/2} \leq \delta_n' n^{-1/2} \notag
\end{equation}
and this gives the upper bound on $\lambda'_n$ in Assumption 3.2(c) of \cite{Chetal2018} as long as $C_{\epsilon} \geq C_{\epsilon}''$. We used the following inequalities
\begin{eqnarray}
\left\| \mu-\mu_0\right\|_{2} &=& \left\|\mu(d,1,X,M)-\mu_0(d,1,X,M)\right\|_{2} \leq  \left\|\mu(D,S,X,M)-\mu_0(D,S,X,M)\right\|_{2}/\epsilon \notag \\
\left\| \nu-\nu_0\right\|_{2} &=& \left\|\nu(d,1,X)-\nu_0(d,1,X)\right\|_{2} \leq  \left\|\nu(D,S,X)-\nu_0(D,S,X)\right\|_{2}/\epsilon^2 \notag \\
\left\| \pi-\pi_0\right\|_{2} &=& \left\|\pi(d,X)-\pi_0(d,X)\right\|_{2} \leq  \left\|\pi(D,X)-\pi_0(D,X)\right\|_{2}/\epsilon, \notag
\end{eqnarray}
and these can be shown using similar steps as in Assumption 3.1(b).

To verify that $\left|\frac{\partial^2 f(r)}{\partial r^2} \right|  \leq C_{\epsilon}'' \delta_n n^{-1/2}$ holds, note that by the triangular inequality it is sufficient to bound the absolute value of each of the ten terms in (\ref{secondGatDer}) separately. We illustrate it for the first, third, and last terms. For the first term:
\begin{eqnarray}
&& \left| E\Bigg[  I\{D=d\} \cdot S (-2) \frac{(\mu-\mu_0)(p_d - p_{d0})}{\left(p_{d0} + r(p_d -p_{d0})\right)^2\left(\pi_0 + r(\pi -\pi_0)\right)}  \Bigg] \right| \notag \\
&\leq& 2 \left|  E\Bigg[  \frac{(\mu-\mu_0)(p_d - p_{d0})}{\left(p_{d0} + r(p_d -p_{d0})\right)^2(\pi_0 + r(\pi -\pi_0)}  \Bigg] \right| \notag \\
&\leq&  \frac{2}{ \epsilon^3} \left| E\Bigg[  (\mu-\mu_0)(p_d - p_{d0})  \Bigg] \right| \leq  \frac{2}{\epsilon^3} \frac{\delta^{}_N}{\epsilon} n^{-1/2},  \notag
\end{eqnarray}
in the second inequality we used the fact that for  $1 \geq p_{d0} + r(p_d -p_{d0}) =  (1-r)p_{d0} + r p_{d} \geq (1-r)\epsilon + r \epsilon = \epsilon$ and similarly for $\pi$ and in the third Holder's inequality. For the third term, we get
\begin{eqnarray}
&& \left| E\Bigg[  I\{D=d\} \cdot S 2 \frac{(Y - \mu_0 - r(\mu-\mu_0) )(p_d - p_{d0})^2}{\left(p_{d0} + r(p_d -p_{d0})\right)^3\left(\pi_0 + r(\pi -\pi_0)\right)}  \Bigg] \right|  \notag \\
&\leq& \frac{2}{\epsilon^4} \left| E\Bigg[  I\{D=d\} \cdot S(Y - \mu_0 - r(\mu-\mu_0) )(p_d - p_{d0})^2 \Bigg] \right|  \notag \\
&\leq&  \frac{8}{\epsilon^4} \left| E\Bigg[ I\{D=d\} \cdot S(Y - \mu_0) \Bigg] \right|  +  \frac{2}{\epsilon^4}  \left| E\Bigg[r(\mu-\mu_0) (p_d - p_{d0})^2 \Bigg] \right| \notag \\
&\leq& \frac{2 \cdot 2}{\epsilon^4}  \left| E\Bigg[1\cdot(\mu-\mu_0) )(p_d - p_{d0}) \Bigg] \right| \leq \frac{4}{\epsilon^4} \frac{\delta^{}_N}{\epsilon} n^{-1/2}, \notag
\end{eqnarray}
where in addition we made use of conditions (\ref{Tn}).

For the last term, we have
\begin{eqnarray}
&& E\Bigg[  I\{D=d\} 2 \frac{(\mu_0 -  \nu_0  )(p_d - p_{d0})^2}{\left(p_{d0} + r(p_d -p_{d0})\right)^3}  \Bigg] \notag \\
&=& E\Bigg[ \overbrace{ I\{D=d\} \frac{(\mu_0 -  \nu_0 )}{p_{d0}} }^{\int E\big[  \mu_0(d,1,X,M)-\nu_0(d,1,X) \big| D=d,  X=x,M=m \big] dF_{M=m|D=d,X=x}=0} \cdot \frac{2p_{d0}(p_d - p_{d0})^2}{\left(p_{d0} + r(p_d -p_{d0})\right)^3}  \Bigg]  = 0. \notag
\end{eqnarray}

The remaining terms in (\ref{secondGatDer}) are bounded similarly.

\textbf{Assumption 3.2(d)} 
\begin{eqnarray}
E\Big[ (\theta_{d}(W, \eta_0, \Psi_{d0}) )^2\Big] &=& E\Bigg[ \Bigg( \underbrace{ \frac{ I\{D=d\} \cdot S \cdot [Y-\mu_0(d,1,X,M)]}{p_{d0}(X)\cdot \pi_0(d,X,M)} }_{=I_1}  \notag\\
& + & \underbrace{ \frac{I\{D=d\}\cdot  [\mu_0(d,1,X,M)-\nu_0(d,1,X)]}{p_{d0}(X)} }_{=I_2}  \notag\\
& + & \underbrace{ \nu_0(d,1,X) - \Psi_{d0}}_{=I_3} \Bigg)^2 \Bigg]  \notag\\
& = & E[I_1^2 + I_2^2 + I_3^2] \geq E[I^2_1]\notag\\
& = & E\Bigg[  I\{D=d\} \cdot S \cdot \Bigg( \frac{ [Y-\mu_0(d,1,X,M)]}{p_{d0}(X)\cdot \pi_0(d,X,M)} \Bigg)^2 \Bigg] \notag\\
& \geq & \epsilon^2 E\Bigg[ \Bigg( \frac{ [Y-\mu_0(d,1,X,M)]}{p_{d0}(X)\cdot \pi_0(d,X,M)} \Bigg)^2 \Bigg] \notag\\
&\geq& \frac{\epsilon^2 c^2}{(1-\epsilon)^4} > 0, \notag
\end{eqnarray}
because $\Pr(D= d, S=1 |X,M)= p_{d0} (X) \cdot \pi_{0} (d,X,M) \geq \epsilon^2, \  p_{d0}(X) \leq 1-\epsilon$ and $\pi_0(d,X,M) \leq 1-\epsilon$.



The the second equality follows from
\begin{eqnarray}
E\Big[ I_1 \cdot I_2\Big] &=& E\Bigg[  \overbrace{\frac{ I\{D=d\} \cdot S}{ (p_{d0}(X))^2\cdot \pi_0(d,X,M)}  \cdot [Y-\mu_0(d,1,X,M)]}^{E[\cdot|X]=E[E[Y-\mu_0(d,1,X,M)|D=d,S=1,X,M]|D=d,X]= 0}   \cdot  [\mu_0(d,1,X,M)-\nu_0(d,1,X)] \Bigg], \notag\\
E\Big[ I_2 \cdot I_3\Big] &=& E\Bigg[ \overbrace{  \frac{ I\{D=d\}}{p_{d0}(X)} \cdot  [\mu_0(d,1,X,M)-\nu_0(d,1,X)]}^{E[\cdot|X]=\int E\big[  \mu_0(d,1,X,M)-\nu_0(d,1,X) \big| D=d,  X=x,M=m \big] dF_{M=m|D=d,X}=0}  \cdot [ \nu_0(d,1,X) - \Psi_{d0}] \Bigg],\notag\\
E\Big[ I_1 \cdot I_3\Big] &=& E\Bigg[  \overbrace{\frac{ I\{D=d\} \cdot S}{ p_{d0}(X)\cdot \pi_0(d,X,M)}  \cdot [Y-\mu_0(d,1,X,M)]}^{E[\cdot|X]=E[E[Y-\mu_0(d,1,X,M)|D=d,S=1,X,M]|D=d,X]= 0}   \cdot  [\nu_0(d,1,X) - \Psi_{d0}] \Bigg]. \notag
\end{eqnarray}

\bigskip

\section{Derivation of efficient influence functions}\label{effinf}

The proof that our estimators are based on efficient influence functions closely follows \cite{Levy2019}. For deriving the efficient influence functions under the identifying assumptions considered in Sections \ref{ivident} and \ref{marident}, let $\tilde{Y}=Y\cdot S$. Furthermore, denote the observed data by $O=(\tilde{Y}\cdot S, S, D, X)\sim P$, with distribution $P$ having the density  $f(o)=f_{\tilde{Y}}(y|d,s,x)f_{D,S}(d,s|x)f_X(x)$, where $f_A(a)$ is the unconditional density or probability of variable $A$ at value $a$ and $f_A(a|b)$ is the conditional density/probability given variable $B=b$.

Define $\Psi_{d0}=\Psi_{d}(P)=E_P[E_P[Y|D=d,S=1,X]]=E_P[E_P[\tilde{Y}|D=d,S=1,X]]$, where the second equality from the fact that $\tilde{Y}=Y$ for $S=1$. We now consider the derivative of $\Psi_{d^*}(P)$ w.r.t.\ the distribution $P$, with $d^*$ $\in$ $\{0,1,...,Q\}$:
\begin{eqnarray}
&&\frac{\partial}{\partial \epsilon} \Big|_{\epsilon=0} \Psi_{d^*}(P_{\epsilon})=E_{P_{\epsilon}}[E_{P_{\epsilon}}[\tilde{Y}|D=d^*,S=1,X]]\notag\\
&=&\int\int y \frac{\partial}{\partial \epsilon}\Big|_{\epsilon=0} (f_{\tilde{Y},\epsilon}(y|d^*,1,x)dy f_{X,\epsilon}(x))dx\notag\\
&=&\int\int y \frac{\partial}{\partial \epsilon}\Big|_{\epsilon=0} f_{\tilde{Y},\epsilon}(y|d^*,1,x)dy f_{X}(x)dx  + \int\int y  f_{\tilde{Y}}(y|d^*,1,x)dy \frac{\partial}{\partial \epsilon}\Big|_{\epsilon=0} f_{X,\epsilon}(x)dx  \notag\\
&=& \int \int \int y \frac{\partial}{\partial \epsilon}\Big|_{\epsilon=0}   f_{\tilde{Y},\epsilon}(y|d,s,x)dy\frac{ I\{D=d^*,S=1\} f_{D,S}(d,s|x)}{f_{D,S}(d,s|x)} d(d,s) f_{X}(x)dx\label{tobechanged1}\\
&+& \int\int y  f_{\tilde{Y}}(y|d^*,1,x)dy \frac{\partial}{\partial \epsilon}\Big|_{\epsilon=0} f_{X,\epsilon}(x)dx.  \label{tobechanged2}
\end{eqnarray}
Denote by $\mathcal{S}=\frac{\partial}{\partial \epsilon} \log f_{\epsilon} \Big|_{\epsilon=0}$ the score function (i.e.\ the derivative of the log density/likelihood). Applying identity (1) of \cite{Levy2019}, it follows that
\begin{eqnarray}
\frac{\partial}{\partial \epsilon} f_{\tilde{Y},\epsilon}(y|d,s,x) |_{\epsilon=0}&=&[\mathcal{S}(o)-E[\mathcal{S}(O)|d,s,x]]f_{\tilde{Y}}(y|d,s,x),\label{tobepluggedin1}\\
\frac{\partial}{\partial \epsilon} f_{X,\epsilon}(x) |_{\epsilon=0}&=&[E[\mathcal{S}(O)|x]-E[\mathcal{S}(O)]]f_{X}(x).\label{tobepluggedin2}
\end{eqnarray}
Plugging \eqref{tobepluggedin1} and \eqref{tobepluggedin2} into \eqref{tobechanged1} and \eqref{tobechanged2}, respectively, yields
\begin{eqnarray}
&&\int \int \int y \mathcal{S}(o) f_{\tilde{Y}}(y|d,s,x) dy \frac{ I\{D=d^*,S=1\} f_{D,S}(d,s|x)}{f_{D,S}(d,s|x)} d(d,s) f_{X}(x)dx\notag\\
&-&\int \int \int y E[\mathcal{S}(O)|d,s,x] f_{\tilde{Y}}(y|d,s,x) dy \frac{ I\{D=d^*,S=1\} f_{D,S}(d,s|x)}{f_{D,S}(d,s|x)} d(d,s) f_{X}(x)dx\notag\\
&+& \int\int y  f_{\tilde{Y}}(y|d^*,1,x)dy [E[\mathcal{S}(O)|x]-E[\mathcal{S}(O)]] f_{X}(x)dx\notag\\
&=&\int \int \int y \mathcal{S}(o) f_{\tilde{Y}}(y|d,s,x) dy \frac{ I\{D=d^*,S=1\} f_{D,S}(d,s|x)}{f_{D,S}(d,s|x)} d(d,s) f_{X}(x)dx\notag\\
&-&\int \int E[\tilde{Y}|d,s,x] E[\mathcal{S}(O)|d,s,x] \frac{ I\{D=d^*,S=1\} f_{D,S}(d,s|x)}{f_{D,S}(d,s|x)} d(d,s) f_{X}(x)dx\notag\\
&+& \int E[\tilde{Y}|d^*,1,x] [E[\mathcal{S}(O)|x]-E[\mathcal{S}(O)]] f_{X}(x)dx\notag\\
&=&\int \int \int y \mathcal{S}(o) f_{\tilde{Y}}(y|d,s,x) dy \frac{ I\{D=d^*,S=1\} f_{D,S}(d,s|x)}{f_{D,S}(d,s|x)} d(d,s) f_{X}(x)dx\notag\\
&-&\int \int E[\tilde{Y}|d,s,x] \int \mathcal{S}(o) f_{\tilde{Y}}(y|d,s,x) dy \frac{ I\{D=d^*,S=1\} f_{D,S}(d,s|x)}{f_{D,S}(d,s|x)} d(d,s) f_{X}(x)dx\notag\\
&+& \int E[\tilde{Y}|d^*,1,x] \int \mathcal{S}(o) f_{\tilde{Y,D,S}}(y,d,s|x)d(y,d,s)f_{X}(x)dx-\int \mathcal{S}(o) f(o)do\int E[\tilde{Y}|d^*,1,x] f_{X}(x)dx\notag\\
&=& \int y \mathcal{S}(o)  \frac{ I\{D=d^*,S=1\}}{f_{D,S}(d,s|x)} f(o)do-\int  E[\tilde{Y}|d,s,x]  \mathcal{S}(o) \frac{ I\{D=d^*,S=1\}}{f_{D,S}(d,s|x)} f(o)do \notag\\
&+& \int E[\tilde{Y}|d^*,1,x]\mathcal{S}(o)f(o)do-\int \mathcal{S}(o) \Psi_{d^*}(P) f(o)do \notag\\
&=&\int  \mathcal{S}(o) \left[ \frac{ I\{D=d^*,S=1\}}{f_{D,S}(d,s|x)} (y-E[\tilde{Y}|d,s,x])+E[\tilde{Y}|d^*,1,x]- \Psi_{d^*}(P)\right] f(o)do. \label{innerprod}
\end{eqnarray}
\eqref{innerprod} is an $L_0^2(P)$ inner product of the score $\mathcal{S}$ and the following function, which is thus the efficient influence function:
\begin{eqnarray}
&&\frac{ I\{D=d^*,S=1\}}{f_{D,S}(d,s|x)} [Y-E[\tilde{Y}|D,S,X]]+E[\tilde{Y}|D=d^*,S=1,x]- \Psi_{d^*}(P)\notag\\
&=& \frac{ I\{D=d^*,S=1\}\cdot [Y-E[Y|D=d^*,S=1,X]]}{f_{D,S}(d^*,1|x)} +E[Y|D=d^*,S=1,x]- \Psi_{d^*}(P)\notag\\
&=& \frac{ I\{D=d^*\} \cdot S \cdot [Y-\mu(d^*,1,X)]}{p_{d^*}(X)\cdot \pi(d^*,X)} +\mu(d^*,1,X)- \Psi_{d^*}(P)\notag\\
&=& \psi_{d^*}-\Psi_{d^*}(P), \notag
\end{eqnarray}
with $\psi_{d^*}$ corresponding to \eqref{score} for $d^*=d$.

Analogously, one can define $\Psi_{d0}=\Psi_{d}(P)=E_P[E_P[Y|D=d,S=1,X,\Pi]]$ and show that the efficient influence function corresponds to
$\phi_{d}-\Psi_{d}(P)$, with $\phi_{d}$ defined in \eqref{score3}. This follows straightforwardly from replacing $X$ by $X,\Pi$ everywhere in the previous derivations. In a similar manner, one can demonstrate that for $\Psi_{d}^{S=1}(P)=E_P[E_P[Y|D=d,S=1,X, \Pi]|S=1]$, the efficient influence function corresponds to $\phi_{d,S=1}-\Psi_{d}^{S=1}(P)$,  with $\phi_{d,S=1}$ defined in \eqref{score2}. This follows  from considering the derivative in the selected population with $S=1$ (rather than the total population) and replacing $D,S$ by $D$ as well as $X$ by $X,\Pi$ everywhere in the previous derivations. The proofs for these cases are thus omitted for the sake of brevity.

For deriving the efficient influence function under the identifying assumptions of Section \ref{dynsel}, let $\tilde{Y}=Y\cdot S$ and denote the observed data by $O=(\tilde{Y}\cdot S, S, D, X, M)\sim P$, with distribution $P$ having the density  $f(o)=f_{\tilde{Y}}(y|d,s,x,m)f_{S}(s|d,x,m) f_{M}(m|d,x)  f_{D}(d|x) f_X(x)$.

Define $\Psi_{d0}=\Psi_{d}(P)=E_P[E_P[E_P[Y|D=d,S=1,X,M]|D=d,X]]=E_P[E_P[E_P[\tilde{Y}|D=d,S=1,X,M]|D=d,X]]$, where the second equality from the fact that $\tilde{Y}=Y$ for $S=1$. We now consider the derivative of $\Psi_{d^*}(P)$ w.r.t.\ the distribution $P$, with $d^*$ $\in$ $\{0,1,...,Q\}$:

\begin{eqnarray}
&&\frac{\partial}{\partial \epsilon} \Big|_{\epsilon=0} \Psi_{d^*}(P_{\epsilon})=E_{P_{\epsilon}}[E_{P_{\epsilon}}[E_{P_{\epsilon}}[\tilde{Y}|D=d^*,S=1,X, M]|D=d^*,X]]\notag\\
&=&\int\int\int y \frac{\partial}{\partial \epsilon}\Big|_{\epsilon=0} (f_{\tilde{Y},\epsilon}(y|d^*,1,x,m)dy f_{M,\epsilon}(m|d,x)dm f_{X,\epsilon}(x))dx\notag\\
&=&\int\int\int y \frac{\partial}{\partial \epsilon}\Big|_{\epsilon=0} f_{\tilde{Y},\epsilon}(y|d^*,1,x)dy f_{M}(m|d,x)dm f_{X}(x)dx \notag\\
&+& \int\int\int y  f_{\tilde{Y}}(y|d^*,1,x,m)dy  \frac{\partial}{\partial \epsilon}\Big|_{\epsilon=0} f_{M,\epsilon}(m|d,x)dm f_{X}(x)dx  \notag\\
&+& \int\int\int y  f_{\tilde{Y}}(y|d^*,1,x,m)dy  f_{M}(m|d,x)dm\frac{\partial}{\partial \epsilon}dm\Big|_{\epsilon=0} f_{X,\epsilon}(x)dx  \notag\\
&=& \int \int \int \int\int y \frac{\partial}{\partial \epsilon}\Big|_{\epsilon=0}   f_{\tilde{Y},\epsilon}(y|d,s,x,m)dy\frac{ I\{D=d^*,S=1\} f_{S}(s|d,x,m)}{f_{S}(s|d,x,m)}ds f_{M}(m|d,x)dm \frac{f_{D}(d|x)}{f_{D}(d|x)} dd f_{X}(x)dx\notag\\ \label{tobechanged3}\\
&+& \int\int\int\int y  f_{\tilde{Y}}(y|d,1,x,m)dy  \frac{\partial}{\partial \epsilon}\Big|_{\epsilon=0} f_{M,\epsilon}(m|d,x)dm \frac{I\{D=d^*\} f_{D}(d|x)}{f_{D}(d|x)} dd f_{X}(x)dx  \label{tobechanged4}\\
&+& \int\int\int y  f_{\tilde{Y}}(y|d^*,1,x,m)dy  f_{M}(m|d,x)dm \frac{\partial}{\partial \epsilon}dm\Big|_{\epsilon=0} f_{X,\epsilon}(x)dx  \label{tobechanged5}.
\end{eqnarray}

Denote by $\mathcal{S}=\frac{\partial}{\partial \epsilon} \log f_{\epsilon} \Big|_{\epsilon=0}$ the score function (i.e.\ the derivative of the log density/likelihood). Applying identity (1) of \cite{Levy2019}, it follows that
\begin{eqnarray}
\frac{\partial}{\partial \epsilon} f_{\tilde{Y},\epsilon}(y|d,s,x,m) |_{\epsilon=0}&=&[\mathcal{S}(o)-E[\mathcal{S}(O)|d,s,x,m]]f_{\tilde{Y}}(y|d,s,x,m),\label{tobepluggedin3}\\
\frac{\partial}{\partial \epsilon} f_{M,\epsilon} (m|d,x) |_{\epsilon=0} &=&  [E[\mathcal{S}(O)|m,d,x]-E[\mathcal{S}(O)| d,x]] f_{M} (m|d,x),\label{tobepluggedin4}\\
\frac{\partial}{\partial \epsilon} f_{X,\epsilon}(x) |_{\epsilon=0}&=&[E[\mathcal{S}(O)|x]-E[\mathcal{S}(O)]]f_{X}(x).\label{tobepluggedin5}
\end{eqnarray}
Plugging \eqref{tobepluggedin3}, \eqref{tobepluggedin4}, and \eqref{tobepluggedin5} into \eqref{tobechanged3}, \eqref{tobechanged4}, and \eqref{tobechanged5}, respectively, yields
\begin{eqnarray}
&&\int \int \int \int \int y \mathcal{S}(o) f_{\tilde{Y}}(y|d,s,x,m) dy \frac{ I\{D=d^*,S=1\} f_{S}(s|d,x,m)}{f_{S}(s|d,x,m)}ds f_{M}(m|d,x)dm \frac{f_{D}(d|x)}{f_{D}(d|x)} dd f_{X}(x)dx\notag\\
&-& \int \int \int \int \int y E[\mathcal{S}(O)|d,s,x,m] f_{\tilde{Y}}(y|d,s,x,m) dy \frac{ I\{D=d^*,S=1\} f_{S}(s|d,x,m)}{f_{S}(s|d,x,m)}ds f_{M}(m|d,x)dm \frac{f_{D}(d|x)}{f_{D}(d|x)} dd f_{X}(x)dx \notag\\
&+& \int\int\int \int y  f_{\tilde{Y}}(y|d,1,x,m)dy [E[\mathcal{S}(O)|m,d,x]-E[\mathcal{S}(O)| d,x]] f_{M} (m|d,x) dm \frac{I\{D=d^*\} f_{D}(d|x)}{f_{D}(d|x)} dd f_{X}(x)dx \notag\\
&+& \int\int \int y  f_{\tilde{Y}}(y|d^*,1,x,m) dy f_{M}(m|d,x) dm  [E[\mathcal{S}(O)|x]-E[\mathcal{S}(O)]] f_{X}(x)dx\notag\\
&=&\int \int \int \int \int y \mathcal{S}(o) f_{\tilde{Y}}(y|d,s,x,m) dy \frac{ I\{D=d^*,S=1\} f_{S}(s|d,x,m)}{f_{S}(s|d,x,m)}ds f_{M}(m|d,x)dm \frac{f_{D}(d|x)}{f_{D}(d|x)} dd f_{X}(x)dx\notag\\
&-&\int \int \int \int  E[\tilde{Y}|d,s,x,m] E[\mathcal{S}(O)|d,s,x,m] \frac{ I\{D=d^*,S=1\} f_{S}(s|d,x,m)}{f_{S}(s|d,x,m)}ds f_{M}(m|d,x)dm \frac{f_{D}(d|x)}{f_{D}(d|x)} dd f_{X}(x)dx \notag\\
&+& \int\int \int E[\tilde{Y}|d,1,x,m] [E[\mathcal{S}(O)|m,d,x]-E[\mathcal{S}(O)| d,x]] f_{M} (m|d,x) dm \frac{I\{D=d^*\} f_{D}(d|x)}{f_{D}(d|x)} dd f_{X}(x)dx \notag\\
&+& \int\int E[\tilde{Y}|d^*,1,x,m] f_{M}(m|d,x) dm  [E[\mathcal{S}(O)|x]-E[\mathcal{S}(O)]] f_{X}(x)dx\notag\\
&=&\int \int \int \int \int y \mathcal{S}(o) f_{\tilde{Y}}(y|d,s,x,m) dy \frac{ I\{D=d^*,S=1\} f_{S}(s|d,x,m)}{f_{S}(s|d,x,m)}ds f_{M}(m|d,x)dm \frac{f_{D}(d|x)}{f_{D}(d|x)} dd f_{X}(x)dx\notag\\
&-&\int \int \int \int   E[\tilde{Y}|d,s,x,m] \int S(o) f_{\tilde{Y}}(y|d,s,x,m)dy  \frac{ I\{D=d^*,S=1\} f_{S}(s|d,x,m)}{f_{S}(s|d,x,m)}ds f_{M}(m|d,x)dm \frac{f_{D}(d|x)}{f_{D}(d|x)} dd f_{X}(x)dx \notag\\
&+& \int\int \int E[\tilde{Y}|d,1,x,m]  \int S(o) f_{\tilde{Y}}(y,s|d,x,m) d(y,s) f_{M} (m|d,x) dm \frac{I\{D=d^*\} f_{D}(d|x)}{f_{D}(d|x)} dd f_{X}(x)dx\notag\\ &-&  \int \int\int E[\tilde{Y}|d,1,x,m]  f_{M} (m|d,x) dm   \int S(o) f_{\tilde{Y}}(y,s, m|d,x) d(y,s,m) \frac{I\{D=d^*\} f_{D}(d|x)}{f_{D}(d|x)} dd f_{X}(x)dx \notag\\
&+& \int\int E[\tilde{Y}|d^*,1,x,m] f_{M}(m|d,x) dm  \int S(o) f_{\tilde{Y}}(y,d,s, m|x) d(y,d,s,m) f_{X}(x)dx\notag\\ &-& \int \mathcal{S}(o)f(o)do \int\int E[\tilde{Y}|d^*,1,x,m] f_{M}(m|d,x) dm f_{X}(x)dx  \notag\\
&=&\int y\mathcal{S}(o) \frac{ I\{D=d^*,S=1\}}{f_{S}(s|d,x,m)\cdot f_{D}(d|x)} f(o)do - \int E[\tilde{Y}|d,s,x,m] \frac{ I\{D=d^*,S=1\}}{f_{S}(s|d,x,m)\cdot f_{D}(d|x)} f(o)do \notag\\
&+& \int E[\tilde{Y}|d,1,x,m]  S(o) \frac{I\{D=d^*\}}{f_{D}(d|x)} f(o)do -  \int  E[E[\tilde{Y}|d,1,x,m]| d,x] S(o) \frac{I\{D=d^*\}}{f_{D}(d|x)}  f(o)do \notag\\
&+& \int E[E[\tilde{Y}|d^*,1,x,m]| d^*,x] S(o) f(o)do -\int \mathcal{S}(o) \Psi_{d^*}(P) f(o)do \notag\\
&=&  \int \mathcal{S}(o) \left[\frac{ I\{D=d^*,S=1\}}{f_{S}(s|d,x,m)\cdot f_{D}(d|x)} (y- E[\tilde{Y}|d,s,x,m] ) \right.\notag\\
&+&\left.  \frac{I\{D=d^*\}}{f_{D}(d|x)} (E[\tilde{Y}|d,1,x,m]- E[E[\tilde{Y}|d,1,x]| d,x]) + E[E[\tilde{Y}|d^*,1,x]| d^*,x]- \Psi_{d^*}(P) \right]   f(o)do  .   \label{innerprod2}
\end{eqnarray}
\eqref{innerprod2} is an $L_0^2(P)$ inner product of the score $\mathcal{S}$ and the following function, which is thus the efficient influence function:
\begin{eqnarray}
&&\frac{ I\{D=d^*,S=1\}}{f_{S}(s|D,X,M)\cdot f_{D}(d|X)} (y- E[\tilde{Y}|D,S,X,X] ) \notag\\
&+& \frac{I\{D=d^*\}}{f_{D}(d|X)} (E[\tilde{Y}|D,S=1,X,M]- E[E[\tilde{Y}|D,S=1,X]| D,X]) + E[E[\tilde{Y}|D=d^*,S=1,X]| D=d^*,X]- \Psi_{d^*}(P) \notag\\
&=&  \frac{I\{D=d^*\} \cdot S \cdot [Y- \mu(d^*,1,X,M)] }{ p_{d^*}(X)\cdot  \pi (d^*,X,M)} + \frac{I\{D=d^*\}\cdot  [\mu (d^*,1,X,M)-\nu(d^*,1,X)]}{p_{d^*}(X)} +\nu(d^*,1,X) ]- \Psi_{d^*}(P)\notag\\
&=&\psi_{d^*}-\Psi_{d^*}(P),\notag
\end{eqnarray}
with $\psi_{d^*}$ corresponding to \eqref{scoredyn} for $d^*=d$.


}

\section{Descriptive Statistics}\label{desstat}
In the following tables we report descriptive statistics, namely the number of observations and means by treatment groups,  for a selected set of pre-treatment covariates $X$ measured at Job Corps assignment and post-treatment covariates $M$ measured in the second and third year after assignment.\footnote{Additional descriptive statistics are available upon request from the authors.} 
\pagebreak

{\tiny
\begin{tabular}{@{\extracolsep{-1pt}}lcc}
\\[-1.8ex]\hline
\hline \\[-1.8ex]
variable name ($X$) & \multicolumn{1}{c}{N} & \multicolumn{1}{c}{any train} \\
\hline \\[-1.8ex]
no child at random assignment & 1,673 & 0.699  \\
pregnant at random assignment & 1,673 & 0.008  \\
sample member does not contribute to rent & 1,673 & 0.686 \\
 & & \\
\textit{diploma at random assignment} & & \\
no HS diploma  & 1,673 & 0.757 \\
no GED & 1,673 & 0.954  \\
no other degree & 1,673 & 0.980  \\
& & \\
\textit{job and training in the past year}  &  &  \\
no training & 1,673 & 0.325  \\
full time or part Time  & 1,673 & 0.230  \\
no job & 1,673 & 0.393  \\
stayed in most recent job & 1,673 & 0.188  \\
& & \\
\textit{public assistance in the past year } &  &  \\
no public assistance & 1,673 & 0.302  \\
no AFDC  & 1,673 & 0.536  \\
no other welfare  & 1,673 & 0.673 \\
no food stamps & 1,673 & 0.445 \\
 did not have health problems that limited work & 1,673 & 0.944  \\
 & & \\
 \textit{use of drugs and alcohol in the past year} &  &  \\
 no use of alcohol & 1,673 & 0.470  \\
 no use of marijuana & 1,673 & 0.671  \\
 no use of cocaine & 1,673 & 0.962  \\
 no use of crack & 1,673 & 0.980  \\
 no use of heroin & 1,673 & 0.985  \\
  & & \\
\textit{arrested or convicted in past years (at least once)} &  &  \\
 arrested & 1,673 & 0.836  \\
 convicted for a crime & 1,673 & 0.895 \\
 convicted for murder or Assault & 1,673 & 0.952  \\
 convicted for robbery & 1,673 & 0.959  \\
 convicted for burglary & 1,673 & 0.958  \\
 convicted for larceny & 1,673 & 0.920  \\
 convicted for drug violation & 1,673 & 0.959  \\
 & & \\
  &  &  \\
 \textit{no arrest charges pending at random assignment} & 1,673 & 0.969 \\
\hline \\[-1.8ex]
\end{tabular}
\quad
\begin{tabular}{@{\extracolsep{-1pt}}lcc}
\\[-1.8ex]\hline
\hline \\[-1.8ex]
\multicolumn{1}{c}{N} & \multicolumn{1}{c}{no train}  \\
\hline \\[-1.8ex]
 200 & 0.600 \\
 200 & 0.015  \\
 200 & 0.640  \\
  & & \\
 &  &  \\
 200 & 0.755  \\
200 & 0.960  \\
 200 & 0.965  \\
 & & \\
 &  &  \\
200 & 0.400  \\

200 & 0.210 \\
 200 & 0.355 \\
 200 & 0.190  \\
 & & \\
 &  &  \\
 200 & 0.250  \\
200 & 0.500 \\
 200 & 0.700  \\
200 & 0.375  \\
 200 & 0.920 \\
 & & \\
 &  &  \\
 200 & 0.395 \\
 200 & 0.700 \\
 200 & 0.965  \\
200 & 0.985 \\
 200 & 0.990  \\
 & & \\
 &  &  \\

  200 & 0.830  \\
200 & 1.000  \\

 200 & 0.915 \\
200 & 0.915  \\
 200 & 0.915  \\
200 & 0.965  \\
 200 & 0.970  \\
 & & \\
 &  &  \\
 200 & 0.970 \\
\hline \\[-1.8ex]
\end{tabular}
\quad
\begin{tabular}{@{\extracolsep{-1pt}}lcc}
\\[-1.8ex]\hline
\hline \\[-1.8ex]
\multicolumn{1}{c}{N} & \multicolumn{1}{c}{acad.}  \\
\hline \\[-1.8ex]
 830 & 0.700  \\
 830 & 0.007  \\
 830 & 0.699 \\
  & & \\
 &  &  \\
 830 & 0.853 \\
 830 & 0.970  \\

 830 & 0.980  \\
  & & \\
 &  &  \\
830 & 0.301 \\
830 & 0.270  \\
 830 & 0.430 \\
 830 & 0.184  \\
  & & \\
 &  &  \\
   830 & 0.278  \\
 830 & 0.520  \\
 830 & 0.645  \\
 830 & 0.413  \\
 830 & 0.940  \\
  & & \\
 &  &  \\
 830 & 0.522  \\
830 & 0.671  \\
 830 & 0.964  \\
 830 & 0.977  \\
 830 & 0.981  \\
  & & \\
 &  &  \\
 830 & 0.810  \\
830 & 0.964  \\

 830 & 0.881 \\
 830 & 0.881  \\
 830 & 0.881  \\
 830 & 0.936  \\
830 & 0.948 \\
 & & \\
 &  &  \\
  830 & 0.946  \\

 \hline \\[-1.8ex]
\end{tabular}
\quad
\begin{tabular}{@{\extracolsep{-1pt}}lcc}
\\[-1.8ex]\hline
\hline \\[-1.8ex]
\multicolumn{1}{c}{N} & \multicolumn{1}{c}{voc.}  \\
\hline \\[-1.8ex]
 843 & 0.698  \\
 843 & 0.009  \\

 843 & 0.674  \\
  & & \\
 &  &  \\
843 & 0.662 \\
 843 & 0.938 \\
 843 & 0.966  \\
& & \\
 &  &  \\
 843 & 0.348  \\
 843 & 0.190 \\
 843 & 0.357 \\
 843 & 0.192  \\
 & & \\
 &  &  \\
 843 & 0.326  \\
 843 & 0.550 \\
 843 & 0.701 \\
 843 & 0.476  \\
843 & 0.948 \\
& & \\
 &  &  \\
 843 & 0.419  \\
843 & 0.671  \\
 843 & 0.960  \\
 843 & 0.982 \\
 843 & 0.989  \\
  & & \\
 &  &  \\
 843 & 0.861  \\
  843 & 0.985  \\

 843 & 0.910  \\
843 & 0.910  \\
 843 & 0.910 \\
 843 & 0.967  \\
 843 & 0.970  \\
& & \\
 &  &  \\
 843 & 0.973  \\
\hline \\[-1.8ex]
\end{tabular}
}

\pagebreak

{\tiny
\begin{tabular}{@{\extracolsep{-1pt}}lcc}
\\[-1.8ex]\hline
\hline \\[-1.8ex]
variable name ($X$) & \multicolumn{1}{c}{N} & \multicolumn{1}{c}{any train} \\
\hline \\[-1.8ex]
\textit{reason for joining JC program}& &  \\
 to get  away from home & 1,673 & 0.611  \\
 to get  away from community & 1,673 & 0.644 \\
 to be trained & 1,673 & 0.987  \\
 for career & 1,673 & 0.995 \\
 to get GED & 1,673 & 0.962 \\
 because unemployed & 1,673 & 0.929  \\
 other  reason & 1,673 & 0.759 \\
  & &  \\
 \textit{expectations about JC program} & &  \\
JC improves math & 1,673 & 0.745 \\
JC improves reading & 1,673 & 0.586 \\
JC improves network & 1,673 & 0.640  \\
JC improves self control & 1,673 & 0.595  \\
JC improves self esteem & 1,673 & 0.634  \\
JC expected to train for specific jobs & 1,673 & 0.957  \\
JC leads to new friendship & 1,673 & 0.715  \\
& &  \\
training received last week before random assignment & 1,673 & 0.015 \\
worked last week before random assignment & 1,673 & 0.209  \\
& &  \\
\textit{no welfare receipt}  & & \\
month1 - year before random assignment  & 1,673 & 0.345 \\
month2 - year before random assignment  & 1,673 & 0.340 \\
month3 - year before random assignment  & 1,673 & 0.345 \\
month4 - year before random assignment  & 1,673 & 0.348  \\
month5 - year before random assignment  & 1,673 & 0.344 \\
month6 - year before random assignment & 1,673 & 0.343  \\
month7 - year before random assignment & 1,673 & 0.340  \\
month8 - year before random assignment & 1,673 & 0.338  \\
month9 - year before random assignment & 1,673 & 0.337  \\
month10 - year before random assignment & 1,673 & 0.338  \\
month11 - year before random assignment & 1,673 & 0.340  \\
month12 - year before random assignment & 1,673 & 0.340 \\
had a job at random assignment & 1,673 & 0.791  \\
 & &  \\
\textit{support received by friends and parents for attending JC} & &  \\
encouraged by parents to attend JC & 1,673 & 0.983 \\
encouraged by relatives to attend JC & 1,673 & 0.981  \\
encouraged by friends to attend JC & 1,673 & 0.956  \\
encouraged by teacher to attend JC & 1,673 & 0.991  \\
encouraged by case worker to attend JC & 1,673 & 0.999  \\
encouraged by officer to attend JC & 1,673 & 0.999  \\
\hline \\[-1.8ex]
\end{tabular}
\quad
\begin{tabular}{@{\extracolsep{-1pt}}lcc}
\\[-1.8ex]\hline
\hline \\[-1.8ex]
\multicolumn{1}{c}{N} & \multicolumn{1}{c}{no train}  \\
\hline \\[-1.8ex]
 & &  \\
 200 & 0.585  \\
 200 & 0.605  \\
200 & 0.980 \\
200 & 1.000  \\
 200 & 0.965 \\
 200 & 0.915  \\
200 & 0.745 \\
& &  \\
 & &  \\
 200 & 0.770  \\
 200 & 0.595 \\
 200 & 0.645 \\
 200 & 0.595  \\
200 & 0.605  \\
200 & 0.970 \\
 200 & 0.690  \\
 & &  \\
200 & 0.015  \\
 200 & 0.215  \\
 & &  \\
   & & \\
 200 & 0.280  \\
  200 & 0.28 \\
200 & 0.280 \\
 200 & 0.280  \\
 200 & 0.280  \\
 200 & 0.275  \\
 200 & 0.275  \\
 200 & 0.270 \\
 200 & 0.270 \\
 200 & 0.270  \\
200 & 0.275  \\
 200 & 0.290  \\
 200 & 0.785  \\
  & &  \\
  & &  \\
 200 & 0.965 \\
 200 & 0.975 \\
 200 & 0.965  \\
 200 & 0.985  \\
200 & 0.995 \\
 200 & 1.000  \\

\hline \\[-1.8ex]
\end{tabular}
\quad
\begin{tabular}{@{\extracolsep{-1pt}}lcc}
\\[-1.8ex]\hline
\hline \\[-1.8ex]
\multicolumn{1}{c}{N} & \multicolumn{1}{c}{acad.}  \\
\hline \\[-1.8ex]
& &  \\
 830 & 0.623  \\
 830 & 0.710  \\
 830 & 0.987  \\
 830 & 0.994  \\
 830 & 0.039  \\
 830 & 0.929  \\
 830 & 0.761  \\
 & &  \\
  & &  \\
 830 & 0.801  \\
830 & 0.649 \\
 830 & 0.664 \\
 830 & 0.625 \\
 830 & 0.655  \\
 830 & 0.954  \\
  830 & 0.700  \\
 & &  \\
 830 & 0.724 \\
 830 & 0.018  \\
& &  \\
& & \\
830 & 0.313  \\
 830 & 0.313  \\
 830 & 0.310 \\
 830 & 0.312  \\
830 & 0.308  \\
 830 & 0.312 \\
 830 & 0.307 \\
 830 & 0.310  \\
 830 & 0.310  \\
 830 & 0.311  \\
 830 & 0.307  \\
830 & 0.311  \\
 830 & 0.787 \\
  & &  \\
  & &  \\
 830 & 0.980  \\
 830 & 0.981  \\
 830 & 0.947  \\
 830 & 0.988  \\
 830 & 0.998  \\
 830 & 1.000  \\
\hline \\[-1.8ex]
\end{tabular}
\quad
\begin{tabular}{@{\extracolsep{-1pt}}lcc}
\\[-1.8ex]\hline
\hline \\[-1.8ex]
\multicolumn{1}{c}{N} & \multicolumn{1}{c}{voc.}  \\
\hline \\[-1.8ex]
& &  \\
 843 & 0.600  \\
 843 & 0.580 \\
 843 & 0.988 \\
843 & 0.996 \\
 843 & 0.963  \\
 843 & 0.929 \\
 843 & 0.756  \\
 & &  \\
  & &  \\
 843 & 0.690 \\
 843 & 0.523  \\
 843 & 0.616 \\
 843 & 0.566  \\
843 & 0.612  \\
 843 & 0.960  \\
 843 & 0.706  \\
 & &  \\
 843 & 0.012  \\
 843 & 0.205  \\
 & &  \\
   & & \\
 843 & 0.376  \\
 843 & 0.375  \\
 843 & 0.377  \\
 843 & 0.383  \\
 843 & 0.378 \\
 843 & 0.374  \\
843 & 0.372 \\
 843 & 0.367 \\
 843 & 0.363  \\
 843 & 0.365  \\
843 & 0.371 \\
843 & 0.368  \\
 843 & 0.795  \\
 & &  \\
  & &  \\
 843 & 0.986  \\
 843 & 0.982 \\
 843 & 0.964  \\
 843 & 0.994 \\
 843 & 1.000 \\
 843 & 0.999  \\
\hline \\[-1.8ex]
\end{tabular}
}

\pagebreak

{\tiny
\begin{tabular}{@{\extracolsep{-1pt}}lcc}
\\[-1.8ex]\hline
\hline \\[-1.8ex]
variable name ($M$)  & \multicolumn{1}{c}{N} & \multicolumn{1}{c}{any train} \\
\hline \\[-1.8ex]
\textit{did not get unemployment benefits}& & \\
week1 (1 year after assignment) & 1,673 & 0.997 \\
week8 & 1,673 & 0.998 \\
week18 & 1,673 & 0.998 \\
week28 & 1,673 & 0.997 \\
week38 & 1,673 & 0.998 \\
week48 & 1,673 & 0.998 \\
week52 & 1,673 & 0.999 \\
& & \\
\textit{not in JC} & & \\
week1 (1 year after assignment)  & 1,673 & 0.860 \\
week8 & 1,673 & 0.946 \\
week18 & 1,673 & 0.931 \\
week28 & 1,673 & 0.910 \\
week38 & 1,673 & 0.899 \\
week48  & 1,673 & 0.884 \\
week52  & 1,673 & 0.880 \\
& & \\
\textit{not in a drug treatment program} & & \\
week1 (1 year after assignment)  & 1,673 &  0.999\\
week8 & 1,673 & 0.997 \\
week18 & 1,673 & 0.996 \\
week28 & 1,673 & 0.997\\
week38 & 1,673 & 0.997 \\
week48  & 1,673 & 0.995\\
week52  & 1,673 & 0.994 \\
& & \\
\hline \\[-1.8ex]
\end{tabular}
\quad
\begin{tabular}{@{\extracolsep{-1pt}}lcc}
\\[-1.8ex]\hline
\hline \\[-1.8ex]
\multicolumn{1}{c}{N} & \multicolumn{1}{c}{no train} \\
\hline \\[-1.8ex]
& & \\
200 & 1.000 \\
200 & 0.995 \\
200 & 0.995 \\
200 & 0.995 \\
200 & 0.995 \\
200 & 0.995 \\
200 & 0.995 \\
& & \\
& & \\
200 & 0.830 \\
200 & 0.835 \\
200 & 0.785 \\
200 & 0.750 \\
200 & 0.780 \\
200 &  0.800 \\
200 & 0.785 \\
& & \\
& & \\
200 & 1.000 \\
200 & 0.995 \\
200 & 1.000 \\
200 & 1.000 \\
200 & 1.000 \\
200 & 1.000 \\
200 & 1.000 \\
& & \\
\hline \\[-1.8ex]
\end{tabular}
\quad
\begin{tabular}{@{\extracolsep{-1pt}}lcc}
\\[-1.8ex]\hline
\hline \\[-1.8ex]
\multicolumn{1}{c}{N} & \multicolumn{1}{c}{acad.} \\
\hline \\[-1.8ex]
& & \\
830 & 0.996 \\
830 & 0.999 \\
830 & 0.999 \\
830 & 0.999 \\
830 & 0.999 \\
830 & 0.996 \\
830 & 0.999 \\
& & \\
& & \\
830 & 0.847 \\
830 & 0.931 \\
830 & 0.905 \\
830 & 0.867 \\
830 & 0.866 \\
830 & 0.852 \\
830 & 0.853 \\
& & \\
& & \\
830 & 1.000 \\
830 & 0.997 \\
830 & 0.996 \\
830 & 0.997\\
830 & 0.996 \\
830 & 0.993\\
830 & 0.991\\
& & \\
\hline \\[-1.8ex]
\end{tabular}
\quad
\begin{tabular}{@{\extracolsep{-1pt}}lcc}
\\[-1.8ex]\hline
\hline \\[-1.8ex]
\multicolumn{1}{c}{N} & \multicolumn{1}{c}{voc.} \\
\hline \\[-1.8ex]
& & \\
843 & 0.998 \\
843 & 0.998 \\
843 & 0.998 \\
843 & 0.996 \\
843 & 0.999 \\
843 & 1.000 \\
843 & 1.000 \\
& & \\
& & \\
843 & 0.873 \\
843 & 0.962 \\
843 & 0.957 \\
843 & 0.951 \\
843 & 0.931 \\
843 & 0.916 \\
843 & 0.906 \\
& & \\
& & \\
843 & 0.998 \\
843 & 0.997\\
843 & 0.996 \\
843 & 0.996 \\
843 & 0.998 \\
843 & 0.997\\
843 & 0.996 \\
& & \\
\hline \\[-1.8ex]
\end{tabular}
}

\pagebreak

{\tiny
\begin{tabular}{@{\extracolsep{-1pt}}lcc}
\\[-1.8ex]\hline
\hline \\[-1.8ex]
variable name ($M$)  & \multicolumn{1}{c}{N} & \multicolumn{1}{c}{any train} \\
\hline \\[-1.8ex]
\textit{earnings}& & \\
week53 (2 years after assignment) & 1,673 &  73.875\\
week60 & 1,673 & 87.341\\
week70 & 1,673 & 100.763\\
week80 & 1,673 & 114.549\\
week90 & 1,673 &  127.213\\
week100 & 1,673 &  131.921\\
week104 & 1,673 &  136.584 \\
& & \\
\textit{hours worked} & & \\
week53 (2 years after assignment)  & 1,673 &  12.952\\
week60 & 1,673 & 14.339\\
week70 & 1,673 & 15.764\\
week80 & 1,673 & 17.552\\
week90 & 1,673 & 19.345\\
week100  & 1,673 & 20.145\\
week104  & 1,673 & 20.699\\
& & \\
\textit{employed} & & \\
week53 (2 years after assignment)  & 1,673 & 0.341\\
week60 & 1,673 & 0.354 \\
week70 & 1,673 & 0.385 \\
week80 & 1,673 & 0.429\\
week90 & 1,673 & 0.468 \\
week100  & 1,673 & 0.481\\
week104  & 1,673 & 0.497\\
& & \\
\hline \\[-1.8ex]
\end{tabular}
\quad
\begin{tabular}{@{\extracolsep{-1pt}}lcc}
\\[-1.8ex]\hline
\hline \\[-1.8ex]
\multicolumn{1}{c}{N} & \multicolumn{1}{c}{no train} \\
\hline \\[-1.8ex]
& & \\
200 &  70.782\\
200 & 83.211\\
200 &  97.380\\
200 & 95.989\\
200 & 93.454\\
200 & 100.158\\
200 & 102.975\\
& & \\
& & \\
200 & 12.778\\
200 & 14.600\\
200 & 15.646\\
200 & 15.450\\
200 & 16.177\\
200 &  16.925\\
200 &  17.418\\
& & \\
& & \\
200 & 0.335 \\
200 & 0.350 \\
200 & 0.415\\
200 & 0.380 \\
200 & 0.400 \\
200 & 0.430 \\
200 & 0.440 \\
& & \\
\hline \\[-1.8ex]
\end{tabular}
\quad
\begin{tabular}{@{\extracolsep{-1pt}}lcc}
\\[-1.8ex]\hline
\hline \\[-1.8ex]
\multicolumn{1}{c}{N} & \multicolumn{1}{c}{acad.} \\
\hline \\[-1.8ex]
& & \\
830 & 64.641\\
830 &  74.509\\
830 & 86.827\\
830 & 99.868\\
830 & 104.594\\
830 & 111.555\\
830 & 115.530\\
& & \\
& & \\
830 & 11.592\\
830 & 12.388\\
830 & 13.742\\
830 & 15.698\\
830 & 16.319\\
830 & 17.596\\
830 & 18.248\\
& & \\
& & \\
830 & 0.308 \\
830 & 0.304 \\
830 & 0.337 \\
830 & 0.380\\
830 & 0.395 \\
830 & 0.413\\
830 & 0.432\\
& & \\
\hline \\[-1.8ex]
\end{tabular}
\quad
\begin{tabular}{@{\extracolsep{-1pt}}lcc}
\\[-1.8ex]\hline
\hline \\[-1.8ex]
\multicolumn{1}{c}{N} & \multicolumn{1}{c}{voc.} \\
\hline \\[-1.8ex]
& & \\
843 & 82.965\\
843 & 99.974\\
843 & 114.484\\
843 & 129.003\\
843 &  149.483\\
843 & 151.973\\
843 & 157.313\\
& & \\
& & \\
843 & 14.291\\
843 & 16.259\\
843 & 17.755\\
843 & 19.378\\
843 & 22.325\\
843 & 22.654\\
843 & 23.112\\
& & \\
& & \\
843 & 0.374 \\
843 & 0.403\\
843 & 0.432 \\
843 & 0.476 \\
843 & 0.539 \\
843 & 0.548\\
843 & 0.561 \\
& & \\
\hline \\[-1.8ex]
\end{tabular}
}

\end{appendix}

\end{document}